\documentclass[superscriptaddress, twocolumn, prx, longbibliography]{revtex4-2}
\usepackage{libertine}
\usepackage{color, amsfonts, amsmath, graphicx, tikz, amssymb, mathtools}
\usepackage{amsthm}
\usepackage[libertine, cmintegrals, bigdelims, vvarbb]{newtxmath}
\usepackage{appendix}
\usepackage{comment}
\usepackage{upgreek}
\definecolor{linkcolor}{rgb}{0,0,0.6}

\definecolor{lgreen} {RGB}{180,210,100}
\definecolor{dblue}  {RGB}{20,66,129}
\definecolor{jblue}  {RGB}{20,50,100}
\definecolor{nblue}  {RGB}{0,120,200}
\definecolor{dgreen} {RGB}{78,138,21}
\definecolor{ngreen} {RGB}{98,158,31}
\definecolor{lred}   {RGB}{220,0,0}
\definecolor{nred}   {RGB}{224,0,0}
\usepackage[pdftex,colorlinks=true,
	pdfstartview = FitV,
	linkcolor    = red,
	citecolor    = blue,
	urlcolor     = blue,	
	hyperindex   = true,
	hyperfigures = false]{hyperref}
\usepackage{cleveref}
\newcommand*{\fullref}[1]{\hyperref[{#1}]{\Cref*{#1} \nameref*{#1}}}

\usepackage{array}
\newcolumntype{P}[1]{>{\centering\arraybackslash}p{#1}}

\newcommand{\anh}{\hat{\eta}_{i}}

\newcommand{\dens}{N_{i}^\#}

\newcommand{\bra}{|\dens \rangle} 
\newcommand{\ket}{\langle \dens|}

\newcommand{\braplus}{|\dens + 1\rangle} 
 
\newcommand{\braminus}{|\dens - 1 \rangle}

\newcommand{\crtlong}[2]{{(\hat{\eta}_{#1}^{#2})}^\dag} 
\newcommand{\anhlong}[2]{\hat{\eta}_{#1}^{#2}}

\newcommand{\densoplong}[1]{\hat{N}_{#1}} 
\newcommand{\densoplongg}[2]{{{\hat{N}}_{#1}}^{#2}}

\patchcmd{\subsubsection}{\itshape}{\bfseries}{}{}
\usepackage{multirow}
\usepackage{natbib}
\usepackage{orcidlink}
\usepackage{wasysym}
\usepackage{starfont}

\begin{document}

\title{ Thermodynamically Consistent
Coarse-graining: 
\\ from Interacting Particles
to Fields via Second Quantization }
\author{Atul Tanaji Mohite\,\orcidlink{0009-0004-0059-1127}}
\email{atul.mohite@uni-saarland.de}
\affiliation{Department of Theoretical Physics and Center for Biophysics, Saarland University, Saarbrücken, Germany}

\author{Heiko Rieger\,\orcidlink{0000-0003-0205-3678}}
\affiliation{Department of Theoretical Physics and Center for Biophysics, Saarland University, Saarbrücken, Germany}
\begin{abstract}
We systematically derive an exact coarse-grained description for interacting particles with thermodynamically consistent stochastic dynamics, applicable across different observation scales, the mesoscopic and macroscopic scales. We implement the coarse-graining procedure using the Doi--Peliti field theory, which preserves microscopic noise effects on the mesoscopic/macroscopic scale. The exact mapping reveals the key role played by the Poissonian particle occupancy statistics. We show the implications of the exact coarse-graining method using a prototypical flocking model, namely the active Ising model, which exhibits a mismatch between the microscopic and macroscopic mean-field coarse-grained descriptions. Our analysis shows that the high- and low-density regimes are governed by two different coarse-grained equations. In the low-density regime, noise effects play a prominent role, leading to a first-order phase transition. In contrast, a second-order phase transition occurs in the high-density regime. Due to the exact coarse-graining method, our work also opens up the applicability to systematically analyze noise-induced phase transitions in other models of reciprocally and non-reciprocally interacting particles.
\end{abstract}
\date{\today}
\maketitle
\section{Introduction}
\subsection{Motivation}
The equilibrium mean-field dynamics of microscopic particles have been qualitatively well understood using a coarse-grained macroscopic/mesoscopic field-theoretical description for the order parameter \cite{Chaikin_lubensky_1995,Hohenberg_1977,Cross_1993,Bray_2002,Ramaswamy_2010}. This methodology is based on delineating the order parameter, and it relies on symmetries and conservation laws: \textit{a top-down approach} towards the mean-field dynamics of the order parameter. The field-theoretical physical description of many-body particle systems has been overwhelmingly successful in encapsulating the universal physical properties of different microscopic systems. For example, the critical phenomena in equilibrium systems \cite{Hohenberg_1977}, non-equilibrium mean-field systems such as reaction-diffusion systems \cite{Cross_1993}, and mean-field chemical reaction networks \cite{Beard_2008}. The path-integral formulations that incorporate fluctuations have been studied, namely the Martin–Siggia–Rose functional \cite{martin_1973} and the Bausch–Janssen–Wagner–de Dominicis functional \cite{bausch_1976,janssen_1976,dedominicis_1976}. However, they incorporate close-to-equilibrium Gaussian fluctuations around the mean-field dynamics of the order parameter: again \textit{a top-down approach} towards fluctuations.

Despite its success, the field-theoretical description has major drawbacks \cite{Chaikin_lubensky_1995,Hohenberg_1977,Cross_1993,Bray_2002,Ramaswamy_2010}. First, a field theory contains multiple control parameters without a systematic and transparent connection to the microscopic control parameters of the system under consideration, which are usually only a few. Second, an effective coarse-grained description lumps together microscopic degrees of freedom that do not necessarily have the same dynamical and thermodynamic properties. Therefore, these field theories are usually not thermodynamically consistent because the exact connection to the microscopic system is missing, which impedes the field-theoretic formulation of the stochastic thermodynamics of the system under consideration, in particular, the computation of the \textit{thermodynamic} entropy production \cite{atm_st_nr_2024}. Third, the coarse-grained phenomenological description utilizes the mean-field assumption, which completely ignores the microscopic noise effects. This approximation has been shown to exhibit a large qualitative and quantitative mismatch between the microscopic and coarse-grained descriptions. For example, the prototypical flocking model, the active Ising model (AIM) \cite{Solon2013,Solon2015}. The AIM has a first-order phase transition from the disordered to the ordered phase \cite{Solon2013,Solon2015}. In contrast, the macroscopic coarse-grained description predicts a second-order phase transition from the disordered to the ordered phase \cite{Solon2013,Solon2015}. Incorporating noise effects for the particle occupancy has been shown to successfully bridge the microscopic and coarse-grained physical descriptions in the low-density regime; however, it fails in the high-density regime \cite{Mourtaza2018}. In addition, noise effects play a key role not only in the correctness of the phase diagram but also in the exact quantification of the thermodynamic dissipation across different observation scales. This highlights the importance of incorporating noise effects for a dynamical and thermodynamically consistent description across different observable scales. Fourth, the focus has been on the non-equilibrium dynamics of non-interacting (ideal) particles modelled using ideal reaction-diffusion systems or chemical reaction networks, but the majority of real-world particles are interacting (non-ideal). This implies that a tool similar to reaction-diffusion systems and chemical reaction networks for studying the dynamics and thermodynamics of out-of-equilibrium interacting particles is missing. \textit{Bottom-up approaches} to the coarse-grained description have been explored \cite{Kawasaki_1994,Dean_1996,Bertin_2013,Grossmann_2013,Peshkov_2014}, but are susceptible to one or more of the aforementioned drawbacks.

To remedy the aforementioned four drawbacks, \textit{a bottom-up approach} is necessary, which we present here for a generic class of interacting particle systems, comprising non-equilibrium systems with reciprocal and non-reciprocal interactions and an externally enzymatic/colloidal driven bath generating chemical/self-propulsion driving \cite{atm_st_nr_2024}, that preserves the microscopic noise and the thermodynamical consistency on the macroscale/mesoscale. Here, we systematically elaborate on the technical aspects of the thermodynamically--consistent coarse-graining procedure for the interacting particles and obtain the corresponding mesoscopic/macroscopic Langevin description for the stochastic particle number/density fields.

{ Throughout this work, we define the notions of mesoscopic particle numbers and macroscopic particle densities as coarse-grained degrees of freedom, denoted by $N_i$ and $\rho_i$, respectively. We define the particle densities $\rho_i$ with respect to the parameter $\Omega$ such that $N_i = \Omega \rho_i$; this definition is consistent with large deviation theory \cite{Touchette_2009}. Physically, $\Omega$ quantifies the average number of particles per lattice site. In addition, it serves as a model parameter that ensures the intensiveness of thermodynamic quantities. Therefore, fluctuations of $\rho_i$ and $N_i$ due to microscopic transitions scale as $O(1/\Omega)$ and $O(1)$, respectively. Hence, larger values of $\Omega$ dictate the thermodynamic limit, in which microscopic fluctuations are suppressed; consequently, the macroscopic coarse-grained description correctly captures the physics of the microscopic system. 
In contrast, for smaller values of $\Omega$, microscopic fluctuations play a major role. For this reason, the macroscopic coarse-grained description is insufficient, and one is required to consider an intermediate mesoscopic coarse-grained description obtained for $N_i$ (or equivalently $\Omega = 1$). The necessity of a mesoscopic description is particularly relevant for active matter models and population dynamics models, where the average number of particles per lattice site (defined here as $\Omega$) is finitely small. The definition of $\Omega$ offers the flexibility to define a coarse-grained description at an intermediate scale between the mesoscopic and macroscopic descriptions. In this sense, $\Omega$ is the non-equilibrium coarse-graining analogue of the inverse temperature $\beta$ in equilibrium systems. The limit $\Omega \to \infty$ corresponds to the thermodynamic limit. 
Existing hydrodynamic approaches tend to use the large system volume $\mathcal{V}$ as a parameter equivalent to $\Omega$. However, this overlooks the regime in which the number of particles per lattice site ($\Omega$) remains finitely small, and the impact of microscopic transition fluctuations does not decay as $O(1/\mathcal{V})$. As a result, the resulting coarse-grained hydrodynamic description becomes redundant and fails to correctly capture the underlying microscopic physics.
}

Thermodynamically consistent coarse-graining enables us to identify the Local detailed balance condition (LDB) on the mesoscale/macroscale, which preserves the thermodynamically consistent formulation of the microscopic particles \cite{atm_st_nr_2024}. Moreover, in contrast to simulating the microscopic Master equation, simulating the coarse-grained mesoscopic/macroscopic Langevin equations for the particle number/density greatly decreases the computational requirements, emphasizing the practical application of the coarse-graining methodology. We implement and derive the coarse-graining procedure for the thermodynamically consistent non-reciprocally interacting particles. However, it is easily modified to any system of interacting particles that does not necessarily satisfy the thermodynamic consistency condition, but requires the microscopic noise effects to be exactly' incorporated in the coarse-grained description.
\subsection{Summary of results}
%
%
{
The notion of the mean-field approximation is usually referred to as vanishing `transition noise', which should be more precisely defined as the `mean-field deterministic dynamics approximation'. However, there is also a second notion of the mean-field approximation, which is unnoticed and unexplored. In particular, the treatment of the particle number (in the coarse-grained representation) as a deterministic variable, which fundamentally differs from its microscopic representation as a stochastic variable; the validity of such a mean-field approximation breaks down in the low particle number regimes, which emphasizes a careful treatment. Throughout this work, we predominately focus on addressing the mean-field approximation by exactly incorporating the `occupancy noise'. Particularly, we will focus on the interacting particle systems, for which addressing this approximation is of paramount importance, compared to the non-interacting ideal particle counterparts, where the `occupancy noise' does not matter; therefore, the differences due to this approximation become irrelevant. Further, using the \textit{bottom-up approach} to the microscopic `transition noise', an exact path-integral formulism for the discrete state systems is derived in Refs.\cite{atm_2024_var_epr,atm_2025_var_epr_derivation,atm_2025_gftoc}, and the dynamic and thermodynamic implications of Poisssonian transitions for far-from-equilibrium systems are detailed. The analysis of the `transition noise' is out of the scope of this work.
}

We aim to use the Doi–Peliti field theory (DPFT) technique to implement the coarse-graining procedure \cite{Doi_1976,Doi_1976_2,ZelDovich_1978,rose_1979,grassberger_1980,Mikhailov_1981,Mikhailov_1981_2,Mikhailov_1985,Peliti,Cardy_2008,Wiese_2016,Weber_2017}, which has the potential to exactly quantify the effects of the `occupancy noise' in the coarse-grained description due to its second-quantized nature. Since the DPFT is an exact methodology, our use of the terminology `DPFT for the coarse-graining procedure' sounds contradictory, but should be understood as the DPFT being a `necessary and important' step in the coarse-graining procedure that can quantify the effects of the `occupancy noise' in a coarse-grained description.

\begin{table*}[t!]
\begin{tabular}{ |m{1.8cm}|m{2.1cm}|m{2.3cm}|m{1.1cm}|m{1.9cm}|m{1.7cm}|m{1.5cm}|m{1.5cm}|m{1.8cm}| }
     \hline
     \centering
     \textbf{Level}
     & 
     \multicolumn{2}{|c|}{\textbf{Dynamics}} 
     &
     \centering
     {\textbf{LDB}}
     & \multicolumn{5}{|c|}{\textbf{Thermodynamics}}
     \\
     \hline
     \textbf{} 
     &
     \centering
     \textbf{State}
     \vspace{5pt}
     &
     \centering
     \textbf{Equations of Motion}
     \vspace{2pt}
     & 
     \centering
     & 
     \centering
     \textbf{Interaction coefficients}
     \vspace{2pt}
     & 
     \centering
     \textbf{Boltzmann weight}
     \vspace{2pt}
     & 
     \centering
     \textbf{Reciprocal part } 
     \vspace{2pt}
     &\centering
     \textbf{Non-reciprocal part}
     & 
     \textbf{External reservoir driving}
     \vspace{0pt}
     \\
     \hline
     \centering
     \textbf{Microscopic description}
     \vspace{10pt}
     & 
     \centering
     Multi-particle state probability $P_{ \{ N \} }$ over the discrete-state occupancy $\{N\}$
     & 
     \centering
     Master equation
     \cref{eq:microscopic_master_equation}
     \vspace{10pt}
     &
     \centering
     \cref{eq:microscopic_transition_rate} 
     \vspace{15pt}
     & \centering
     $v_{ij}$
     \vspace{20pt}
     & \centering
     $\epsilon_i^\#$ in \cref{eq:microscopic_reciprocal_interaction_energy_particle}
     \vspace{15pt}
     & \centering
     $\epsilon_i^r$ in \cref{eq:microscopic_reciprocal_interaction_energy_particle}
     \vspace{5pt}
     & \centering 
     $f_i^{nr}$ in \cref{eq:microscopic_reciprocal_interaction_energy_particle}
     \vspace{10pt}
     &
     $f_{\gamma\gamma'}^{ch}$ and $\vec{f}_{i}^{sp}$ in \cref{eq:microscopic_transition_rate}
     \vspace{10pt}
     \\
     \hline
     \centering
     \textbf{Mesoscopic description $\Omega = 1$} 
     \vspace{10pt}
     & \centering
     Particle number $N_i^\#$ is a continuous extensive stochastic variable
     \vspace{0pt}
     & \centering
     Langevin dynamics \cref{eq:mesoscopic_eom_mft}
     \vspace{15pt}
     & \centering
     \cref{eq:mesoscopic_local_detailed_balance_reactive}
     \vspace{20pt}
     & \centering
     $\mathcal{V}_{ij} = g(v_{ij})$ \cref{eq:mesoscopic_interaction_coefficients}
     \vspace{20pt}
     & \centering
     $\upmu_i^\#$ \cref{eq:mesoscopic_boltzmann_weight} 
     \vspace{20pt}
     & \centering
     $\upmu_i^r$ \cref{eq:mesoscopic_boltzmann_weight_decomposition}
     \vspace{20pt}
     & \centering
     $\mathcal{F}_i^{nr}$ in \cref{eq:mesoscopic_boltzmann_weight_decomposition}
     \vspace{15pt}
     & 
     $\mathcal{F}_{\gamma \gamma'}^{ch}$ and $\vec{\mathcal{F}}_i^{sp}$ in \cref{eq:mesoscopic_local_detailed_balance_reactive}
     \vspace{15pt}
     \\
     \hline
     \centering
     \textbf{Macroscopic description $\Omega >> 1$} 
     \vspace{10pt}
     & \centering
     Particle density $\Omega \rho_i = N_i^\#$ is an intensive continuous stochastic variable
     \vspace{0pt}
     & \centering
     Langevin dynamics \cref{eq:macroscopic_eom_mft}
     \vspace{15pt}
     & \centering
     \cref{eq:macroscopic_local_detailed_balance_reactive}
     \vspace{20pt}
     & \centering
     $V_{ij} = g(v_{ij}, \Omega)$ \cref{eq:macroscopic_interaction_coefficients}
     \vspace{20pt}
     & \centering
     $\mu_i$ \cref{eq:macroscopic_boltzmann_weight} 
     \vspace{25pt}
     &
     \centering
     $\mu_i^r$ \cref{eq:macrosocpic_boltzmann_weight_decomposition}
     \vspace{25pt}
     &
     \centering
     $F_i^{nr}$ in \cref{eq:macrosocpic_boltzmann_weight_decomposition}
     \vspace{20pt}
     &
     $F_{\gamma \gamma'}^{ch}$ and $\vec{F}_i^{sp}$ in \cref{eq:macroscopic_local_detailed_balance_reactive}
     \vspace{20pt}
     \\
     \hline
\end{tabular}
\caption{ \textit{Coarse-graining diagram}. \textendash  
The diagram illustrates three distinct regimes: microscopic, mesoscopic, and macroscopic. At the microscopic level, the probability of a lattice configuration $\{N\}$ is required to define the evolution of particle dynamics. The microscopic configuration space $P(\{N\})$ scales exponentially with the total number of particles. The stochastic mesostate $N_i^\#$ is obtained by counting all microscopic particle configurations that contribute to $N_i^\#$ and are thermodynamically identical due to the particle type $i$ and local lattice site index $\#$. Consequently, the mesoscopic configuration space scales with the number of lattice sites and particle types. The large parameter $\Omega$ governs the mesoscopic-to-macroscopic scaling through $N_i^\# = \Omega \rho_i(\vec{\mathbf{r}})$. In this scaling, $O(1)$ fluctuation effects arising from microscopic transitions that generate $N_i^\# \to N_i^\# + 1$ are suppressed at the macroscopic level, since they become $O(1/\Omega)$ for $\rho_i(\vec{\mathbf{r}})$. The limit $\Omega \to \infty$ corresponds to the thermodynamic limit. The dimension of the macroscopic state space is equal to that of the mesoscopic state space. $\Omega$ is physically equivalent to the average number of particles per lattice site, defined as $\Omega = N^{tot} /N^{lattice}$, the ratio of total microscopic particles divided by the number of lattice points. }
\label{table:summary}
\end{table*}
%

{
However, the existing work on the application of the DPFT methodology has three drawbacks. First, thermodynamic consistency is usually ignored in the application of the DPFT methodology, making it susceptible to the second aforementioned drawback. Second, the DPFT technique has been mainly implemented on non-interacting or ideal particles \cite{Doi_1976,Doi_1976_2,ZelDovich_1978,Peliti,rose_1979,grassberger_1980,Mikhailov_1981,Mikhailov_1981_2,Mikhailov_1985,Cardy_2008,Wiese_2016,Weber_2017}, making it susceptible to the fourth aforementioned drawback. The extension of the DPFT methodology to interacting systems is a technical challenge; therefore, our work is a major technical development of the application of the DPFT methodology that addresses this void. Third, due to a fundamentally incorrect understanding of the DPFT methodology, it has been criticized for its physical interpretation and applicability to classical systems, which is unfortunately the state of the art in the existing works. We address these fundamental misconceptions associated with the DPFT technique.
}

The key novel results obtained with our framework are summarized as follows:
\begin{itemize}
    \item[{\Hygiea}]
    {
    We define the class of interacting microscopic particles that satisfy thermodynamically-consistent dynamics \cref{sec:microscopic} [see \cref{eq:microscopic_reciprocal_interaction_energy_particle,eq:microscopic_transition_rate} below], which is the novelty of Ref.\cite{atm_st_nr_2024}, but serves as the starting point of this work. \Cref{sec:cg_dp_field_theory} addresses the first drawback of the application of DPFT methodology and formulates the corresponding `correct' second-quantized description within the DPFT methodology for the thermodynamically-consistent microscopic dynamics of the interacting particles.
    }
    
    \item[{\Hygiea}]
    {
    \Cref{sec:cg_doi_peliti_action} addresses the second drawback of the application of DPFT methodology. In particular, \cref{eq:lagrangian_defination,eq:doi_peliti_path_integral,eq:hamiltonian_reactive_jump_sigle} formulates the `exact' coherent-state path-integral mesoscopic representation for the thermodynamically-consistent microscopic dynamics of the interacting particles.
    }

    \item[{\Hygiea}]
    {
    \Cref{sec:cg_stochastic_action} and \cref{sec:example_equivalence} focus on the third drawback of the application of DPFT methodology, addressing the technical (mathematical) and fundamental (physical) aspects, respectively, and are related to the fundamental physical interpretation of the Cole-Hopf transform [\cref{eq:cole-hopf_transform}], which is a form of `gauge fixing', but is poorly understood in the existing works. In particular, \cref{eq:hamiltonian_reactive_transition_density_noise,eq:relation_hamiltonian_lagrangian_density_noise,eq:doi_peliti_path_integral_density_noise} are the stochastic path-integral formulism (SPIF) of the coherent-state path-integral counterparts \cref{eq:lagrangian_defination,eq:doi_peliti_path_integral,eq:hamiltonian_reactive_jump_sigle}. This exact mapping between the DPFT and the SPIF also solves the open problem previously quoted in Ref.\cite{Weber_2017} (see the discussion below equation (194) in Ref.\cite{Weber_2017}). Fundamentally, this equivalence is related to the `gauge fixing', see \cref{sec:example_equivalence}; therefore, it is a `model-independent' principle that holds for any physical system modeled using the DPFT or the SPIF \cite{Sudarshan_1963}, and is not related to the `thermodynamically-consistent interacting particle' addressed here.
    }
    
    \item[{\Hygiea}]
    {
    \Cref{sec:cg_meso_to_macro} deals with the transition from the mesoscopic description to the macroscopic description and corresponds to the coarse-graining step in our work. Here, the thermodynamic aspects of the interacting fields are highlighted. The macroscopic thermodynamic cost of interactions is quantified by \cref{eq:macroscopic_boltzmann_weight} and \cref{eq:macroscopic_energy_functional} and is physically analogous to the chemical potential and the energy functional, respectively. The macroscopic LDB condition \cref{eq:macroscopic_local_detailed_balance_reactive} ensures the thermodynamic consistency on the macroscale and connects the transition rates between macrostates to the required thermodynamic cost \cref{eq:macroscopic_boltzmann_weight}. Then, the stochastic evolution of the macrostate is governed by the stochastic Langevin \cref{eq:macroscopic_eom_mft} with multiplicative noise, and is the central result of the thermodynamically-consistent coarse-graining procedure that incorporates the `occupancy noise'. \Cref{eq:mesoscopic_boltzmann_weight,eq:mesoscopic_energy_funational,eq:mesoscopic_local_detailed_balance_reactive,eq:mesoscopic_eom_mft} are the mesoscopic counterparts of the macroscopic \cref{eq:macroscopic_boltzmann_weight,eq:macroscopic_energy_functional,eq:macroscopic_local_detailed_balance_reactive,eq:macroscopic_eom_mft}, respectively. We obtain an exact large-deviation rate functional for the stochastic dynamics of the interacting particles.
    }

    \item[{\Hygiea}]
    {
    The mesoscopic and macroscopic interaction coefficients for fields are exactly non-linearly related to the microscopic interaction coefficients through \cref{eq:mesoscopic_interaction_coefficients,eq:macroscopic_interaction_coefficients}. This non-linear relation is the manifestation of exactly incorporating the microscopic occupancy noise on the mesoscale/macroscale, and addresses the `mean-field' approximation of the `occupancy noise'. The mean-field' approximation of the `occupancy noise' is only satisfied in the thermodynamic limit ($\Omega \to \infty$). This also formulates a transparent and systematic connection between the microscopic and macroscopic control parameters of the system.
    }
    
    \item[{\Hygiea}]
 In addition to its primary applications to `thermodynamically-consistent nonreciprocal particles' detailed in Ref.\cite{atm_st_nr_2024}, \cref{sec:example} addresses a few secondary applications of the methodology developed here. In particular, compared to existing coarse-graining methodologies such as the Kawasaki–Dean approach for diffusive dynamics \cite{Kawasaki_1994,Dean_1996} and the classical path-integral approach for reactive dynamics \cite{Lefevre_2007}, we demonstrate that our methodology offers a systematic qualitative improvement by exactly incorporating `occupancy noise', microscopic interactions, and thermodynamic consistency. Using the coarse-grained equations for the dynamics of the AIM. we show that incorporating `occupancy noise' suffices for correctly predicting the phase diagram for the microscopic dynamics of AIM.

\end{itemize}
%
%

\Cref{table:summary} summarizes the relationship between the vertical connections in the observation scale description and the horizontal thermodynamically-consistent connection between dynamics and thermodynamics via the LDB condition.
\section{Microscopic Description}\label{sec:microscopic}
%
%
{
We consider a lattice gas model with lattice spacing $l$, continuous-time dynamics, and the total number of particles of the lattice $N_{tot}$. The lattice spacing is assumed to be one, unless stated otherwise. Physically, this corresponds to choosing the microscopic diffusive length scale as a measurement unit. Here, we study the discrete-space description; however, the continuous-space limit is obtained by taking an infinitesimally small lattice spacing. Therefore, throughout this work, although the discrete-space representation is preferred, the equivalence between the discrete space and continuous space is assumed, and they are used interchangeably with the relevant definitions of the gradient and Laplace operators. In addition, we use the notion of a lattice site to define the range of the microscopic particle interaction (subsequently defined below in the model setup). However, these two parameters can be decoupled and defined independently for other systems that require incorporating the nearest-neighbour interactions or the long-range interactions. Each particle has a type index $i$ and a lattice index $\#$. $N_i^\#$ denotes the number of particles of type $i$ at the lattice site index $\#$. ${N}$ denotes the lattice configuration specified by the particle occupancy vector. The dimension of ${N}$ is the product of the number of lattice sites and the particle types.
}
\subsection{Thermodynamics}\label{sec:microscopic_thermodynamics}
%
%
{
We define the microscopic interaction coefficients $v_{ij}$ that quantify the thermodynamic energetic cost of inserting a type $i$ particle in the presence of a type $j$ particle. $v_{ij} > 0$ ($v_{ij} < 0$) signify a repulsive (an attractive) interaction experienced by a type $i$ particle in the presence of a type $j$ particle. 
We consider a general class of interacting particles that explicitly break the `actio=reactio' symmetry, termed non-reciprocal. This implies $v_{ij} = v_{ij}$ is not necessarily satisfied, and $v_{ij} \neq v_{ji}$ holds. 
}

{A particle on a lattice site simultaneously interacts with multiple other particles. Therefore, the microscopic Boltzmann weight $\epsilon_i^\#$ quantifies the total thermodynamic cost of inserting a type $i$ particle at lattice site $\#$ due to all particles on that lattice site.} Throughout this work, we define the Boltzmann weight, which is the analogue (generalization) of a chemical potential but defined for non-reciprocal particles/systems that explicitly break the `actio=reactio' symmetry. It is divided into its reciprocal and non-reciprocal parts, $\epsilon_i^r$ and $f_i^{nr}$, respectively, such that $\epsilon_i^\# = \epsilon_i^r + f_i^{nr}$,
\begin{equation}\label{eq:microscopic_reciprocal_interaction_energy_particle}
\begin{split}
    \epsilon_i^r = \beta \sum_{j \neq i} v_{ij}^r N_j^\# + \beta v_{ii}^r \left( N_i^\# - 1 \right),\hspace{0.5cm}
    f_{i}^{nr} = \beta \sum_{j} v_{ij}^{nr} N_j^\#.
\end{split}    
\end{equation}
$v_{ij}^r$ quantifies the reciprocal interaction between the particle types $i$ and $j$ and is defined as $v_{ij}^{r} = (v_{ij} + v_{ji})/2$ \cite{atm_st_nr_2024}. Similarly, $v_{ij}^{nr}$ quantifies the non-reciprocal interaction between a type $i$ particle due to a type $j$ particle and is defined as $v_{ij}^{nr} = (v_{ij} - v_{ji})/2$. Importantly, by construction, the symmetry $v_{ij}^r = v_{ji}^r$ and the anti-symmetry $v_{ij}^{nr} = -v_{ji}^{nr}$ are satisfied, which physically correspond to the decomposition of the thermodynamic cost associated with the `actio=reactio' symmetry-preserving and symmetry-breaking terms, respectively. In \cref{eq:microscopic_reciprocal_interaction_energy_particle}, the vanishing thermodynamic cost $\epsilon_i^\# = 0$ corresponds to the non-interacting ideal particles.

{Summing over $\epsilon_i^\#$ for all particles on the lattice, the reciprocal part of the microscopic Boltzmann weight in \cref{eq:microscopic_reciprocal_interaction_energy_particle} can be derived from the change of a global interaction energetic function defined for the whole lattice system as,
\begin{equation}\label{eq:microscopic_energy_functional}
\begin{split}
    \varepsilon^{int} = \frac 1 2 \beta \sum_{\#, \: i} v_{ii}^{r} N_i^\# \left( N_i^\# - 1 \right) + \beta \sum_{\#, \: i \neq j} v_{ij}^{r} N_i^\# N_j^\#,
\end{split}
\end{equation}
which requires the constraint of reciprocity $v_{ij}^r =v_{ji}^r$ to be satisfied. Therefore, the non-reciprocal part $f_i^{nr}$ of $\epsilon_i^\#$ cannot be written as a change of a global interaction energy functional of the system. Consequently, this implies that non-reciprocal interactions inherently drive the system out-of-equilibrium.
}
\subsection{Dynamics}\label{sec:microscopic_dynamics}
\subsubsection{Transition rates}

To avoid confusion with the microscopic interaction coefficients $v_{ij}$ denoted by Latin indices, we denote reactive transitions that generate the change in particle type by Greek indices $\gamma \gamma'$. $\Delta_{\gamma \gamma'}^\#$ denotes a reactive transition of type $\gamma'$ to $\gamma$ at the lattice site $\#$. The reactive transition changes the lattice occupancy vector $\{N\} \to \{N + \Delta_{\gamma \gamma'}^\# N  \}$. Similarly, $\Delta_i^{\vec{\mathcal{D}} \#}$ denotes a diffusive transition of a particle of type $i$ at the lattice site $\#$ along the direction $\vec{\mathcal{D}}$. The diffusive transition changes the lattice occupancy vector $\{N\} \to \{N + \Delta_{i}^{\vec{\mathcal{D}}\#} N  \}$. $k_{\gamma'\gamma}^\#$ denotes the reactive transition rate to change the particle type from $\gamma$ to $\gamma'$ at the lattice site $\#$. Similarly, $k_{i}^{\vec{\mathcal{D}} \#}$ denotes the diffusive transition rate of a particle of type $i$ along the direction vector $\vec{\mathcal{D}}$. Reactive and diffusive transition rates are related to the microscopic Boltzmann weight \cref{eq:microscopic_reciprocal_interaction_energy_particle}. The exact expressions read:
\begin{equation}\label{eq:microscopic_transition_rate}
\begin{split}
    k_{\gamma' \gamma}^{\#} = d_{\gamma' \gamma} e^{\epsilon_\gamma^{\#}  + \frac 1 2 f_{\gamma' \gamma}^{ch} }, \hspace{0.5cm}
    k_{i}^{\vec{\mathcal{D}} \#} = d_i^{\mathcal{D}} e^{\epsilon_i^{\#} + \frac 1 2 \vec{\mathcal{D}} \cdot \vec{f}_{i}^{sp} }.
\end{split}    
\end{equation}
%
%
{
Here, $d_{\gamma' \gamma}$ and $d_i^{\mathcal{D}}$ are the constants that quantify the reactive and diffusive transition rates, respectively. To ensure the thermodynamic consistency and time-reversal asymmetry, the notion of a forward and corresponding unique backward transition is required. $\{ \Delta_{\gamma \gamma'}^\# \}$ and $\{ \Delta_i^{\vec{\mathcal{D}} \#} \}$ denote the set of all reactive and diffusive bidirectional transitions (pairs of forward and the corresponding backward transition), respectively.
}

{ In addition to the thermodynamic cost of particle interaction quantified by the microscopic Boltzmann weight $\epsilon_i^\#$ in \cref{eq:microscopic_transition_rate}, we have introduced external chemical (self-propulsion) driving forces $f_{\gamma \gamma'}^{ch}$ ($\vec{f}_{i}^{sp}$), which are supported by a reservoir of large number of enzymatic (Hydrorgenperoxide) particles and correspond to a directional asymmetry generated between the forward and backward reactive (diffusive) transitions that explicitly breaks time-reversal symmetry. $\vec{\mathcal{D}} \cdot \vec{f}_{i}^{sp}$ takes the projection of the self-propulsion force along the direction vector $\vec{\mathcal{D}}$ for the diffusive transition on the lattice. 
To ensure the thermodynamic consistency of the microscopic transition dynamics, the transition rates \cref{eq:microscopic_transition_rate} are an exponential function of the thermodynamic cost given by the microscopic Boltzmann weight \cref{eq:microscopic_reciprocal_interaction_energy_particle}; thereby, the system satisfies the microscopic Local Detailed Balance condition.
Since $\epsilon_\gamma^\# = 0$ corresponds to the non-interacting ideal particles, the transition rates in \cref{eq:microscopic_transition_rate} are constants that are independent of the instantaneous state of the lattice.
}

{
\Cref{eq:microscopic_transition_rate}, written for the forward and backward reactive (diffusive) transition rates, constructs the LDB condition in a familiar form.
\begin{equation}\label{eq:microscopic_local_detailed_balance_final}
\begin{split}
    \frac{k_{\gamma \gamma'}^\#}{k_{\gamma' \gamma}^\#} 
    = e^{ \epsilon_{\gamma'}^\# - \epsilon_{\gamma}^\# + f_{\gamma \gamma'}^{ch}}, \hspace{0.5cm}
    \frac{ k_{ i }^{ \vec{\mathcal{D}} \# } } { k_{ { i } }^{ (\vec{\mathcal{D} } \#)^{-1} } }
    = e^{ \epsilon_{i}^\# - \epsilon_{i}^{\vec{\mathcal{D}} \#} + \vec{\mathcal{D}} \cdot \vec{f}_{i}^{sp} }.
\end{split}
\end{equation}
Since, in comparison to \cref{eq:microscopic_transition_rate}, the time-symmetric parts of the transition rates ($d_{\gamma' \gamma}$ and $d_i^{\mathcal{D}}$) become irrelevant in \cref{eq:microscopic_local_detailed_balance_final}, the different microscopic timescales associated with different transition rates are lost in \cref{eq:microscopic_local_detailed_balance_final}. However, \cref{eq:microscopic_transition_rate} avoids this issue; hence, we will define microscopic models using \cref{eq:microscopic_transition_rate}. To highlight the connection to existing formulations, we consider the sub-case $f_{\gamma\gamma'}^{ch} = 0$ for the reactive transition in \cref{eq:microscopic_local_detailed_balance_final}. 
We decompose $\epsilon_{\gamma'}^\# - \epsilon_{\gamma}^\# = \epsilon_{\gamma'}^r - \epsilon_{\gamma}^r + f_{\gamma'}^{nr} - f_{\gamma}^{nr}$ into its reciprocal and non-reciprocal contributions. Here, the change in the reciprocal part of the Boltzmann weight can be written as the change in the global interaction energy of the lattice \cref{eq:microscopic_energy_functional}, in particular, $\epsilon_{\gamma'}^r - \epsilon_{\gamma}^r = -\Delta_{\gamma \gamma'} \varepsilon^{int}$. If the particles are reciprocal, i.e., $f_i^{nr} = 0, \forall i$, one recovers the equilibrium detailed balance condition for Markov chains.
}

{
The microscopic notion of the particle number is a stochastic variable, denoted by ${N}$, and defined for any microscopic configuration. 
In comparison, a coarse-grained description substitutes this particle number with a deterministic variable $\langle N \rangle$. Therefore, in the context of this work, we loosely define the mean-field approximation of the `occupancy noise' as the mathematical substitution $N = \langle N \rangle$; however, in general, $N \neq \langle N \rangle$ holds. Subsequently, throughout this work, we aim to systematically address such a `mean-field' approximation. The importance of addressing this approximation for the thermodynamically-consistent dynamics of the interacting particles is particularly evident from \cref{eq:microscopic_transition_rate}, due to the exponential and non-linear dependence of the transition rates on the particle occupancy through \cref{eq:microscopic_reciprocal_interaction_energy_particle,eq:microscopic_transition_rate}, the stochastic effect of particle occupancy is further non-linearly amplified. For a non-interacting ideal particle $\epsilon_i^\# = 0$, microscopic transition rates are constant and independent of particle number, making this issue less relevant to tackle.
}

{
To demonstrate the setup and notation, let us consider a prototypical single-lattice site system with two particle types denoted by $+$ and $-$ with initial lattice configuration $\{N\} = (N_+, N_{-})$, with external driving $f_{+-}^{ch} = 0$. Model parameters for microscopic interactions are $v_{++} = v_{--} = 0, v_{+-} = a \neq v_{-+} = b$, therefore, $v_{+-}^r = (a+b)/2$ and $v_{+-}^{nr} = (a-b)/2$. A reactive transition for the change of a $-$ particle to a $+$ particle is defined as $\Delta_{+-}$ and leads to the final lattice configuration $\{ N + \Delta_{+-} N \} = (N_{+} + 1, N_{-} - 1)$. Using \cref{eq:microscopic_reciprocal_interaction_energy_particle,eq:microscopic_transition_rate}, the forward single-particle reactive transition rate is $k_{+-} = d_{+-} e^{ \beta b N_+ }$  from $\{N\} = (N_+, N_{-})$ to $\{N + \Delta_{+-} N\} = (N_+ + 1, N_{-} - 1)$ and the backward single-particle reactive transition rate is $k_{-+} = d_{+-} e^{ \beta a (N_- - 1) }$ from $\{N + \Delta_{+-} N\} = (N_+ + 1, N_{-} - 1)$ to $\{N\} = (N_+, N_{-})$. 
}
\subsubsection{Master equation}
The probability of a configuration $\{N\}$ at time $t$ is denoted by $P_{\{N\}}(t)$. The Master Equation for the evolution of $P_{\{N\}}(t)$ reads:
\begin{widetext}
\begin{equation}\label{eq:microscopic_master_equation}
\begin{split}
    \partial_t P_{\{N\}}(t)
    & = \sum_{\{ \Delta_{\gamma \gamma'}^\# \}}
    \underbrace{\left( k_{ \gamma' \gamma}^\# (\{ N + \Delta_{\gamma \gamma'}^\# N \}) 
    P_{ \{ N + \Delta_{\gamma \gamma'}^\# \} } 
    - k_{ \gamma \gamma' }^\#( \{ N \} ) 
    P_{ \{ N \} } \right)}_{-j_{\gamma \gamma'}^\# ( \{ N \} )}
    + \sum_{ \{ \Delta_i^{\vec{\mathcal{D}} \#} \} } 
    \underbrace{ \left( k_i^{ {( \vec{\mathcal{D}} \# )}^{-1} } ( \{ N + \Delta_i^{ \vec{\mathcal{D}} \# } N  \} ) 
    P_{ \{ N + \Delta_i^{ \vec{\mathcal{D}} \# } \} } 
    - k_i^{ \vec{\mathcal{D}} \# } ( \{ N \} ) 
    P_{ \{ N \} }  \right) }_{ -j_i^{{ \vec{\mathcal{D}} \# }} ( \{ N \} ) }, 
\end{split}    
\end{equation}
\end{widetext}
Here, $j_{\gamma \gamma'}^\# ( \{ N \} )$ and $j_i^{{ \vec{\mathcal{D}} \# }}( \{ N \} )$ quantify the net probability outflow due to the transitions $\Delta_{\gamma \gamma'}^\#$ and $\Delta_i^{\vec{\mathcal{D}} \#}$, respectively.
\section{Coarse-Graining: Microscopic to Mesoscopic} \label{sec:coarse_graining_micro_to_meso}
The microscopic stochastic description requires tracking all possible configurations in the $P_{\{N\}}$ space, which scales exponentially with the number of microscopic particles. We aim to formulate a coarse-grained mesoscopic Langevin description of the average particle occupancy $\{N\}$ of the lattice, which fluctuates stochastically due to microscopic transitions. This is achieved by summing over the $P_{\{N\}}$ space. This reduces the complexity of tracking the mesoscopic dynamics given by a Langevin equation compared to the microscopic Master equation. Since, the number of mesoscopic order parameters required is equal to the dimension of the lattice multiplied by the number of particle types.
\subsection{Doi-Peliti field theory for microscopic dynamics }\label{sec:cg_dp_field_theory}
\subsubsection{Introduction}\label{sec:dp_introduction}
DPFT uses a second-quantised formulation for classical many-body systems \cite{Doi_1976,Doi_1976_2,Peliti,ZelDovich_1978,grassberger_1980,rose_1979,Mikhailov_1981,Mikhailov_1981_2,Mikhailov_1985,Cardy_2008,Wiese_2016,Weber_2017}. Because of its second-quantised formulation, DPFT incorporates the discreteness of the microscopic system into the mesoscopic description. Moreover, DPFT has obtained a mesoscopic description using a Poissonian measure over the microscopic particle occupancy. The phenomenological coarse-grained mean-field description does not account for Poissonian fluctuations or microscopic discreteness. Poissonian occupancy has prominent importance in the low-particle density limit. In this regime, the coarse-grained mean-field description fails due to the importance of Poissonian noise \cite{Solon2015, Mourtaza2018}. In addition, it qualitatively and quantitatively affects the thermodynamic dissipation. Hence, it has prominent importance for the thermodynamically consistent coarse-graining.
\subsubsection{Doi representation}\label{sec:cg_doi_reperesentation}
The second-quantized state for the $N_i^\#$ number of type $i$ particles at the lattice site $\#$ is denoted by $\bra$, and the corresponding dual by $\ket$ \cite{Doi_1976,Doi_1976_2,Peliti,ZelDovich_1978,grassberger_1980,rose_1979,Mikhailov_1981,Mikhailov_1981_2,Mikhailov_1985,Cardy_2008,Wiese_2016,Weber_2017}. The second-quantized states satisfy the following ladder operator relations,
\begin{equation} \label{eq:doi_representation}
\begin{split}
    \crtlong{i}{\#} \bra = \braplus, \hspace{0.6cm}
    \anhlong{i}{\#} \bra = \dens \braminus,
\end{split}
\end{equation}
The creation operators $\crtlong{i}{\#}$ acting on $\bra$ increases the occupancy state by 1. The annihilation operators $\anhlong{i}{\#}$ acting on $\bra$ decreases the occupancy state by 1. Analogously, the number operator $\densoplong{i}^{\#} = \crtlong{i}{\#} \anhlong{i}{\#}$ satisfies $\crtlong{i}{\#} \anhlong{i}{\#} \bra = \dens \bra$. In contrast to the quantum notation, the classical notation asymmetrically distributes the eigenvalue $N_i^\#$ over the creation and annihilation operators. The consequence of combinatorics is that there are $N_i^\#$ different possibilities of annihilating a particle, but only a single possibility of creation.
\subsubsection{Second Quantized Hamiltonian}\label{sec:cg_hamiltonian_quantised}
The second-quantized Hamiltonian for $\Delta_{\gamma \gamma'}^\#$ and $\Delta_{i}^{\Vec{\mathcal{D}} \#}$ is given by \cite{Doi_1976,Doi_1976_2,Peliti,ZelDovich_1978,grassberger_1980,rose_1979,Mikhailov_1981,Mikhailov_1981_2,Mikhailov_1985,Cardy_2008,Wiese_2016,Weber_2017}:
\begin{equation}\label{eq:hamiltonian_quantised_single_reactive_jump_1}
\begin{split}
\hat{H}_{i}^{\Vec{\mathcal{D}} \#}
&= d_i \left[ \crtlong{ i }{\#} - \crtlong{ i }{\Vec{\mathcal{D}} \#} \right]
\left[ \anhlong{ i }{ \# } e^{ \epsilon_{ i }^{ \# } + \frac 1 2 \vec{\mathcal{D}} \cdot \vec{f}_{i}^{sp} }  
- 
\anhlong{ i }{ \Vec{\mathcal{D}} \#}
e^{\epsilon_{ i }^{\Vec{\mathcal{D}} \#} - \frac 1 2 \vec{\mathcal{D}} \cdot \vec{f}_{i}^{sp} } \right],  
\\
\hat{H}_{\gamma \gamma'}^{\#}
& = d_{\gamma \gamma'} \left[ \crtlong{\gamma}{\#} - \crtlong{\gamma'}{\#} \right] \left[ \anhlong{\gamma}{\#}  e^{\epsilon_{\gamma}^\# - \frac 1 2 f_{\gamma \gamma'}^{ch} } - \anhlong{\gamma'}{\#} e^{\epsilon_{\gamma'}^\# + \frac 1 2 f_{\gamma \gamma'}^{ch} } \right].
\end{split}    
\end{equation}
The total second-quantized Hamiltonian $\hat{H}$ for all possible transitions is:
\begin{equation}\label{eq:hamiltonian_quantized_total}
    \hat{H} = \hat{H}^\mathcal{R} + \hat{H}^\mathcal{D}.
\end{equation}
Here, $\hat{H}^\mathcal{R} = \sum_{ \{ \Delta_{\gamma \gamma'}^\# \} } \hat{H}_{\gamma \gamma'}^\#$ and $\hat{H}^\mathcal{D} = \sum_{\{ \Delta_i^{ \vec{\mathcal{D}} \# }\}} \hat{H}_{i}^{ \vec{\mathcal{D}} \# }$ are the reactive and diffusive contributions, respectively.
\subsubsection{Summary of the section}
%
%
{
Using the existing framework of the DPFT methodology, we have formulated a second--quantized description of thermodynamically--consistent microscopic transition dynamics for the interacting particles, \cref{eq:hamiltonian_quantized_total,eq:hamiltonian_quantised_single_reactive_jump_1}, which is a necessary step towards a technically correct formulation of the problem statement. Up to this point, the physical quantities are defined for a single microscopic transition and particle. Therefore, the DPFT methodology is exactly equivalent to the microscopic dynamics. 
}
\subsection{The mesoscopic coherent state path integral}\label{sec:cg_doi_peliti_action}
\subsubsection{The coherent state and master equation}\label{sec:cg_coherant_state}
The coherent state $ | \phi_i^\# \rangle $ for a type $i$ particle at $\#$ and its dual counterpart $ \langle (\phi_i^\#)^* |$ are defined as:
\begin{equation}\label{eq:coherant_state_single_species}
\begin{split}
    | \phi_i^\# \rangle = e^{\phi_i^\# \crtlong{i}{\#} } | 0 \rangle, \hspace{0.7cm}
    \langle (\phi_i^\#)^* | = \langle 0 | e^{ (\phi_i^\#)^* \anhlong{i}{\#} }. 
\end{split}
\end{equation}
The Taylor series expansions of it are $| \phi_i^\# \rangle =  \sum_{l} \frac{ \left( \phi_i^\# \right)^{l} \left( \crtlong{i}{\#} \right)^{l}}{ l !}  | 0 \rangle$ and $\langle (\phi_i^\#)^* | = \langle 0 | \sum_{l} \frac{ \left( (\phi_i^\#)^* \right)^{l} \left( \anh \right)^{l}}{ l!}$. It has an inherent probabilistic interpretation, namely $| \phi_i^\# \rangle = \sum_l P(l) \bra $. Therefore, $ P( l ) = \left( \phi_i^\# \right)^l/l! $ gives a realization of a Poissonian distribution for the particle occupancy \footnote{Except for the normalization prefactor $e^{ -\phi_i^\#(\phi_i^\#)^* }$, which is circumvented by taking the inner product of the coherent state in the denominator, the convention we will adhere to compute expectation values throughout this work.}.
{
Due to this special property of the coherent state, it is also known as a classical approximation of a quantum state, where each microstate $| N_i^\#\rangle$ contributes to $|\phi_i^\#\rangle$ in proportion to its probability. Importantly, $ \crtlong{i}{\#} | \phi_i^\# \rangle =  \phi_i^\# | \phi_i^\# \rangle$ and $ \langle (\phi_i^\#)^* | \anhlong{i}{\#} = \langle (\phi_i^\#)^* | (\phi_i^\#)^*$, therefore, $\phi_i^\#$ is the eigenvalue of the coherent state. 
Using only two eigenvalues $\phi_i^\#$ and $(\phi_i^\#)^*$ of the coherent state, the coherent state representation allows a simplified description of the microscopic system that otherwise would have required the set of all particle occupancy states $|N_i^\# \rangle$.
The inner product of any microscopic operator $\hat{O}$ with the coherent state is equivalent to the Poisson probability measure of the operator over the microscopic particle occupancy. Then, the mesoscopic expectation value of any microscopic operator $\hat{O}$ over the Poissonian occupancy probability measure is defined as $ \langle \hat{O} \rangle = \langle  \{ \phi^* \} | \hat{O} | \{ \phi \} \rangle / \langle \phi^* | \phi \rangle $. Here, $\langle \hat{O} \rangle$ now defines the mesoscopic coarse-grained representation of a microscopic operator $\hat{O}$.
}

The composite coherent state for the whole lattice is defined as:
\begin{equation}\label{eq:coherant_state}
\begin{split}
    | \{ \phi \} \rangle = \prod_{i,\#} \otimes | \phi_i^\# \rangle, \hspace{0.7cm}
    \langle \{ \phi \} | = \prod_{i,\#} \otimes \langle \phi_i^\# |. 
\end{split}    
\end{equation}
In \cref{eq:coherant_state}, $\otimes$ denotes the tensor product of the coherent states for all types of particles and lattice indices. The ladder operators commute for particles of different types; thus, $|\{ \phi \}\rangle$ is the tensor product of $| \phi_i \rangle$ over all types of particles and lattice indices. 

The time-dependent Schrödinger equation $\partial_t | \{ \phi \} \rangle = - \hat{H} | \{ \phi \} \rangle $ for the coherent state is equivalent to the master \cref{eq:microscopic_master_equation} for $P_{\{N\}}(t)$, if the dynamics of each linearly independent basis vector of $| \{ \phi \} \rangle$ are considered. This establishes the analogy between classical many-body systems and their second-quantized quantum representations for thermodynamically consistent dynamics of interacting particles.

\subsubsection{Normal ordering}\label{sec:cg_normal_ordering}
Computing the expectation values for the operators (relevant physical quantities) becomes a cumbersome task using the second-quantized methods. To simplify this computation, one needs to obtain the normal-ordered form $:\hat{O}:$ of the operator $\hat{O}$. In the normal-ordered form, the creation and annihilation operators are replaced by the eigenvalues of $|\{ \phi \} \rangle$, in particular $\crtlong{i}{\#} \to \phi_i^\#$ and $\anhlong{i}{\#} \to (\phi_i^\#)^*$. The reason is that the coherent state is an eigenvector of the creation operator with eigenvalue $\phi_i^\#$. Thus, $\langle \hat{O} \rangle = \langle  \{ \phi^* \} | :\hat{O}: | \{ \phi \} \rangle / \langle \phi^* | \phi \rangle$ is computed trivially using the normal-ordered form. For non-interacting particles, \cref{eq:hamiltonian_quantised_single_reactive_jump_1} is normal-ordered. In contrast, the interacting nature of the particles poses a challenge due to the exponential dependence of $\hat{H}$ on the number operator $\hat{N}_i^\# = \crtlong{i}{\#} \anhlong{i}{\#}$. The normal ordering of \cref{eq:hamiltonian_quantised_single_reactive_jump_1} derived in \cref{app:cg} reads \footnote{ We use the notation $\exp{(*)}  = e^{*}$, to simplify the representation of the double exponential obtained subsequently throughout the paper. }:
\begin{widetext}
\begin{equation}\label{eq:hamiltonian_quantised_single_reactive_jump_normal_ordered}
\begin{split}
    :\hat{H}_{\gamma \gamma'}^\#: 
    & = d_{\gamma \gamma'} \left[ \crtlong{\gamma}{\#} - \crtlong{\gamma'}{\#} \right]
    \bigg[ 
    \exp{ \left( \sum_{j} \densoplongg{j}{\#}  \left( e^{\beta v_{\gamma j}} - 1 \right) - \frac 1 2 f_{\gamma \gamma'}^{ch} \right) }
    \anhlong{\gamma}{\#} 
    - 
    \exp{ \left( \sum_{j} \densoplongg{j}{\#}  \left( e^{\beta v_{\gamma' j}} - 1 \right) + \frac 1 2 f_{\gamma \gamma'}^{ch} \right) }
    \anhlong{\gamma'}{\#}
    \bigg],
    \\
    :\hat{H}_{i}^{ \vec{\mathcal{D}} \#}: 
    & = d_{i} \left[ \crtlong{i}{\#} - \crtlong{i}{\vec{\mathcal{D}}\#} \right]
    \bigg[ 
    \exp{ \left( \sum_{j} \densoplongg{j}{\#}  \left( e^{\beta v_{ij}} - 1 \right) - \frac 1 2 f_{\gamma \gamma'}^{ch} \right) }
    \anhlong{i}{\#} 
     - 
    \exp{ \left( \sum_{j} \densoplongg{j}{ \vec{\mathcal{D}} \# } \left( e^{\beta v_{ij}} - 1 \right) + \frac 1 2 f_{\gamma \gamma'}^{ch} \right) }
    \anhlong{i}{ \vec{\mathcal{D}} \#}
    \bigg]. 
\end{split}    
\end{equation}
\end{widetext}
%
%

{
We elaborate on the physical aspects of the normal-ordering procedure. Normal ordering accounts for \textit{all microscopic configurations} that contribute to the same mesostate \cite{blasiak2007}. Importantly, different microscopic operators $\hat{O}_1 \neq \hat{O}_2$ might lead to the same normal-ordered form, $:\hat{O}_1: \, = \, :\hat{O}_2:$, and hence $\langle \hat{O}_1 \rangle = \langle \hat{O}_2 \rangle$. Subsequently, the normal ordering of an operator is unique but does not define a one-to-one mapping. This implies that the microscopic-to-mesoscopic mapping $\crtlong{i}{\#} \to \phi_i^\#$ and $\anhlong{i}{\#} \to (\phi_i^\#)^*$ from operators to numbers is one-way. The inverse mapping, $\phi_i^\# \to \crtlong{i}{\#}$ and $(\phi_i^\#)^* \to \anhlong{i}{\#}$, does not necessarily hold if one attempts to invert the mesoscopic description to obtain the microscopic description. Therefore, the normal-ordering procedure can be interpreted as a form of coarse-graining, where the information on the microscopic combinatorics is lost. This issue is irrelevant for ideal (non-interacting) particle systems, where the microscopic single-particle transition dynamics \cref{eq:hamiltonian_quantised_single_reactive_jump_1} depend linearly on the number operator; therefore, the inverse mapping from the mesoscopic to the microscopic description can be implemented. However, in general, this is not true. This fundamental intricacy of the DPFT methodology should be handled carefully, particularly for thermodynamically consistent interacting particle systems, where it becomes especially evident.
}

{
In summary, due to the mapping from a microscopic single-particle and single-transition description to a mesoscopic coherent-state path integral, two different notions of coarse-graining emerge. First, the coherent state projects the microscopic occupancy states in proportion to their Poissonian probabilities. Therefore, using the coherent state, the mesoscopic representation $\langle \hat{O} \rangle$ yields the average of a microscopic observable $\hat{O}$ over the Poissonian probabilities of the microscopic occupancy states. Second, to ensure the correctness of this procedure, all microscopic configurations have to be correctly counted using normal ordering; thereby, the resulting mesoscopic description $\langle \hat{O} \rangle$ of the operator is non-invertible, if one aims to recover the microscopic description.
}
\subsubsection{The mesocscopic Doi-Peliti Lagrangian}\label{sec:cg_doi_peliti_lagrangian}
In DPFT, the mesoscopic description of \cref{eq:microscopic_master_equation} is obtained using the mesoscopic Doi-Peliti action $\mathcal{S}_{DP}$. It is constructed using the coherent-state path integral approach \cite{Doi_1976,Doi_1976_2,ZelDovich_1978,Peliti,grassberger_1980,rose_1979,Mikhailov_1981,Mikhailov_1981_2,Mikhailov_1985,Cardy_2008,Weber_2017,Wiese_2016}. $\mathcal{S}_{DP}$ is obtained by computing an expectation value of the transition Hamiltonian in \cref{eq:hamiltonian_quantized_total} over the coherent state. $\mathcal{S}_{DP}$ is rewritten as an integral of a Lagrangian, $\mathcal{S}_{DP} \left[ \{ \phi^*, \phi \} \right] = \int_{t_i}^{t_f} dt \mathcal{L} \left[ \{ \phi^*, \phi \} \right]$. The exact expression for $\mathcal{L} \left[ \{ \phi^*, \phi \} \right]$ obtained using \cref{eq:hamiltonian_quantized_total} reads \cite{Doi_1976,Doi_1976_2,Peliti,ZelDovich_1978,grassberger_1980,rose_1979,Mikhailov_1981,Mikhailov_1981_2,Mikhailov_1985,Cardy_2008,Weber_2017,Wiese_2016}:
\begin{equation}\label{eq:lagrangian_defination}
\begin{split}
    \mathcal{L} \left[ \{ \phi^*, \phi \} \right]
    & = \left \{ - \langle \partial_t \{ \phi^* \} | \{ \phi \} \rangle + \frac{ \langle \{ \phi^* \} | \hat{H} | \{ \phi \}\rangle }{ \langle \{ \phi^* \} | \{ \phi \} \rangle } \right \}.
\end{split}
\end{equation}
In \cref{eq:lagrangian_defination}, the first term corresponds to the evolution of the fields in time, the left-hand side of the master equation. The second term encapsulates the mesoscopic transition jumps, the right-side of the master equation summed over Poisssonian state occupancy. Defining the mesoscopic transition Hamiltonians for reactive $\mathcal{H}_{\gamma \gamma'}^\# \left[ \{ \phi^*, \phi \} \right] = -\frac{ \langle \{ \phi^* \} | \hat{H}_{\gamma \gamma'}^\# | \{ \phi \}\rangle }{ \langle \{ \phi^* \} | \{ \phi \} \rangle }$, and diffusive $\mathcal{H}_i^{\vec{\mathcal{D}} \#} \left[ \{ \phi^*, \phi \} \right] = -\frac{ \langle \{ \phi^* \} | \hat{H}_i^{\vec{\mathcal{D}} \#} | \{ \phi \}\rangle }{ \langle \{ \phi^* \} | \{ \phi \} \rangle }$, and total $\mathcal{H} \left[ \{ \phi^*, \phi \} \right] = -\frac{ \langle \{ \phi^* \} | \hat{H} | \{ \phi \}\rangle }{ \langle \{ \phi^* \} | \{ \phi \} \rangle }$ transitions. Using \cref{eq:hamiltonian_quantised_single_reactive_jump_normal_ordered}, the mesoscopic Hamiltonian for $\Delta_{\gamma \gamma'}^\#$ and $\Delta_{i}^{\Vec{\mathcal{D}} \#}$ are as follows:
\begin{widetext}
\begin{equation}\label{eq:hamiltonian_reactive_jump_sigle}
\begin{split}
   \mathcal{H}_{\gamma \gamma'}^\# \left[ \{ \phi^*, \phi \} \right] 
   & = -d_{\gamma \gamma'} \left[ (\phi_\gamma^\#)^* - (\phi_{\gamma'}^\#)^* \right] 
   \bigg[ 
   \phi_{\gamma} 
   \exp{ \left( \sum_{j} (\phi_j^\#)^* \phi_j^\# \left( e^{\beta v_{\gamma j}} - 1 \right) - \frac 1 2 f_{\gamma \gamma'}^{ch} \right) }
   - \phi_{\gamma'} 
   \exp{ \left( \sum_{j} (\phi_j^\#)^* \phi_j^\#  \left( e^{\beta v_{\gamma' j}} - 1 \right) + \frac 1 2 f_{\gamma \gamma'}^{ch} \right) }
   \bigg],
   \\
   \mathcal{H}_i^{\vec{\mathcal{D}} \#} \left[ \{ \phi^*, \phi \} \right] & = -d_{i} \left[ (\phi_i^\#)^* - (\phi_{i}^{ \vec{\mathcal{D}} \# })^* \right] 
   \bigg[ \phi_{i} 
   \exp{ \left( \sum_{j} (\phi_j^{ \#})^* \phi_j^\# \left( e^{\beta v_{\gamma j}} - 1 \right)  \right) }
   - \phi_{ i }^{ \vec{\mathcal{D}} \# } 
   \exp{ \left( \sum_{j} (\phi_j^{\vec{\mathcal{D}} \#})^* \phi_j^{ \vec{\mathcal{D}} \# }  \left( e^{\beta v_{i j}} - 1 \right) \right) }
   \bigg].
\end{split}
\end{equation}
\end{widetext}
The transition probability measure $\mathcal{P} \left[ \left\{ \phi^*, \phi \right\} \right]$, obtained using $\mathcal{S}_{DP}$, reads,
\begin{equation}\label{eq:doi_peliti_path_integral}
\begin{split}
    \mathcal{P} \left[ \left\{ \phi^*, \phi \right\} \right] 
    = e^{ -\mathcal{S}_{DP} \left[ \left\{ \phi^*, \phi \right\} \right] }.
\end{split}
\end{equation}
The normalization factor for \cref{eq:doi_peliti_path_integral} is obtained by imposing probability conservation, $\int \mathbb{D} \{ \phi^* \} \mathbb{D} \{ \phi \} \mathcal{P} \left[ \left\{ \phi^*, \phi \right\} \right] = 1$.
Here, $\mathbb{D}$ in \cref{eq:doi_peliti_path_integral} represents the path integral over all realizations of the coherent-state eigenvalues $\{ \phi \}$. \Cref{eq:doi_peliti_path_integral} formulates the mesoscopic path integral representation for stochastic dynamics.
\subsubsection{Summary of the section}\label{sec:cg_doi_peliti_action_summary}
%
%
{
Using the existing DPFT methodology, we have managed to formulate an `exact' coherent-state path integral mesoscopic description of thermodynamically-consistent dynamics for the interacting particles. The equivalence between the microscopic transition Hamiltonian  \cref{eq:hamiltonian_quantised_single_reactive_jump_1} and the mesoscopic transition Hamiltonian  \cref{eq:hamiltonian_reactive_jump_sigle} characterized by `coherent-state' eigenvalues is exact over a poissonian distribution of particle occupancy, such that all microscopic configurations that contribute to the same eigenvalue of the coherent state are grouped,
which results in the double exponential in \cref{eq:hamiltonian_reactive_jump_sigle}.
In this aspect, this procedure is equivalent to counting the statistical degeneracy of the mesostate. Therefore, the normal-ordering procedure could be interpreted as a form of coarse-graining that accounts for microscopic combinatorial aspects on the mesoscale \cite{blasiak2007}. To highlight this point, let us consider a very simple prototypical system that consists of two particle types denoted by $+$ and $-$ on a single lattice with a reactive transition $+-$ and a total number of particles equal to $15$. Then, the microscopic configuration space consists of $16$ configurations of $\{ N_+, N_-\}$. In comparison, the mesoscopic second-quantized description requires only two eigenvalues, $\phi_+, \phi_-$, and their conjugates, $\phi_+^*, \phi_-^*$. For the small system example discussed above, the differences between microscopic and mesoscopic representations appear to be less relevant. However, for more sophisticated models, the microscopic state-space scales exponentially with the total number of microscopic particles; in comparison, the mesoscopic second-quantized description requires all independent eigenvalues of the coherent state, which is equal to the number of particle types multiplied by the dimension of the lattice. The analytical and computational ease offered by the mesoscopic description becomes important, when the number of particles is sufficiently large. 
}
\subsection{Mesoscopic stochastic path integral representation}\label{sec:cg_stochastic_action}
\subsubsection{Cole-Hopf Transform}\label{sec:cg_cole_hopf_transform}
The ladder operators are not a natural realization of the change of a physically quantifiable observable of the classical many-body system, in contrast to the quantum case. Thus, the coherent-state path integral formalism leads to a scalar Lagrangian as a function of the eigenvalues \(\phi\) and \(\phi^*\) of the coherent state, which is not directly related to a physical observable of classical dynamics of many-body systems. To address this issue, one needs to deploy a canonical transformation called the Cole-Hopf transform \cite{itakura_2010}. It defines the relationship between the `mesoscopic' mean occupancy of the particles \(N_i^\#\) and \(\phi_i^\#\) and \((\phi_i^\#)^*\). The Cole-Hopf transform is defined as:
\begin{equation}\label{eq:cole-hopf_transform}
\phi_i^\# = N_i^\# e^{-\chi_i^\#}, \hspace{1.5cm} (\phi_i^\#)^* = e^{\chi_i^\#}.
\end{equation}
The particle number is the eigenvalue of the number operator, therefore \(N_i^\# = (\phi_i^\#)^* \phi_i^\#\) is rather trivial. \(\chi_i^\#\) is referred to as a conjugate field, noise field, or bias field. It realizes the generator for the stochastic transition between mesostates. For example, the transition \(\Delta_{\gamma\gamma'}^\#\) implies \(\crtlong{\gamma}{\#} \anhlong{\gamma'}{\#} \to N_{\gamma'}^\# e^{\chi_\gamma^\# - \chi_{\gamma'}^\#}\), where \(\chi_{\gamma}^\# - \chi_{\gamma'}^\#\) signifies the conjugate field of the generator for the transition \(\Delta_{\gamma\gamma'}^\#\). The change in the conjugate field due to the transition (stochastic or driven) represents the most-likelihood force that generates the transition. Note that \(\chi_i\) is an intensive field. 

%
%
{
Since, $\phi_i$ and $\phi_i^*$ are continuous variables in  \cref{eq:doi_peliti_path_integral}, $N_i^\#$ and $\chi_i^\#$ are also continuous variables, in contrast to it's microscopic operator representation. Therefore, the `mesoscopic' particle number is a continuous stochastic variable in comparison to the `microscopic' particle number defined in \cref{sec:microscopic,sec:cg_dp_field_theory} as a discrete variable, highlighting the difference between the `microscopic' and `mesoscopic' formulations of the particle number. This is because the coherent state path integral formulism obtains $\phi_i^\#,(\phi_i^\#)^{*}$ (and subsequently, $N_i^\#$) by summing over all microscopic configurations, as highlighted previously in \cref{sec:cg_doi_peliti_action_summary}.
} In addition, the Cole-Hopf transform also addresses imaginary noise for coarse-grained fields, which was identified as a problem associated with obtaining a Langevin equation using DPFT \cite{itakura_2010}. Since, mesoscopic mean particle occupancy $N_i^\#$ is a physically well defined quantity for classical many-body systems. 

\subsubsection{The mesoscopic Doi-Peliti Lagrangian in the occupancy-noise picture}\label{sec:cg_lagrangian_density_noise}
Inserting the Cole-Hopf transform \cref{eq:cole-hopf_transform} into \cref{eq:hamiltonian_reactive_jump_sigle}, the mesoscopic transition Hamiltonian is expressed in occupancy-noise fields.
\begin{widetext}
\begin{equation}\label{eq:hamiltonian_reactive_transition_density_noise}
\begin{split}
    \mathcal{H}_{\gamma \gamma'}^\# \left[ \{ N, \chi \} \right] &= d_{\gamma \gamma'}
   \bigg[ 
   \left( e^{ \chi_{\gamma'}^\# - \chi_{\gamma}^\# } - 1 \right) e^{ \upmu_\gamma^\# - \frac{1}{2} \mathcal{F}_{\gamma \gamma'}^{ch} } + \left( e^{\chi_{\gamma}^\# - \chi_{\gamma'}^\# } - 1 \right) e^{\upmu_{\gamma'}^\# + \frac{1}{2} \mathcal{F}_{\gamma \gamma'}^{ch} }
   \bigg], 
   \\
   \mathcal{H}_i^{\vec{\mathcal{D}} \#} \left[ \{ N, \chi \}\right]
   & = d_{i}
   \bigg[ 
   \left( e^{ \chi_i^{\vec{\mathcal{D}} \#} - \chi_i^\#} - 1 \right) 
   e^{ \upmu_i^\# + \frac{1}{2} \vec{\mathcal{D}} \cdot \vec{\mathcal{F}}_{i}^{sp} } + \left( e^{ \chi_i^\# - \chi_i^{\vec{\mathcal{D}} \#} } - 1 \right) 
   e^{ \upmu_i^{\vec{\mathcal{D}} \#} - \frac{1}{2} \vec{\mathcal{D}} \cdot \vec{\mathcal{F}}_{i}^{sp} }
   \bigg].
\end{split}
\end{equation}
\end{widetext}
%
%
{
Where, $\mathcal{F}_{\gamma \gamma'}^{ch} = f_{\gamma \gamma'}$ and $\vec{\mathcal{F}}^{sp} = \vec{{f}}^{sp}$. The structre of 
\cref{eq:hamiltonian_reactive_transition_density_noise} reveals that the transition dynamics between the mesostate are generated by the mesoscopic Boltzmann weights $\upmu_i^\#$ (in addition to the external driving forces $\mathcal{F}_{\gamma \gamma'}^{ch}$ and $\vec{\mathcal{F}}^{sp}$ supported by the thermodynamic reservoir)
, where, the mesoscopic Boltzmann weight of the mesostate \(N_i^\#\) is defined as,  
\begin{equation}\label{eq:mesoscopic_boltzmann_weight}
    \upmu_i^\# = \ln N_i^\# + \sum_{j} \mathcal{V}_{ij} N_j^\#,
\end{equation}
and quantifies the mesoscopic thermodynamic cost,  
with the mesoscopic interaction coefficients, 
\begin{equation}\label{eq:mesoscopic_interaction_coefficients}
    \mathcal{V}_{ij} = \left(e^{\beta v_{ij}} - 1\right), 
\end{equation}
that quantifies the interaction experienced by the mesostate $N_i^\#$ due to $N_j^\#$, and is also equal to the second virial coefficient.  The first term of $\upmu_i^\#$ ($\ln N_i^\#$) quantifies the statistical degeneracy of the mesostate and corresponds to the Boltzmann entropy and is equal to the chemical potential defined for an ideal (non-interacting) mesostate. The non-linear dependence (renormalization) of \(\mathcal{V}_{ij}\) on \(v_{ij}\) in \cref{eq:mesoscopic_interaction_coefficients} is attributed to the Poissonian occupancy statistics. Importantly, for the strongly interacting systems, incorporating the Poissonian mesostate occupancy is qualitatively and quantitatively important. In this aspect, a naive `mean-field' approximation of the microscopic interactions implies $\mathcal{V}_{ij} \approx \beta v_{ij}$, which hold only for small $\beta v_{ij}$ approximation of \cref{eq:mesoscopic_interaction_coefficients}.  $\mathcal{V}_{ij} > 0$ ( $\mathcal{V}_{ij} < 0$) implies that the interaction are repulsive (attractive). Therefore, repulsive (attractive) interactions of the mesostate increases (decreases) the mesoscopic Boltzmann weight compared to the non-interacting (ideal) counterpart. Physically, this implies that the repulsive (attractive) interactions with other mesostates make the mesostate thermodynamically unfavourable (favourable).  
}

The simplification of \cref{eq:lagrangian_defination} leads to the Lagrangian in the occupancy-noise picture.
\begin{equation}\label{eq:relation_hamiltonian_lagrangian_density_noise}
\begin{split}
    \mathcal{L} \left[ \{ N, \chi \} \right] = \vec{\chi} \cdot \partial_t \vec{N} - \mathcal{H} \left[ \{ N, \chi \} \right].
\end{split}
\end{equation}
\Cref{eq:relation_hamiltonian_lagrangian_density_noise} reveals the more familiar structure between the Hamiltonian and the Lagrangian, justifying their previous definitions. The transition probability measure \cref{eq:doi_peliti_path_integral} is reduced to,
\begin{equation}\label{eq:doi_peliti_path_integral_density_noise}
\begin{split}
    \mathcal{P} \left[ \left\{ N, \chi \right\} \right] 
    = e^{ -\mathcal{S}_{DP} \left[ \left\{ N, \chi \right\} \right] },
\end{split}
\end{equation}
with a normalization constraint \(\int \mathbb{ D } \{ N \} \: \mathbb{ D } \{ \chi \} \:  \mathcal{P} \left[ \left\{ N, \chi \right\} \right]  = 1\) for the probability conservation.
\(\mathbb{D}\) represents the path integral over all realizations of occupancy \(\{ N \}\) and conjugate noise fields \(\{ \chi \}\). Thus, the Cole-Hopf transformation \(\{\phi, \phi^*\} \to \{N, \chi\}\) illuminates the underlying relevant physical structure for the dynamics of classical many-body systems.
\subsubsection{Mesoscopic Energy Functional and Interaction Coefficients}\label{sec:mesoscopic_interaction_energy}
%
%
{
$\upmu_i^\#$ in \cref{eq:mesoscopic_boltzmann_weight} is further decomposed into its reciprocal and non-reciprocal contributions such that $\upmu_i^\# = \upmu_i^r + \mathcal{F}_i^{nr}$, where,  
\begin{equation}\label{eq:mesoscopic_boltzmann_weight_decomposition}
\begin{split}
    \upmu_i^r = \ln N_i^\# + \sum_{j} \mathcal{V}_{ij}^r N_j^\#, \hspace{0.8cm}
    \mathcal{F}_i^{nr} = \sum_{j} \mathcal{V}_{ij}^{nr} N_j^\#.
\end{split}    
\end{equation}
Similar to the microscopic decomposition, the decomposition of the interaction coefficient $\mathcal{V}_{ij} = \mathcal{V}_{ij}^r + \mathcal{V}_{ij}^{nr}$, by construction, satisfies the `actio=reactio' symmetry $\mathcal{V}_{ij}^r = \mathcal{V}_{ji}^r$ and anti-symmetry $\mathcal{V}_{ij}^{nr} = -\mathcal{V}_{ji}^{nr}$, here, $\mathcal{V}_{ij}^r = (\mathcal{V}_{ij} + \mathcal{V}_{ji})/2$, and $\mathcal{V}_{ij}^{nr} = (\mathcal{V}_{ij} - \mathcal{V}_{ji})/2$ leads to, 
\begin{equation}\label{eq:mesoscopic_interaction_coefficients_decomposition}
\begin{split}
    \mathcal{V}_{ij}^r = \left( \cosh{\left(\beta v^{nr}_{ij}\right)} e^{\beta v_{ij}^r} - 1 \right), \hspace{0.4cm}
    \mathcal{V}_{ij}^{nr} = \sinh{ \left( \beta v_{ij}^{nr} \right) } e^{\beta v_{ij}^r}.
\end{split}    
\end{equation}
Since, $\mathcal{V}_{ij}^r$ satisfies the `actio=reactio' symmetry, the reciprocal part of the interaction between the mesostate $N_i^\#$ and $N_j^\#$ is derived from the variation of a single energy functional of the lattice, which reads,
\begin{equation}\label{eq:mesoscopic_energy_funational}
    \mathcal{E}^{int} = \frac{1}{2} \sum_{i,j, \#} \mathcal{V}_{ij}^r N_i^\# N_j^\#. 
\end{equation}
The decomposition of $\mathcal{V}_{ij}$ into its reciprocal $\mathcal{V}_{ij}^{r}$ and non-reciprocal $\mathcal{V}_{ij}^{nr}$ parts is not necessarily unique. Subsequently, $\mathcal{E}^{int}$ is also not unique. Choosing a particular decomposition is a gauge-fixing, i.e., $\mathcal{V}_{ij}^r = \mathcal{V}_{ji}^r$ and $\mathcal{V}_{ij}^{nr} = - \mathcal{V}_{ji}^{nr}$. Utilizing the reciprocal and non-reciprocal parts as the symmetric and antisymmetric interactions is the unique gauge fixing corresponding to the orthogonal decomposition. This particular choice of gauge fixing uniquely preserves the microscopic orthogonal symmetry ($v_{ij}^r = v_{ji}^r$ and $v_{ij}^{nr} = -v_{ji}^{nr}$) on the mesoscale ($\mathcal{V}_{ij}^r = \mathcal{V}_{ji}^r$ and $\mathcal{V}_{ij}^{nr} = -\mathcal{V}_{ji}^{nr}$). Therefore, the microscopic orthogonal decomposition of interaction coefficients $v_{ij}$ ensures a unique orthogonal decomposition of the mesoscopic interaction coefficients $\mathcal{V}_{ij}$. Therefore, rthe mesoscopic interaction energy functional \cref{eq:mesoscopic_energy_funational} has a physical connection to the microscopic counterpart \cref{eq:microscopic_energy_functional}. 
}
\subsubsection{Mesoscopic Local detailed balance condition}
From the structure of the transition Hamiltonian \cref{eq:hamiltonian_reactive_transition_density_noise} on the mesoscopic level, it is evident that a microscopic reactive transition $\Delta_{\gamma\gamma'}^\#$ generates a mesoscopic reactive transition $N_{\gamma'}^\# \to N_{\gamma}^\#$ with transition rate $\mathcal{K}_{\gamma \gamma'}^\#$. Similarly, a microscopic diffusive transition $\Delta_i^{\vec{\mathcal{D}} \#}$ generates a mesoscopic diffusive transition $N_{i}^\# \to N_{i}^{\vec{\mathcal{D}} \#}$ with transition rate $\mathcal{K}_i^{\vec{\mathcal{D}} \#}$. The mesoscopic Local Detailed Balance condition is obtained using the mesoscopic transition Hamiltonian \cref{eq:hamiltonian_reactive_transition_density_noise}.
\begin{equation}\label{eq:mesoscopic_local_detailed_balance_reactive}
\begin{split}
    \frac{\mathcal{K}_{\gamma\gamma'}^\#}{\mathcal{K}_{\gamma'\gamma}^\#} = e^{ \upmu_{\gamma'}^\# - \upmu_{\gamma}^\# + \mathcal{F}_{\gamma \gamma'}^{ch} },
    \hspace{0.5cm}
    \frac{\mathcal{K}_i^{\vec{\mathcal{D}} \#}}{\mathcal{K}_i^{(\vec{\mathcal{D}} \#)^{-1}}} 
    = e^{ \upmu_{i}^\# - \upmu_i^{\vec{\mathcal{D}}\#} + \vec{\mathcal{D}} \cdot \vec{\mathcal{F}}_{i}^{sp} }.
\end{split}
\end{equation}
\Cref{eq:mesoscopic_local_detailed_balance_reactive} is the thermodynamically consistent identification of the microscopic Local detailed balance condition on the mesoscale. Hence, all transition dynamics for mesostates $\{N_i^\#\}$ are constrained/generated by their mesoscopic Boltzmann weights $\{\upmu_i^\#\}$.
\subsubsection{Summary of the section}
%
%
{
We summarize the progress done in this section.
First, using the transition Hamiltonian \cref{eq:hamiltonian_reactive_transition_density_noise}, we extracted the mesoscopic Boltzmann weight \cref{eq:mesoscopic_boltzmann_weight}, which quantifies the mesoscopic thermodynamic cost and connects the mesoscopic transition rates through \cref{eq:mesoscopic_local_detailed_balance_reactive}, thereby formulating a thermodynamically-consistent description of the mesostate dynamics. Second, \cref{eq:mesoscopic_interaction_coefficients} gives the exact non-linear renormalization of microscopic interaction coefficient that connects to the mesoscopic interaction coefficients. Third, the exact equivalence between the `coherent-state path integral' mesoscopic description
\cref{eq:lagrangian_defination,eq:hamiltonian_reactive_jump_sigle,eq:doi_peliti_path_integral} and the `stochastic path integral' description \cref{eq:hamiltonian_reactive_transition_density_noise,eq:doi_peliti_path_integral_density_noise,eq:relation_hamiltonian_lagrangian_density_noise} for thermodynamically-consistent interacting particle systems is related by the Cole-Hopf canonical transform \cref{eq:cole-hopf_transform}. Therefore, the suitable choice of path integral representation depends on the physical requirement: `coherent-state path integral' for quantum systems and `stochastic path integral' for classical stochastic systems. In the context of the example setup discussed in \cref{sec:cg_doi_peliti_action_summary}, the microscopic description is characterized by $16$ configuration variables; however, the mesoscopic description requires only $N_+, N_-$, which track the mean mesoscopic particle occupancy, and the conjugate fields $\chi_+, \chi_-$, which track transitions between mesostates.
}
\section{Coarse-graining: Mesoscopic to Macroscopic}\label{sec:cg_meso_to_macro}
%
%
{
The mesoscopic description is suitable for systems that exhibit a finitely small number of particles per lattice site. In addition, an increasing number of particles suppresses the importance of microscopic fluctuations in the observable mesoscopic quantities. Thus, the particle number and the transition current scale as $O(N)$, whereas the fluctuations scale as $O(1)$. In the limit of a large number of particles per lattice site, we define the macrostate density field $\rho_i(\vec{\mathbf{r}}) = N_i^\#/\Omega$. Here, the density $\rho_i$ is defined in accordance with the Large Deviation Theory \cite{Touchette_2009}, where the intensive variable $\rho_i$ is $O(1)$ and its fluctuations are $O(1/\Omega)$. For $\rho_i$, the lattice index is denoted using the spatial position vector $\vec{\mathbf{r}}$. In the following, all macroscopic physical quantities depend on their spatial location vector $\vec{\mathbf{r}}$, but their spatial field index is dropped. For example, $\rho_i$ instead of $\rho_i(\vec{\mathbf{r}})$, or $\mu_i$ instead of $\mu_i[\{\rho_i (\vec{\mathbf{r}}) \}]$. Similarly, the sum over the lattice site index $\sum_{\#}$ is replaced by its continuous counterpart $ \int \vec{\mathbf{r}}$, equivalent to fixing the lattice constant to $l=1$. The equivalence beween the discrete and continuous space description follows from using respective operator counterparts for the gradient, Laplacian and difference operator.
}

{
$\Omega$ quantifies the average number of particles per lattice site and is defined as $\Omega = N^{tot} / N^{lattice}$. Therefore, $\Omega$ should be interpreted as a free parameter that dictates the hydrodynamic scaling and thermodynamic limit. By construction, $\Omega \geq 1$, since $\Omega < 1 $ contradicts the mesoscopic-to-macroscopic coarse-graining. In the regime $N^{tot} < N^{lattice}$, the mesoscopic or microscopic description of the system becomes more appropriate. Thus, $\rho_i$ is defined as an intensive finite variable, whereas $N_i^\#$ is an extensive variable and becomes ill-defined in the thermodynamic limit of an infinite number of particles. This macroscopic limit is an inherent assumption in the formulation of chemical reaction networks and active matter models, which relies on the van Kampen closure expansion to formulate a coarse-grained description \cite{van_kampen}. Here, however, we will explicitly demonstrate in this \cref{sec:Lyapunov_functional} as a coarse-graining step from the mesoscopic to the macroscopic description, with a free parameter $\Omega$ that can be chosen depending on the system. This is particularly important for systems that require a description scale intermediate between the macroscale and mesoscale.
}
\subsection{Scale-invariance and renormalization of microscopic control parameters}\label{sec:scaling}
\subsubsection{Equilibrium Thermodynamics: Energy functional and Boltzmann weights}\label{sec:scaling_energy_functional}
%
%
{Incorporating the mesoscopic to macroscopic scaling $\Omega \rho_i = N_i^\#$ and \cref{eq:mesoscopic_boltzmann_weight},
the macroscopic Boltzmann weight $\mu_i$ for the density field $\rho_i$ is,
\begin{equation}\label{eq:macroscopic_boltzmann_weight}
    \mu_i = \ln{(\rho_i)} + \sum_{j} V_{ij} \rho_j,
\end{equation}
with the macroscopic interaction coefficient $V_{ij}$ that quantifies the interaction experienced by the macrostate $\rho_i$ due to the macrostate $\rho_j$, which reads,
\begin{equation}\label{eq:macroscopic_interaction_coefficients}
    V_{ij} = \Omega \left( e^{\beta v_{ij}} - 1 \right).
\end{equation}
\Cref{eq:macroscopic_interaction_coefficients} quantifies the non-linear relation between the microscopic and macroscopic interaction coefficients and correspond to the exact renormalization of microscopic interaction coefficients on macroscale.}

{The macroscopic interaction energy functional obeys the extensive scaling in $\Omega$. Thus, it satisfies $\Omega E^{int} = \mathcal{E}^{int}$, similar to the scaling between $\rho_i$ and $N_i$. It leads to the scaling between the macroscopic $V_{ij}^r$ ($V_{ij}^{nr}$) and the mesoscopic $\mathcal{V}_{ij}^r$ ($\mathcal{V}_{ij}^{nr}$) interaction coefficients, $V_{ij}^r = \Omega \mathcal{V}_{ij}^r$ and $V_{ij}^{nr} = \Omega \mathcal{V}_{ij}^{nr}$, hence, 
\begin{equation}\label{eq:macroscopic_interaction_coefficients_decomposition}
    V_{ij}^r = \Omega \left( \cosh{\left(\beta v^{nr}_{ij}\right)} e^{\beta v_{ij}^r} - 1 \right), \hspace{0.3cm} {V}_{ij}^{nr} = \sinh{ \left( \beta v_{ij}^{nr} \right) } e^{\beta v_{ij}^r},
\end{equation}
where, $V_{ij} = V_{ij}^r + V_{ij}^{nr}$, and $V_{ij}^r$ and $V_{ij}^{nr}$ inherit the `actio=reactio' symmetry and anti-symmetry of $\mathcal{V}_{ij}^r$ and $\mathcal{V}_{ij}^{nr}$, since, by definition, $V_{ij}^r = (V_{ij} + V_{ji}) / 2$ and $V_{ij}^{nr} = (V_{ij} - V_{ji}) / 2$. To enforce the microscopic Boltzmann weight $\epsilon_i^\#$ as an intensive physical quantity, $v_{ij}$ are $O(1/\Omega)$. Subsequently, this ensures that $V_{ij}\propto O(1), \mathcal{V}_{ij} \propto O(1/\Omega)$, physically this ensures the extensive scaling of the interaction energy functional across the observable scales; the microscopic, mesoscopic, and macroscopic. From a thermodynamic point of view, the thermodynamically consistent models need to ensure the intensive Boltzmann weights \cref{eq:macroscopic_boltzmann_weight} that lead to the extensive interaction energy \cref{eq:macroscopic_energy_functional}. This is an important feature to ensure the thermodynamic-consistency of the interactions, which has been overlooked in lattice gas models, where the interaction energy might scale super-/sub-extensively \cite{Mourtaza2018,martin2023nr_aim}. Here, a thermodynamically consistent framework across scales requires careful treatment. Importantly, in the high-temperature limit $\beta \to 0$ and hydrodynamics limit $\Omega \to \infty$, \cref{eq:macroscopic_interaction_coefficients_decomposition,eq:macroscopic_interaction_coefficients} satisfies the linear relationship $V_{ij} = \beta v_{ij}$, $V_{ij}^r = \beta v_{ij}^r$, and $V_{ij}^{nr} = \beta v_{ij}^{nr}$. The linear relation between $V_{ij}$ and $v_{ij}$ signifies the `mean-field' approximation of the microscopic interaction coefficients, which is valid due to the thermodynamic limit that suppresses the effect of the microscopic fluctuations and interactions.
}

{
Further, $\mu_i$ \cref{eq:macroscopic_boltzmann_weight} is decomposed into the reciprocal and non-reciprocal parts of it, such that $\mu_i = \mu_i^r + F_i^{nr}$ and are defined as,
\begin{equation}\label{eq:macrosocpic_boltzmann_weight_decomposition}
    \mu_i^r = \ln{(\rho_i)} + \sum_{j} V_{ij}^r \rho_j, \hspace{1cm} F_i^{nr} = \sum_j V_{ij}^{nr} \rho_j,
\end{equation}
$\mu_i^r$ and $F_i^{nr}$ physically quantify the total reciprocal and non-reciprocal thermodynamic interaction cost experienced by the macrostate $\rho_i$. Here, the first and second terms of $\mu_i^r$ are contributions due to the degeneracy of the macrostate (arising from the mapping of different microscopic particles to the same macrostate) and reciprocal interactions, respectively. Writing $\mu_i^r = { \delta E}/{\delta \rho_i}$ as the functional derivative of the energy functional, we decompose the total energy functional into its interaction part $E^{int}$ and the Boltzmann entropic contribution $S_b$. Then, the macroscopic energy functional reads,
\begin{equation}\label{eq:macroscopic_energy_functional}
\begin{split}
    E & = E^{int} - S^b,
    \\
    E^{int} = \frac 1 2 \int_{\vec{\mathbf{r}}} d \mathbf{r} \sum_{i,j} V_{ij}^r \rho_i \rho_j, & \hspace{0.5cm} S_b = -  \int_{\vec{\mathbf{r}}} \sum_i \rho_i \ln{ \left( {\rho_i}/{e} \right) } 
\end{split}
\end{equation}
The macroscopic external chemical/self-propulsion driving forces satisfy ${F}_{\gamma \gamma'}^{ch} = \mathcal{F}_{\gamma \gamma'}^{ch}$ and $\vec{F}_i^{sp} = \vec{\mathcal{F}}_i^{sp}$.
}
\subsubsection{Non-equilibrium dynamics: Dynamical rate functional for the hydrodynamic description}\label{sec:Lyapunov_functional}
Similar to $E$, the macroscopic Doi-Peliti action $\mathcal{S}_{DP} \left[ \{ \rho, \chi \} \right]$ follows the macroscopic scaling in $\Omega$. The scaling is $\Omega \mathcal{S}_{DP} \left[ \{ \rho, \chi \} \right] = \mathcal{S}_{DP} \left[ \{ N, \chi \} \right], \; \Omega \mathcal{L} \left[ \{ \rho, \chi \} \right] = \mathcal{L} \left[ \{ N, \chi \} \right]$ and $\Omega \mathcal{H} \left[ \{ \rho, \chi \} \right] = \mathcal{H} \left[ \{ N, \chi \} \right]$. This is trivially verified by substituting $\Omega \rho_i(\vec{\mathbf{r}}) = N_i^\#$ in \cref{eq:relation_hamiltonian_lagrangian_density_noise,eq:hamiltonian_reactive_transition_density_noise}. Hence, the macroscopic Lagrangian and Hamiltonian are obtained by $N_i^\# \to \rho_i$, $\chi_i^\# \to \chi_i$, $\upmu_i^\# \to \mu_i$ in \cref{eq:relation_hamiltonian_lagrangian_density_noise,eq:hamiltonian_reactive_transition_density_noise}. Their exact expressions read,
\begin{widetext}
\begin{equation}\label{eq:hamiltonian_reactive_transition_macroscopic}
\begin{split}
    \mathcal{L} \left[ \{ N, \chi \} \right] &= \vec{\chi} \cdot  \partial_t \vec{N} - \mathcal{H} \left[ \{ N, \chi \} \right],
    \\
    \mathcal{H}_{\gamma \gamma'} \left[ \{ \rho, \chi \} \right] &= d_{\gamma \gamma'}
   \bigg[ 
   \left( e^{ \chi_{\gamma'} - \chi_{\gamma} } - 1 \right) e^{ \mu_\gamma - \frac{1}{2} {F}_{\gamma \gamma'}^{ch} } 
   + \left( e^{\chi_{\gamma} - \chi_{\gamma'} } - 1 \right) e^{\mu_{\gamma'} + \frac{1}{2} \mathcal{F}_{\gamma \gamma'}^{ch} }
   \bigg], 
   \\
   \mathcal{H}_i^{\vec{\mathcal{D}} } \left[ \{ N, \chi \}\right]
   & = d_{i}
   \bigg[ 
   \left( e^{ \chi_i^{\vec{\mathcal{D}} } - \chi_i } - 1 \right) 
   e^{ \mu_i + \frac{1}{2} \vec{\mathcal{D}} \cdot \vec{{F}}_{i}^{sp} } + \left( e^{ \chi_i - \chi_i^{\vec{\mathcal{D}} } } - 1 \right) 
   e^{ \mu_i^{\vec{\mathcal{D}} } - \frac{1}{2} \vec{\mathcal{D}} \cdot \vec{{F}}_{i}^{sp} }
   \bigg].
\end{split}
\end{equation}
\end{widetext}
Then, the macroscopic transition probability measure $\mathcal{P} \left[ \left\{ \rho, \chi \right\} \right]$ is reduced to the following,
\begin{equation}\label{eq:macroscopic_doi_peliti_path_integral}
\begin{split}
    \mathcal{P} \left[ \left\{ \rho, \chi \right\} \right] 
    = e^{ -\Omega \mathcal{S}_{DP} \left[ \left\{ \rho, \chi \right\} \right] }.
\end{split}
\end{equation}
Here, $\Omega$ plays the role of the large deviation parameter. Hence, $\mathcal{P} \left[ \left\{ \rho, \chi \right\} \right]$ converges to the path of the most-likelihood obtained by extremizing $\mathcal{S}_{DP} \left[ \left\{ \rho, \chi \right\} \right]$. Importantly, $\mathcal{L} [\{ \rho,\chi \}]$ is the same rate functional as that previously obtained
to delineate the orthogonal decomposition symmetry of the EPR
\cite{Maes_2008,Mielke_2014_ldp,Mielke_2017,Kaiser_2018,Renger_2021,Peletier_2022,kobayashi_2022_hessian_geometry,kobayashi_2023_information_graphs_hypergraphs,Peletier_2023,Patterson_2024,Renger_2024,Kobayashi_2022,Sughiyama_2022,Renger_2023,duong_2023,Mizohata_2024,Loutchko_2023_geometry_tur}. Therefore, our novel formulation generalizes its existence to non-reciprocal systems and systematically extends the results from Refs. \cite{Maes_2008,Mielke_2014_ldp,Mielke_2017,Kaiser_2018,Renger_2021,Peletier_2022,kobayashi_2022_hessian_geometry,kobayashi_2023_information_graphs_hypergraphs,Peletier_2023,Patterson_2024,Renger_2024,Kobayashi_2022,Sughiyama_2022,Renger_2023,duong_2023,Mizohata_2024,Loutchko_2023_geometry_tur}. The thermodynamically consistent framework of non-reciprocal interacting systems enables correct identification of the entropy production rate discussed in Ref. \cite{atm_st_nr_2024}. For externally driven systems, the observable time-integrated current scales with the observation time $\tau$, in addition to the hydrodynamic scaling factor $\Omega$. This further leads to the dynamical scaling $\mathcal{S}_{DP} \left[ \left\{ \rho, \chi \right\} \right] = \tau \tilde{\mathcal{S}}_{DP} \left[ \left\{ \rho, \chi \right\} \right]$ and $\mathcal{L} \left[ \left\{ \rho, \chi \right\} \right] = \tau \tilde{\mathcal{L}} \left[ \left\{ \rho, \chi \right\} \right]$ \cite{atm_2024_var_epr,atm_2025_var_epr_derivation}, which quantifies the dynamical rate functional for the transition dynamics \cite{Maes_2020,Touchette_2009}.
\subsubsection{Non-equilibrium dynamics: Macrosocpic local detailed balance condition}\label{sec:macrosocpic_LDB}
Analogous to the mesoscopic LDB \cref{eq:mesoscopic_local_detailed_balance_reactive}, the macroscopic LDB is obtained using the macroscopic transition Hamiltonian $\mathcal{H}[\{ \rho, \chi\}]$. It reads,
\begin{equation}\label{eq:macroscopic_local_detailed_balance_reactive}
\begin{split}
    \frac{{K}_{\gamma\gamma'}}{{K}_{\gamma'\gamma}} = e^{ \mu_{\gamma'} - \mu_{\gamma} + F_{\gamma \gamma'}^{ch} },
    \hspace{0.5cm}
    \frac{{K}_i^{\vec{\mathcal{D}}}}{{K}_i^{(\vec{\mathcal{D}} )^{-1}}} 
    = e^{ \mu_{i} - \mu_i^{\vec{\mathcal{D}}} + \vec{\mathcal{D}} \cdot \vec{F}_{i}^{sp} }.
\end{split}
\end{equation}
where ${K}_{\gamma\gamma'}$ is the reactive transition rate $\rho_{\gamma'} \to \rho_\gamma$ and its reverse transition rate ${K}_{\gamma'\gamma}$ for $\rho_{\gamma} \to \rho_{\gamma'}$, defined between the density macrostates. Similarly, ${K}_i^{\vec{\mathcal{D}}}$ is the diffusive transition rate for the density field $\rho_i$ in the direction $\vec{\mathcal{D}}$. The transition affinities for the reactive and diffusive transitions are defined as $A_{\gamma \gamma'} = \mu_{\gamma'} -\mu_{\gamma} + F_{\gamma\gamma'}^{ch}$ and $A_{i}^{\vec{\mathcal{D}}} =  \mu_{i} - \mu_i^{\vec{\mathcal{D}}} + \vec{\mathcal{D}} \cdot \vec{F}_{i}^{sp}$, respectively. Where, $ \mu_i^{\vec{\mathcal{D}}}$ is the macroscopic Boltzmann weight defined at the spatial location after the diffusive transition in the direction $\vec{\mathcal{D}}$. 
\begin{table*}[t!]
\begin{tabular}{|m{1.5cm}|m{6.3cm}|m{8.7cm}|}
     \hline
     \centering
     \textbf{Noise's nature}
     \vspace{0pt}
     & \centering
     \textbf{Poissonian}
     \vspace{5pt}
     & \textbf{Gaussian} 
     \vspace{5pt}
     \\
     \hline
     Occupancy noise 
     \vspace{0pt}
     & The non-linear renormalization of $\mathcal{V}_{ij}$ \cref{eq:mesoscopic_interaction_coefficients}
     \vspace{5pt}
     & $\mathcal{V}_{ij}$ expanded for small values of $\beta, v_{ij}$ up to $O( (\beta v_{ij})^2 )$
     \vspace{5pt}
     \\
     \hline
     Transition noise
     \vspace{20pt}
     & The exact $\mathcal{S}_{DP}[\{N,\chi\}]$, without invoking the quadratic approximation in the small transition noise $\chi$ in \cref{eq:hamiltonian_reactive_transition_density_noise,eq:hamiltonian_reactive_transition_macroscopic}, captures the full `non-Gaussian' transition fluctuation effects, which are studied in detail in Ref.\cite{atm_2024_var_epr,atm_2025_var_epr_derivation,atm_2025_gftoc}.
     \vspace{0pt}
     & This work expands $\mathcal{S}_{DP}$ up to quadratic order in $\chi$ (transition noise), leading to the macroscopic (mesoscopic) stochastic EOMs \cref{eq:macroscopic_eom_mft} (\cref{eq:mesoscopic_eom_mft}) with multiplicative Gaussian noise for the macrostate (mesostate). This corresponds to the regime relevant for the hydrodynamic coarse-grained description pursued in this work.
     \vspace{0pt}
     \\
     \hline
\end{tabular}
\caption{ This table summarizes the implications of noise in the coarse-grained description. The noise in the occupancy statistics and in the transition statistics can be either Poissonian or Gaussian. For interacting systems, the Poissonian occupancy noise manifests through the non-linear dependence (renormalization) of the microscopic interaction coefficients $v_{ij}$ into the mesoscopic interaction coefficients $\mathcal{V}_{ij}$. The Gaussian approximation of the transition noise leads to the Langevin equations \cref{eq:macroscopic_eom_mft,eq:mesoscopic_eom_mft} for the fluctuating macro- or mesostate. In contrast, the thermodynamic implications of Poissonian transition noise — beyond the Gaussian approximation — are discussed in detail in Ref.~\cite{atm_2024_var_epr,atm_2025_var_epr_derivation,atm_2025_gftoc}. }
\label{table:noise_types}
\end{table*}
\subsection{Hamilton-Jacobi equation: Minimum action principle}\label{sec:cg_hamilton_jacobi_equation}
\Cref{eq:macroscopic_doi_peliti_path_integral} implies convergence to the most likely path in the limit $\Omega \to \infty$. The first-order variation of $\mathcal{S}_{DP}[\{\rho, \chi\}]$ gives $\delta \mathcal{S}_{DP}[\{\rho, \chi\}]$ gives the most likelihood path. The instanton equation, obtained by the variational extremization of \cref{eq:hamiltonian_reactive_transition_macroscopic}, reads, 
\begin{equation}\label{eq:hamilton_jacobi_instanton}
\begin{split}
    &\partial_t \rho_i 
    = \partial_{\chi_i} \mathcal{H} \left[ \{\rho, \chi \}\right]
    ,
    \\
    &\partial_t \chi_i 
    = - \partial_{\rho_i} \mathcal{H} \left[ \{\rho, \chi \}\right].
\end{split}    
\end{equation}
\Cref{eq:hamilton_jacobi_instanton} with $\chi_i = 0$ gives the deterministic continuity equation for $\rho_i$. The non-trivial solution of \cref{eq:hamilton_jacobi_instanton} ($\chi_i \neq 0$) corresponds to the instanton. Here, the instanton is defined as the minimum-action path that describes the transition trajectory between two metastable attractors. Due to the violation of time-reversal symmetry, the instanton does not necessarily coincide with the gradient-descent dynamics of the energy functional.

We are interested in the fluctuating dynamics of the macrostate within the basin of attraction of the fixed point. Thus, we aim to incorporate the Gaussian fluctuations of the transitions. The second-order variation $\delta^2 \mathcal{S}_{DP}[\{\rho, \chi\}]$ around the minimal-action path encodes Gaussian fluctuations from microscopic transitions. The amplitude of the Gaussian fluctuations equals the curvature of the Hamiltonian, hence,
\begin{equation}\label{eq:fluctuations_instanton}
\begin{split}
    T_{\gamma \gamma'}
    & = -\partial_{\chi_{\gamma}} \partial_{\chi_{\gamma'}} \mathcal{H} \left[ \{\rho, \chi \}\right]|_{\chi = 0}.
\end{split}    
\end{equation}  
$T_{\gamma \gamma'}$ and $T_{i}^{\mathcal{D}}$ correspond to the variance of the noise due to transitions $\Delta_{\gamma\gamma'}$ and $\Delta_i^{\mathcal{D}}$, respectively. The deterministic transition current in \cref{eq:hamilton_jacobi_instanton} can further be decomposed into its individual transition currents $J_{\gamma \gamma'} = \partial_{ \chi_{\gamma} - \chi_{\gamma'} } \mathcal{H} \left[ \{\rho, \chi \}\right]|_{\chi = 0}$. The derivation of \cref{eq:hamilton_jacobi_instanton,eq:fluctuations_instanton} is detailed in \cref{app:cummulants_transition_current}. This formulates the stochastic continuity equation for the macrostate/mesostate. Its further analysis is detailed in \cref{sec:generalized_mft}. 

The higher-order variations of the Doi-Peliti action satisfy $\delta^n \mathcal{S}_{DP} [\{ \rho, \chi\}] \propto O(1/\Omega^{n-1})$. Thus, $\Omega \to \infty$ corresponds to the deterministic limit, where the deterministic continuity \cref{eq:hamilton_jacobi_instanton} suffices. For significantly large values of $\Omega$, $\delta^2 \mathcal{S}_{DP} [\{ \rho, \chi\}]$ gives the dominant contribution to the transition noise, effectively incorporating the Gaussian fluctuations due to microscopic transitions. Mesoscopic systems prone to Poissonian transition noise require a careful treatment of the Poissonian noise \cite{atm_2024_var_epr,atm_2025_var_epr_derivation}.

The quadratic approximation of $\mathcal{S}_{DP}$ in $\chi_i$ leads to the Gaussian approximation of the transition noise, which is equivalent to the Onsager-Machlup functional \cite{Onsager_1953,Onsager_1953_2}, the Martin-Siggia-Rose action \cite{martin_1973}, or the Bausch-Janssen-Wagner-de Dominicis action \cite{bausch_1976,janssen_1976,dedominicis_1976}. The Gaussian transition noise underestimates the thermodynamic cost compared to the Poissonian transition noise. Subsequently, it impacts the far-from-equilibrium optimal control formulation of stochastic thermodynamics, in contrast to the close-to-equilibrium formulation \cite{atm_2024_var_epr,atm_2025_var_epr_derivation,atm_2025_gftoc}. \Cref{table:noise_types} summarizes the different regimes for the transition and occupancy noise.
\subsection{Generalized Macroscopic Fluctuating Dynamics}\label{sec:generalized_mft}
Using \cref{eq:hamilton_jacobi_instanton,eq:fluctuations_instanton}, the stochastic equation of motion for the macrostate $\rho_i$ analogous to Langevin equation, derived in \cref{app:cummulants_transition_current,app:cg_diffusive_hamiltonian}, reads,
\begin{equation}\label{eq:macroscopic_eom_mft}
\begin{split}
    \partial_t \rho_i 
    &= -\nabla \cdot \vec{J}_{i}^{\vec{\mathcal{D}}}
    - \sum_{ i \in \{\gamma \gamma'\} } J_{\gamma \gamma'} 
    + \sqrt{ \frac 1 \Omega} \nabla \cdot \left(
    \sqrt{ T_i^{\mathcal{D}} } 
    \: \vec{ \hat{\xi} }_i^{\mathcal{D}} \right)
    \\
    & \hspace{2cm} + \sqrt{ \frac 1 \Omega } \sum_{ i \in \{ \gamma \gamma' \} } \sqrt{ T^{\mathcal{R} }_{\gamma \gamma'} } 
    \: \hat{\xi}^{ \mathcal{R} }_{\gamma \gamma'},
\end{split}    
\end{equation}
where $\vec{\hat{\xi}}_i^{\mathcal{D}}$ and $\hat{\xi}^{\mathcal{R}}_{\gamma \gamma'}$ are white Gaussian noises with unit variance and vanishing mean, and $J_{\gamma \gamma'}$ and $\vec{J}_{i}^{\vec{\mathcal{D}}}$ are the deteministic transition currents for $\Delta_{\gamma \gamma'}: \rho_{\gamma'} \to \rho_\gamma$ and $\Delta_i^{\vec{\mathcal{D}}}: \rho_i \to \rho_i^{\vec{\mathcal{D}}}$, respectively. The exact expressions for the transition currents, derived in \cref{app:cummulants_transition_current,app:cg_diffusive_hamiltonian}, are
\begin{equation}\label{eq:macroscopic_transition_currents}
\begin{split}
     \vec{J}_{i}^{\vec{\mathcal{D}}} 
    & = - D_i^{\vec{{\mathcal{D}}}} \left( \left\{ \rho\right\} \right)
    \nabla^{\vec{\mathcal{D}}} \mu_i
    + \vec{J}_i^{sp}, 
    \\
    J_{\gamma \gamma'} 
    & = 2 D_{\gamma \gamma'} \left( \{ \rho \} \right) 
    \sinh{ \left( \frac 
    { A_{\gamma \gamma'}}{2} \right) },
\end{split}    
\end{equation}
%
%
{
where the mobilities of the diffusive transitions $\Delta_{i}^{\mathcal{D}}$ and reactive transitions $\Delta_{\gamma \gamma'}$ are denoted by $D_i \left( \left\{ \rho\right\} \right)$ and $D_{\gamma \gamma'} \left( \{ \rho \} \right)$, respectively, and are defined as,
\begin{equation}\label{eq:macroscopic_transition_mobility}
\begin{split}
    D_i \left( \left\{ \rho\right\} \right) & = \tilde{d}_i e^{ \mu_i^r + F_i^{nr} }, 
    \\
    D_{\gamma \gamma'} \left( \{ \rho \} \right) & = d_{\gamma \gamma'} e^{ \left( \mu_{\gamma} + \mu_{\gamma'} \right)/2 }. 
\end{split}    
\end{equation}
The transition mobilities quantify the amplitude of the transition currents in \cref{eq:macroscopic_transition_currents}, thereby justifying their nomenclature. The transition mobilities incorporate the reciprocal and non-reciprocal microscopic interactions through $\mu_i^r$ and $F_i^{nr}$. For ideal particles, $\mu_i = \log{(\rho_i)}$ and $F_i^{nr} = 0$, which reduces $D_i \left( \left\{ \rho\right\} \right) = \rho_i$ and $D_{\gamma \gamma'} \left( \{ \rho \} \right) = \sqrt{\rho_\gamma \rho_{\gamma'}}$, corresponding to the diffusive and reactive mobilities for ideal particles. However, due to the thermodynamically consistent coarse-graining, the mobilities depend on the interactions with other fields. $\mu_i$ in \cref{eq:macroscopic_boltzmann_weight} is increased (decreased) due to repulsive (attractive) interactions with other macrostates. This implies that the mobilities are increased (decreased) exponentially due to repulsive (attractive) interactions, which results in increased (decreased) transition currents in \cref{eq:macroscopic_transition_currents} from thermodynamically unfavorable (favorable) macrostates.
}

The diffusive and reactive transition traffic, derived in \cref{app:cummulants_transition_current,app:cg_diffusive_hamiltonian}, are,
\begin{equation}\label{eq:macroscopic_transition_traffic}
\begin{split}
    T_i^{\mathcal{D}} =  2 D_i^{\mathcal{D}} \left( \left\{ \rho\right\} \right),
    \hspace{0.5cm}
    T_{\gamma \gamma'} = 2 D_{\gamma \gamma'} \left( \{ \rho \} \right) 
    \cosh{ \left( \frac 
    { A_{\gamma \gamma'}}{2} \right) },
\end{split}    
\end{equation}
They quantify the variance of the fluctuations for transitions in \cref{eq:macroscopic_eom_mft}. For ideal particles, $\mu_i = \log{(\rho_i)}$ and $F_i^{nr} = 0$ reduces $T_i^{\mathcal{D}} = 2 \rho_i$ and $T_{\gamma \gamma'} = \rho_\gamma + \rho_{\gamma'}$. Using \cref{eq:macroscopic_transition_traffic,eq:macroscopic_transition_currents}, it is shown that the fluctuation-response relation between currents and traffics is satisfied, which reads $\partial J_{\gamma \gamma'} / \partial A_{\gamma \gamma'} = T_{\gamma \gamma'} / 2$ and $\partial \vec{J}_{i}^{\vec{\mathcal{D}}}/ \partial A_i^{\vec{\mathcal{D}}} = T_i^{\mathcal{D}}/2$. In \cref{eq:macroscopic_eom_mft}, the variance of $\rho_i$ converges to $0$ in the thermodynamic limit $\Omega \to \infty$, resulting in the deterministic description of the interacting macrostates corresponding to the limit of chemical reaction networks and reaction-diffusion systems.

Defining $\vec{\mathcal{D}} = \{ \parallel, \perp \}$ for the diffusive transitions in the directions parallel and perpendicular to $\vec{f}_i^{sp}$, the simplified form derived in \cref{app:cg_diffusive_hamiltonian} reads,
\begin{equation}\label{eq:macroscopic_diffusive_current_laplacian}
-\nabla \cdot \vec{J}_i^{\vec{\mathcal{D}}} = d_i^{\parallel} \Delta^{\parallel} e^{\mu_i} + d_i^{\perp} \Delta^{\perp} e^{\mu_i} + \nabla^{\parallel} \cdot \vec{J}_i^{sp},
\end{equation}
where $d_i^{\parallel} = d_i \cosh{(f_i^{sp}/2)}$ and $d_i^{\perp} = d_i$. The first and second terms correspond to the diffusive currents in the directions parallel and perpendicular to the self-propulsion force, respectively. The macroscopic self-propulsion current is given by
$\vec{J}_i^{sp} = 2 d_i e^{\mu_i} \sinh{\left( \vec{f}_i^{sp}/2 \right)}$. 

Defining the diffusion matrix $\mathbb{D}_i = \mathrm{Diag}[d_i^\parallel, d_i^\perp]$, the gradient vector $\vec{\nabla} = (\nabla^\parallel, \nabla^\perp)$, and the mobility matrix $\mathbb{M}_i = \mathrm{Diag}[d_i^\parallel e^{\mu_i}, d_i^\perp e^{\mu_i}]$, a compact shorthand notation for $-\nabla \cdot \vec{J}_i^{\vec{\mathcal{D}}}$ is
\begin{equation}\label{eq:macroscopic_diffusive_current_gradient}
-\nabla \cdot \vec{J}_{i}^{\vec{\mathcal{D}}} = \vec{\nabla} \cdot \mathbb{M}_i \cdot \vec{\nabla} \mu_i
+ \nabla^{\parallel} \cdot \vec{J}_i^{sp}.
\end{equation}
\Cref{eq:macroscopic_diffusive_current_gradient} is the equivalent simplification of \cref{eq:macroscopic_diffusive_current_laplacian} for the diffusive Laplacian term that allows to physically visualize the nomenclature of the diffusive mobility \cref{eq:macroscopic_transition_mobility}. 
{
Therefore, throughout this work, we interchangeably use the equivalent notations $\Delta \: e^{\mu_i} = \nabla \cdot ( e^{\mu_i} \nabla \mu_i )$ depending on the context.
}
The enhanced diffusion coefficient along the self-propulsion direction has been reported previously
\cite{Reimann_2001,Lindner_2016}. Importantly, the distinct transverse and longitudinal diffusion coefficients play a key role in the formation of novel phases \cite{Chatterjee_2020,Mangeat_2020,Chatterjee_2022,Karmakar_2023,Woo_2024}. Moreover, assuming $d_i^\parallel = d_i^\perp$ in the $l \to 0$ limit oversimplifies the coarse-grained diffusive dynamics of the macrostate. For microscopic systems with an inherent finite diffusive length scale, the anisotropic diffusion coefficients should be incorporated into the coarse-grained macroscopic description.

\subsection{Generalized Mesoscopic Fluctuating Dynamics}
Multiplying \cref{eq:macroscopic_eom_mft} by $\Omega$, one obtains the mesoscopic stochastic equation of motion (EOM) for the continuous stochastic variable $N_i^\#$ analogous to the Langevin equation,
\begin{equation}\label{eq:mesoscopic_eom_mft}
\begin{split}
    \partial_t N_i^\# 
    & = -\nabla \cdot \vec{\mathcal{J}}_{i}^{\vec{\mathcal{D}}}
    - \sum_{ i \in \{\gamma \gamma'\} } \mathcal{J}_{\gamma \gamma'}^\# 
    + \nabla \cdot \left(
    \sqrt{ \mathcal{T}_i^{\mathcal{D}} } 
    \: \vec{ \hat{\xi} }_i^{\mathcal{D}} \right)
    \\
    & \hspace{2cm} + \sum_{ i \in \{ \gamma \gamma' \} } \sqrt{ \mathcal{T}_{\gamma \gamma'}^{\#} } 
    \: \hat{\xi}^{ \mathcal{R} }_{\gamma \gamma'}.
\end{split}    
\end{equation}
The mesoscopic currents and traffic are analogous to \cref{eq:macroscopic_transition_currents,eq:macroscopic_transition_traffic}, obtained by replacing $\mu_i \to \upmu_i^\#$ and $\rho_i \to N_i^\#$. In addition, the discrete gradient and Laplacian operators are used in \cref{eq:mesoscopic_eom_mft,app:cg_diffusive_hamiltonian}. In \cref{eq:mesoscopic_eom_mft}, the variance of the noise appears to be $O(1)$; this apparent effect is attributed to considering the EOM for an extensive variable $N_i^\#$. \Cref{eq:mesoscopic_eom_mft} is more useful than \cref{eq:macroscopic_eom_mft} in the low-density regime, where the number of particles is finitely small, making the mean particle number (instead of density) a relevant physical observable to track.
\section{Applications}\label{sec:example}
In this section, we highlight a few secondary applications of `thermodynamically-consistent coarse-graining' beyond the primary application to non-reciprocally interacting particles, in whose context it was developed, detailed in Ref.~\cite{atm_st_nr_2024}.
\subsection{Comparison to other coarse-graining methods}\label{sec:example_other_methods}
\subsubsection{Thermodynamically--consistent coarse--graining method for diffusive dynamics of `interacting' particles}\label{sec:example_other_methods_kawasaki_dean}
%
%
{
We aim to highlight the improvement offered by our exact thermodynamically-consistent coarse-graining method compared to the Kawasaki-Dean coarse-graining method used for the coarse-grained description of diffusive dynamics of microscopically interacting particles \cite{Kawasaki_1994,Dean_1996}. In this setup \cite{Kawasaki_1994,Dean_1996}, there is only one type of particle, but the microscopic interaction coefficient between the particles depends on the distance between the particles within the interaction neighbourhood (lattice site, in our formulation) of the particle. In this context, within our model, we need to relax the assumption that the diffusive length-scale and the interaction length-scale are equal. Therefore, effectively, each microscopic particle at a different spatial location on the lattice site (interaction neighbourhood) needs to be treated as a distinct particle type; equivalently, extending the interactions to infinitesimally small multiple lattice sites defines the interaction neighbourhood. In this context, the equivalence between the different notations is given by $v_{ij} = v(x - x')$, $N_i = \rho(x)$, and $N_j = \rho(x')$. The number of particle types becomes infinite, since the spatial location $x'$ is a continuous variable, which changes the discrete summation over the particle type index $\sum_i$ to an integral over a continuous spatial index $\int dx'$. We highlighted the equivalence between these different notations before discussing the technical aspects.
}

Then, for particles of type $i$ and $j$ with locally confined interactions that depend on the distance between their relative positions on the lattice site (interaction neighbourhood), the `thermodynamically-consistent diffusive dynamics' of the density field \cref{eq:mesoscopic_eom_mft} reads,
\begin{equation}\label{eq:diffusive_mft}
\begin{split}
    \partial_t N_i = \Delta \left( N_i e^{ \sum_{j} N_j ( e^{\beta v_{ij} } - 1)} \right)  
    + \nabla \cdot \left( \sqrt{N_i e^{ \sum_j { N_j ( e^{\beta v_{ij} } - 1) }}} \: \: \vec{ \hat{\xi} }_i^{\mathcal{D}} \right)
\end{split}    
\end{equation}
with diffusive mobility $D_i = N_i e^{ \sum_j N_j ( e^{\beta v_{ij} } - 1)}$ for $N_i$.
%
%
{
Using \cref{eq:macroscopic_diffusive_current_gradient} for the reorganization of the Laplacian term leads to, 
\begin{widetext}
\begin{equation}\label{eq:diffusive_mft_2}
    \partial_t N_i = \nabla \cdot 
    \left[ 
    \underbrace{\left( N_i e^{ \sum_{j} N_j ( e^{\beta v_{ij} } - 1)} \right)}_{N_i} 
    \left( \nabla
    { \log{N_i} } 
    +  \nabla \left[ \sum_{j} N_j
    \underbrace{( e^{\beta v_{ij} } - 1)}_{\beta v_{ij}}  \right] \right)
    \right] 
    + \nabla \cdot \left( 
    \underbrace{
    \sqrt{ N_i e^{ \sum_j { N_j ( e^{\beta v_{ij} } - 1) }} } }_{ \sqrt{N_i}  } \: \: \vec{ \hat{\xi} }_i^{\mathcal{D}} \right),
\end{equation}
\end{widetext}
where, the power-series approximation of \cref{eq:diffusive_mft_2} for small $v_{ij}$ are denoted under under-braces. The $O(1)$ approximation in $v_{ij}$ of the diffusive mobility $D_i$ leads to the diffusive mobility for non-interacting (ideal) fields, $D_i = N_i$; the first term in the under-brace. Incorporating the zeroth-order and first-order [$O(1) + O(\beta v_{ij})$] approximation of the chemical potential gradient, 
$
\nabla \upmu_i = \nabla \log{N_i} + \nabla \left( \sum_j \beta v_{ij} N_j \right) + O([\beta v_{ij}]^2),$ leads to the second under-brace terms. It further simplifies the diffusive currents to $
\vec{J}_i^{\vec{\mathcal{D}}} = - D_i \: \nabla \upmu_i = \nabla N_i + N_i \nabla \left(\sum_{j} \beta v_{ij} N_j \right)
$. The $O(1)$ approximation in $v_{ij}$ of the diffusive mobility $D_i$ leads to the multiplicative noise amplitude for non-interacting (ideal) fields, $\sqrt{D_i} = \sqrt{N_i}$; the third term in the under-brace.
}

Therefore, combining them together, in the small values of $v_{ij}$ regime, \cref{eq:diffusive_mft_2} reduces to the Kawasaki-Dean equation \cite{Kawasaki_1994,Dean_1996}.
\begin{equation}\label{eq:diffusive_dean_equation}
\begin{split}
    \partial_t N_i = \Delta N_i + \nabla \cdot \left[ N_i \nabla \left( \sum_j \beta v_{ij} N_j \right) \right] + \nabla \cdot ( \sqrt{N_i} \: \vec{ \hat{\xi} }_i^{\mathcal{D}} )
\end{split}    
\end{equation}
Substituting $N_i = \rho(x)$, $v_{ij} = v(x-x')$, $N_j = \rho(x')$, and $\sum_j = \int dx'$, summing over all $x'$ in the interaction neighborhood of $x$, \cref{eq:diffusive_dean_equation} is exactly equal to the Kawasaki-Dean equation in its standard notation \cite{Kawasaki_1994,Dean_1996}, it reads,
\begin{equation}\label{eq:diffusive_dean_equation_old}
\begin{split}
    \partial_t \rho(x) & = \Delta \rho (x) + \nabla \cdot \left[ \rho(x) \nabla \left( \int_{x'} dx' \beta v(x-x') \rho(x') \right) \right] \\ 
   & \hspace{1cm} + \nabla \cdot ( \sqrt{\rho(x)} \: \vec{ \hat{\xi} }^{\mathcal{D}}(x) )
\end{split}
\end{equation}

Importantly, this exercise highlights that our methodology naturally incorporates spatially extended interactions by including non-local contributions to the microscopic Boltzmann weight. The conventional formulation of the Kawasaki-Dean equation is holds only in the regime, where interaction coefficients $v_{ij}$ are small, highlighting its shortcomings.

\begin{figure*}[t!]
\centering
\includegraphics[width=\textwidth]{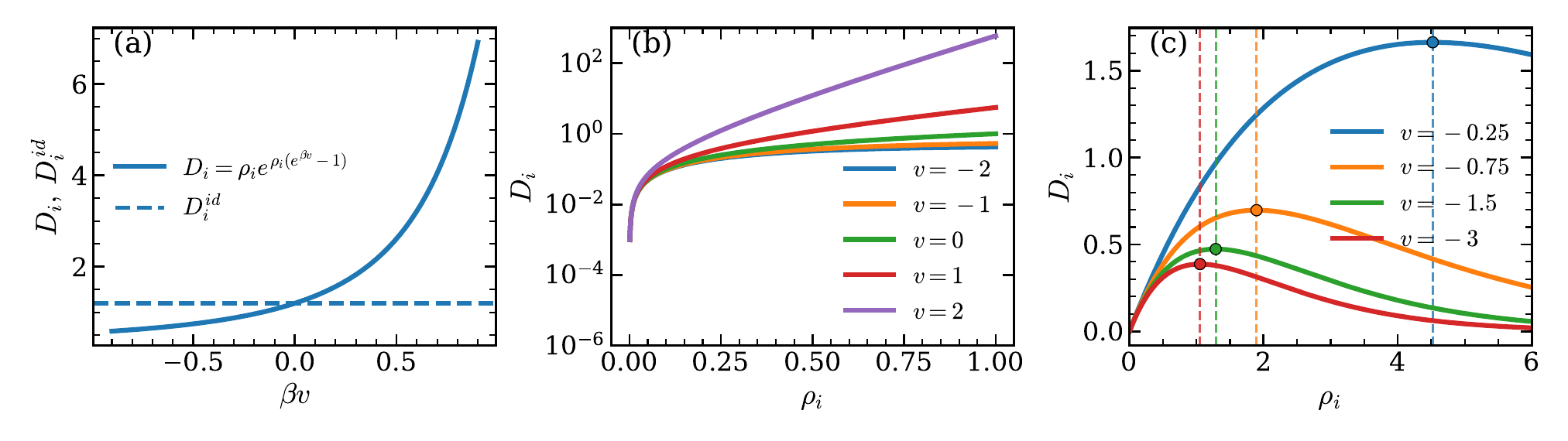}
\caption{ Diffusive mobility: (a) The diffusive mobility $D_i$ for the `thermodynamically-consistent diffusive dynamics' is compared with the Kawasaki--Dean equation $D_i^{id} = \rho_i$ for a fixed $\rho_i = 1.2$ as a function of $\beta v$. The quantitative asymmetry between attractive ($v < 0$) and repulsive ($v > 0$) interactions is evident. (b) The diffusive mobility as a function of $\rho_i$ is plotted for different values of $v = -2, -1, 0, 1, 2$ ($\beta =1$). (c) The attractive interaction regime ($v < 0$) exhibits a non-monotonic dependence of $D_i$ on $\rho_i$, with a maximum mobility $D_i^*$ at a finite density $\rho_i^*$ (dashed vertical lines) ($\beta =1$).}
\label{fig:1}
\end{figure*}

However, \cref{eq:diffusive_mft} represents the Langevin equation for strongly interacting particles, namely, the `thermodynamically-consistent diffusive dynamics', which in the notation of \cite{Kawasaki_1994,Dean_1996} reads,
\begin{widetext}
\begin{equation}\label{eq:generalized_kawasaki_dean_equation}
\begin{split}
    \partial_t \rho(x) & = \Delta \left( \rho(x) e^{ \left[ \int {dx'} \left( e^{\beta v(x-x') } - 1 \right) \rho(x')  \right] } \right)  
    + \nabla \cdot \left( \sqrt{
    \left( \rho(x) e^{ \left[ \int {dx'} \left( e^{\beta v(x-x') } - 1 \right) \rho(x')  \right] } \right)  
    } 
    \: \: \vec{ \hat{\xi} }^{\mathcal{D}} (x) \right)
\end{split}    
\end{equation}
\end{widetext}
%
%
{
In the Kawasaki-Dean equation \cref{eq:diffusive_dean_equation_old}, the fluctuations of the density fields quantified by the amplitude of the third term in \cref{eq:diffusive_dean_equation_old} depend only on the density $\sqrt{\rho(x)}$. Therefore, microscopic interactions (quantified by $v_{ij}$) do not influence density fluctuations, since the amplitude of variance in \cref{eq:diffusive_dean_equation_old} is the same for both ideal particles (non-interacting $v_{ij} = 0$) and strongly interacting particles (interacting $v_{ij} \neq 0$), irrespective of the values of $v_{ij}$. Despite its wide applicability, \cref{eq:diffusive_dean_equation_old} does not address this physical limitation, which is particularly important for strongly interacting particles. Similarly, this shortcoming arises again; even if a fluctuation-response relation is formulated using \cref{eq:diffusive_dean_equation_old}, it remains identical for both non-interacting (ideal) and interacting particles. In contrast, the fluctuation-response relation for interacting fields is naturally accessible using the `thermodynamically-consistent diffusive dynamics' \cref{eq:generalized_kawasaki_dean_equation}. This difference arises fundamentally from the thermodynamically-consistent coarse-graining, which incorporates the effects of microscopic interactions on the macroscopic mobility through \cref{eq:macroscopic_transition_mobility}. The difference between \cref{eq:generalized_kawasaki_dean_equation} and \cref{eq:diffusive_dean_equation_old} is traced back to the microscopic model definitions \cref{eq:microscopic_transition_rate} and \cref{eq:microscopic_local_detailed_balance_final}, respectively. The equivalence class of microscopic models defined by \cref{eq:microscopic_local_detailed_balance_final} inherently omits the time-symmetric part of transitions, which is associated with the transition mobility \cref{eq:macroscopic_transition_mobility}.
}

{
From a technical point of view, there is a major fundamental difference between \cref{eq:generalized_kawasaki_dean_equation} and the Kawasaki--Dean equation in Refs.\cite{Kawasaki_1994,Dean_1996} that deserves to be elaborated. In particular, \cref{eq:generalized_kawasaki_dean_equation} is defined for a continuous stochastic variable $\rho(x)$, a mesoscopic field, in contrast to the Dirac-delta functions $\delta(x)$ that appear in the Kawasaki--Dean equation of Refs.~\cite{Kawasaki_1994,Dean_1996}, where Dirac-delta-type initial-value conditions have been shown to be essential for the mathematically non-trivial solutions of the Kawasaki--Dean equation \cite{Konarovskyi_2019,Konarovskyi_2020}. From a practical standpoint, however, numerically solving the Kawasaki--Dean equation runs into numerical instabilities due to these Dirac-delta functions \cite{Jin_2026}. In contrast, the continuous stochastic variable $\rho(x)$ -- the mesoscopic field defined in \cref{eq:generalized_kawasaki_dean_equation} -- involves no Dirac-delta-type function, owing to the discrete lattice with a finite diffusive length scale, a form of regularization that addresses the numerical divergence \cite{Jin_2026}.
}

To demonstrate the analytical differences between the `thermodynamically-consistent diffusive dynamics' and the Kawasaki--Dean equation, we consider a single particle type confined to a lattice site, i.e., $v(x-x') = v$. The mobility $D_i = \rho_i e^{\rho_i (e^{\beta v} - 1)}$ is plotted against its ideal counterpart $D_i^{id} = \rho_i$ (the entropic contribution), which appears in the Kawasaki--Dean equation; see \cref{fig:1}\textcolor{red}{(a)}. The mobility scales exponentially with $\beta v$, and the sub-case $v = 0$, with $D_i^{id} = \rho_i$, corresponds to the Kawasaki--Dean equation; see \cref{fig:1}\textcolor{red}{(b)}. Interestingly, attractive interactions ($v < 0$) exhibit a non-monotonic dependence of $D_i$ on $\rho_i$ due to the interplay between the entropic and energetic contributions to $D_i$; see \cref{fig:1}\textcolor{red}{(c)}. It follows straightforwardly that the optimum density is $\rho_i^*(\beta v) = 1/(1 - e^{\beta v})$, with the corresponding maximum mobility given by $D_i^*(\beta v) = e^{-1} / (1 - e^{\beta v})$. The physical implications of such a non-monotonic dependence of mobility remain to be explored in future work. 

In summary, this section highlights that `thermodynamically-consistent diffusive dynamics' contributes to a new equivalence class of diffusive dynamics not necessarily captured by the `Kawasaki-Dean equation'.

\subsubsection{Classical Stochastic Path Integral formulism for reactive transition dynamics}\label{sec:example_other_methods_classical_SPI}
We highlight the importance of the exact coarse-graining procedure by comparing our results with the classical stochastic path integral formalism (CSPIF) in Ref.~\cite{Lefevre_2007}, which has been extensively utilized as a preferred coarse-graining method for interacting particles. To this end, we consider only two species of particles that interact within a single volume, which is equivalent to $N$ spins that can take positive ($+$) or negative ($-$) values, ferromagnetically interacting with strength $v/N$ ($v>0$). Particles with the same/opposite spins attract/repel each other. Hence, the microscopic interaction rules are defined as $v_{++} = v_{--} = -v/N$ and $v_{+-} = v_{-+} = v/N$.

We define the total particle number $N = N_+ + N_-$ (which is constant) and the magnetization $M = N_+ - N_-$. For this microscopic system, the exact coarse-grained EOM \cref{eq:mesoscopic_eom_mft} reads,
\begin{equation}\label{eq:ising_spins_reactive_dynamics_dpft}
\begin{split}
    \partial_t N_+ & = \mathcal{J}_{+-} + \sqrt{\mathcal{T}_{+-}} \: \hat{\xi}^{ \mathcal{R} }_{+-}
    \\
    \partial_t N_- & = -\mathcal{J}_{+-} - \sqrt{\mathcal{T}_{+-}} \: \hat{\xi}^{ \mathcal{R} }_{+-}
\end{split}    
\end{equation}
with the spin-flipping transition current $\mathcal{J}_{+-}$ and traffic $\mathcal{T}_{+-}$
\begin{widetext}
\begin{equation}\label{eq:ising_spins_current_dpft}
\begin{split}
    \mathcal{J}_{+-} & = d_{+-} 
    \left( e^{ N \left( \cosh{ \left[ \frac{\beta v}{N} \right] } - 1 \right) } \right)
    \left[ N \sinh{ \left( M \sinh{\left[ \frac{\beta v}{N} \right]} \right)}  - M \cosh{ \left (M \sinh{ \left[ \frac{\beta v}{N} \right]} \right) } \right] 
    \\
    \mathcal{T}_{+-} & = d_{+-}
    \left( e^{N \left( \cosh{ \left( \frac{\beta v}{N}  \right) } - 1 \right)} \right)
    \left[ N \cosh{ \left( M \sinh{\left[ \frac{\beta v}{N}  \right]} \right) }  - M \sinh{ \left( M \sinh{ \left[ \frac{\beta v}{N} \right] } \right) } \right] 
\end{split}    
\end{equation}
\end{widetext}
%
Here, the mesoscopic interaction coefficients $\mathcal{V}_{++} = \mathcal{V}_{--} = e^{-\beta v /N} - 1$ and $\mathcal{V}_{+-} = \mathcal{V}_{-+} = e^{\beta v / N} - 1$ were used to derive the reactive transition current in \cref{eq:ising_spins_current_dpft}.

\begin{figure}[b!]
\includegraphics[width=\linewidth]{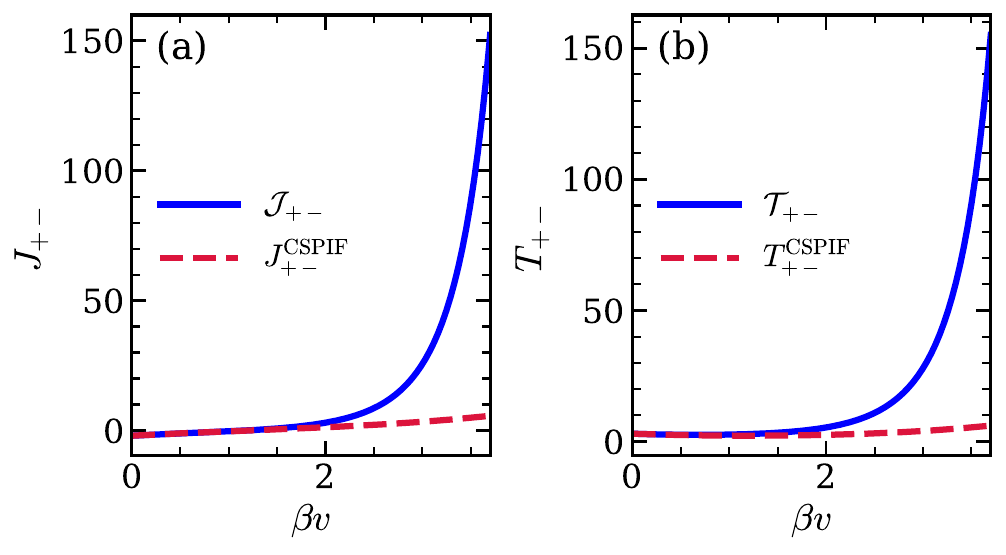}
\caption{ The parameter values $M=2, N=3, d_{+-} = 1$ are used. (a) ${J}_{+-}^{CSPIF}$ in \cref{eq:ising_spins_reactive_dynamics_classical} (red dashed line) and $\mathcal{J}_{+-}$ in \cref{eq:ising_spins_current_dpft} (blue solid line) are plotted. (b) ${T}_{+-}^{CSPIF}$ in \cref{eq:ising_spins_reactive_dynamics_classical} (red dashed line) and $\mathcal{T}_{+-}$ in \cref{eq:ising_spins_current_dpft} (blue solid line) are plotted.}
\label{fig:2}
\end{figure}

Using CSPIF, the coarse-grained EOM for the spin-flipping dynamics has the same structure as \cref{eq:ising_spins_reactive_dynamics_dpft}, but with a different spin-flipping current ($J_{+-}^{CSPIF}$) and traffic ($T_{+-}^{CSPIF}$), which read \cite{Lefevre_2007},
\begin{equation}\label{eq:ising_spins_reactive_dynamics_classical}
\begin{split}
    & {J}_{+-}^{CSPIF} = d_{+-} 
    \left[ N \sinh{ \left( \frac{\beta v M}{N} \right)} - M \cosh{ \left( \frac{\beta v M}{N} \right) } \right]
    \\
    & {T}_{+-}^{CSPIF} = d_{+-} 
    \left[ N \cosh{ \left( \frac{\beta v M}{N} \right)} - M \sinh{ \left( \frac{\beta v M}{N} \right) } \right]
\end{split}    
\end{equation}
%
%
The small $\beta v / N$ approximation for hyperbolic functions, $\sinh{(\beta v / N)} \approx \beta v/N$, allows a mean-field approximation for the particle occupancy. Using this, we find that \cref{eq:ising_spins_reactive_dynamics_classical} corresponds to the deterministic limit of the exact EOM \cref{eq:ising_spins_current_dpft}.

For large values of $\beta v M/N$, which corresponds to small $N$, large $\beta$, large $v$, and $M>0$ (i.e., in the ordered phase of the FM model), the differences between \cref{eq:ising_spins_current_dpft,eq:ising_spins_reactive_dynamics_classical} become significant: \cref{eq:ising_spins_current_dpft} predicts a much larger current and substantially larger fluctuations (traffic) in the vicinity $M \approx N$ (i.e., close to the FM ordered state). The origin of this discrepancy is the mean-field approximation of the particle numbers $N_+$ and $N_-$ in CSPIF, which treats the stochastic variables as deterministic quantities. Within DPFT, this issue is carefully handled by construction due to the requirement of normal-ordering, which simplifies the second-quantized formulism to the coherent state path integral. For non-interacting particles, or for transition rates that do not depend on particle occupancy, the mean-field treatment of transition rates in CSPIF is equivalent to DPFT. This is visually demonstrated by \cref{fig:2}.
\subsection{Active Ising Model}
Flocking models exhibit a first-order transition from an ordered to a disordered phase, accompanied by an intermediate phase-coexistence regime \cite{Solon2013,Solon2015}. The width of the coexistence regime decreases with increasing total density, such that the limit of infinite total density corresponds to a second-order transition. The Active Ising Model (AIM) is a prototypical and highly simplified flocking model that captures the above-mentioned physical characteristics of flocking systems \cite{Solon2013,Solon2015}. However, the coarse-grained description of AIM derived in Ref.~\cite{Solon2013,Solon2015} fails in the low-particle-number regime, incorrectly predicting both the order and onset of the phase transition. Here, we emphasize the importance of microscopic interaction coefficients, Poissonian particle occupancy, and the exact coarse-graining methods developed in this work.

The AIM models flocking with only two types of particles, positive $(+)$ and negative $(-)$, with self-propulsion along the $+$ and $-$ horizontal axes, respectively. The microscopic alignment is modelled via attractive and repulsive interactions between particles of the same and different types. Hence, $v_{++}, v_{--} < 0$ and $v_{+-}, v_{-+} > 0$.
{
In the context of the AIM model definitions in Ref.~\cite{Solon2013,Solon2015}, the interaction coefficients are locally defined intensive variables, $v_{ij} \propto 1/N$, where $N$ is the total number of particles. In contrast, in our framework, $v_{ij} \propto 1/\Omega$ are globally defined intensive variables, independent of the local particle number $N$, and treat the global total density as a control parameter of the model (see the definition of $\Omega$ in \cref{sec:cg_meso_to_macro}). The stochastic nature of particle occupancy $N$ is identified as a key factor affecting the accuracy of coarse-grained descriptions, particularly in low-density regimes. Our exact thermodynamically-consistent coarse-graining explicitly incorporates these occupancy fluctuations, ensuring the correct treatment of interaction-dependent mobilities and transition currents.
}

{
For the AIM, the assumption of constant $v_{ij}$ holds in both very low and very high particle-per-lattice-site limits. When the particle number is low, the minimum number of interacting particles is $1$, giving $v_{ij} \propto 1$. When the particle number is high, $N \gg 1$, $1/N$ is effectively constant and microscopic stochastic fluctuations of $N$ have negligible impact. Therefore, in the high-density limit, the `mean-field' approximation of the microscopic interactions becomes valid, consistent with the scaling relations discussed in \cref{sec:scaling_energy_functional}. Our approach thus systematically resolves the low-density failures of conventional coarse-grained descriptions, such as those in Refs.~\cite{Solon2013,Solon2015}, by properly accounting for the Poissonian occupancy noise and its effect on mesoscopic and macroscopic dynamics.
} 

In the context of regimes where our methodology is applicable, we consider the AIM, in which particles can undergo diffusive or reactive transitions. The reactive transitions are formulated to be thermodynamically consistent. In contrast to the thermodynamically consistent diffusive transitions developed in this work, in the original AIM, the particles have constant diffusive transition rates in addition to the biased self-propulsion. In the following, we elaborate on two regimes corresponding to the low- and high-density limits of the AIM, as different microscopic interaction coefficient rules govern each regime. This distinction plays a crucial role in resolving the physical discrepancies between the low- and high-density regimes.
\subsubsection{Low density regime}
In the low-density regime, the minimum number of interacting particles is one. Hence, the low-density limit corresponds to the microscopic interaction coefficients $v_{+-} = v_{-+} = 1$ and $v_{++} = v_{--} = -1$. This leads to mesoscopic interaction coefficients $\mathcal{V}_{+-} = \mathcal{V}_{-+} = e^{\beta} - 1$ and $\mathcal{V}_{++} = \mathcal{V}_{--} = e^{-\beta} - 1$, with mesoscopic Boltzmann weights
\begin{equation}
\begin{split}
\upmu_+^\# = \log{(N_+^\#)} - \sinh{(\beta)} M^\# + \left( \cosh{(\beta)} - 1 \right) N^\# , \\
\upmu_-^\# = \log{(N_-^\#)} + \sinh{(\beta)} M^\# + \left( \cosh{(\beta)} - 1 \right) N^\# .
\end{split}
\end{equation}
Here, we denote the total number and magnetization of particles by $N^\# = N_+^\# + N_-^\#$ and $M^\# = N_+^\# - N_-^\#$, respectively. The EOM \cref{eq:mesoscopic_eom_mft} for the mean particle number at each lattice site reads:
\begin{widetext}
\begin{equation}\label{eq:aim_low_density}
\begin{split}
    \partial_t N_+^\# 
    & = -\nabla \cdot \vec{\mathcal{J}}_{+}^{\vec{\mathcal{D}}}
    + \mathcal{J}_{+-}^\# 
    + \nabla \cdot \left(
    \sqrt{ \mathcal{T}_+^{\mathcal{D}} } 
    \: \vec{ \hat{\xi} }_+^{\mathcal{D}} \right)
    + \sqrt{ \mathcal{T}^{\mathcal{R} }_{+-} } 
    \: \hat{\xi}^{ \mathcal{R} }_{+-}
    \\
    \partial_t N_-^\# 
    &= -\nabla \cdot \vec{\mathcal{J}}_{-}^{\vec{\mathcal{D}}}
    - \mathcal{J}_{+-}^\# 
    + \nabla \cdot \left(
    \sqrt{ \mathcal{T}_-^{\mathcal{D}} } 
    \: \vec{ \hat{\xi} }_-^{\mathcal{D}} \right)
    - \sqrt{ \mathcal{T}^{\mathcal{R} }_{+-} } 
    \: \hat{\xi}^{ \mathcal{R} }_{+-}
    \\
    \mathcal{J}_{+-}^\# & = 
    d_{+-} \left( e^{N^\#( \cosh{(\beta)} - 1)} 
    \left[ N^\# \sinh{( M^\# \sinh{[\beta]} )}  
    - M^\# \cosh{ (M^\# \sinh{[\beta]} ) } \right] \right)
    \\
    \mathcal{T}_{+-}^\# & = 
    d_{+-} \left( e^{N^\#( \cosh{(\beta)} - 1)} 
    \left[ N^\# \cosh{( M^\# \sinh{[\beta]} )} - M^\# \sinh{ (M^\# \sinh{[\beta]} ) } \right] \right)
\end{split}    
\end{equation}
\end{widetext}
The thermodynamically inconsistent diffusive transition rates involve the modification of the diffusive mobilities $D_+^{\mathcal{D}} = d N_+^\#$ and $D_-^{\mathcal{D}} = d N_-^\#$. Hence, $-\nabla \cdot \vec{\mathcal{J}}_{+}^{\vec{\mathcal{D}}} = d^{\parallel} \Delta^{\parallel} N_+^\# + d^{\perp} \Delta^{\perp} {N_+^\#} - 2 d \sinh{\left( \frac{f^{sp}}{2} \right)} \nabla^{\parallel} N_+^\#$ and $-\nabla \cdot \vec{\mathcal{J}}_{-}^{\vec{\mathcal{D}}} = d^{\parallel} \Delta^{\parallel} N_-^\# + d^{\perp} \Delta^{\perp} {N_-^\#} + 2 d \sinh{\left( \frac{f^{sp}}{2} \right)} \nabla^{\parallel} N_-^\#$. 

The deterministic part of the EOM \cref{eq:aim_low_density} is equivalent to the one obtained in Ref.~\cite{Mourtaza2018}, where it was derived by introducing Poissonian statistics into the density. However, \cref{eq:aim_low_density} is also valid for higher values of $f_{sp}$ and incorporates stochasticity using multiplicative noise, going beyond the formulations in Ref.~\cite{Mourtaza2018}. The linear stability analysis of faster reactive transition currents exhibits a first-order phase transition from the ordered to the disordered phase and an intermediate coexistence regime \cite{Mourtaza2018}. In contrast, the mean-field hydrodynamic EOM fails to capture the ordered-to-disordered first-order phase transition \cite{Solon2015}.
\subsubsection{High density regime}
In the high-density regime, each particle is surrounded by $N >> 1$ particles, where $N$ is the average number of particles per lattice site. Hence, the microscopic interaction coefficients in the high-density limit are $v_{+-} = v_{-+} = 1/N$ and $v_{++} = v_{--} = -1/N$, which ensures that $\mu_+$ and $\mu_-$ are intensive. 
{Note that here the interaction coefficient $v_{ij}$ depends on the total particle number $N$, thus this is an approximation to incorporate the density-dependent $v_{ij}$. For sufficiently large $N$, this is a valid/good approximation, because $v_{ij}$ is infinitesimally small and does not change significantly as $N$ changes. When $N$ is significantly finite, this approximation does not necessarily hold.}

This leads to $V_{+-} = V_{-+} = \beta/\rho$ and ${V}_{+-} = {V}_{-+} = -{\beta}/\rho$. Thus, $\mu_+ = \log{(\rho_+)} + \beta \left( \rho_- - \rho_+ \right)/\rho$ and $\mu_- = \log{(\rho_-)} + \beta \left( \rho_+ - \rho_- \right)/\rho$, where the total density and magnetization are defined as $\rho = \rho_+ + \rho_-$ and $m = \rho_+ - \rho_-$, respectively. The reactive transition currents read $J_{+-} = \rho \sinh{ \left( \beta \frac m \rho \right)} - m \cosh{ \left( \beta \frac m \rho \right) }$. The mean-field EOM for the particle densities reads:
\begin{equation}\label{eq:aim_high_density}
\begin{split}
    \partial_t \rho_+ 
    & = {d}^{\parallel} \Delta^{\parallel} \rho_+ + {d}^{\perp} \Delta^{\perp} {\rho_+} - 2{d} \sinh{\left(\frac{f^{sp}}{2}\right)} \: \nabla^{\parallel} \rho_+
    + {J}_{+-}
    \\
    \partial_t \rho_- 
    & = {d}^{\parallel} \Delta^{\parallel} \rho_- + {d}^{\perp} \Delta^{\perp} {\rho_-} + 2{d} \sinh{\left(\frac{f^{sp}}{2}\right)} \: \nabla^{\parallel} \rho_-
    - {J}_{+-} 
\end{split}    
\end{equation}
The EOM \cref{eq:aim_high_density} are equal to the mean-field equations from Ref.\cite{Solon2013,Solon2015}, with different transverse and longitudinal diffusion coefficients. Taking the $l \to 0$ limit implies $d_i^{\parallel} = d_i^{\perp} = \tilde{d}_i$, which leads to the exact mean-field equations from Ref.\cite{Solon2013}. The linear stability analysis of \cref{eq:aim_high_density} exhibits a second-order phase transition from ordered to disordered phase \cite{Solon2013}. The fluctuations of dominant order ($O(1/\sqrt{\Omega})$) to \cref{eq:aim_high_density} are incorporated using the multiplicative noise $\sqrt{T_{+-}/\Omega} \xi_{+-}^\mathcal{R}$, with $T_{+-} = \rho \cosh{ \left( \beta \frac m \rho \right)} - m \sinh{ \left( \beta \frac m \rho \right) }$.
{As emphasized earlier, the difference between the transverse and longitudinal diffusion coefficients $d^{\parallel}$ and $d^{\perp}$ is relevant for other novel phases in AIM \cite{Chatterjee_2020,Mangeat_2020,Chatterjee_2022,Karmakar_2023,Woo_2024}.}

In conclusion, our analysis reveals the key role that microscopic interaction coefficients play in bridging the low- and high-density regimes of AIM. In particular, the low-density and high-density regimes of the AIM belong to different universality classes of EOM, which results in first-order and second-order phase transitions from the disordered to ordered phase, respectively, studied rigorously in Ref.\cite{Mourtaza2018} and Ref.\cite{Solon2015}. 
{However, the generic applicability of our methodology to lattice gas models (for the intermediate particle number regimes) that follow the local constraint on the intensiveness of the microscopic interactions remains an open problem, such as AIM, or Vicsek model. As highlighted here, the key role is played by the exact renormalization of the microscopic interactions (and the multiplicative noise for the transition dynamics as well). Obviously, for such models, the sophistication associated with the exact coarse-graining is subject to incorporating different microscopic physical effects on the macroscale; therefore, the relevance of a suitable mathematical approximation might be a necessity. For instance, as highlighted in this section, the difference in the order of the phase transition and the onset of the phase transition in the low-density and high-density regimes is sufficiently explained by the non-linear (exponential) and linear renormalization of the microscopic interaction coefficients, respectively. }
\subsection{ The `model-independent' equivalence between the stochastic and coherent-state path integral } \label{sec:example_equivalence}
%
%

{
The Cole-Hopf transform \cref{eq:cole-hopf_transform} defined between the `mesoscopic' coherent-state path integral 
\cref{eq:lagrangian_defination,eq:hamiltonian_reactive_jump_sigle,eq:doi_peliti_path_integral} 
and the `mesoscopic' stochastic path integral 
\cref{eq:hamiltonian_reactive_transition_density_noise,eq:doi_peliti_path_integral_density_noise,eq:relation_hamiltonian_lagrangian_density_noise} 
is a change of variable defined between `the mesoscopic variables (numbers) and not operators'. It should be emphasized that for this statement to be correct, the coherent-state path integral must be derived for normal-ordered transition rate operators, as elaborated in \cref{sec:cg_normal_ordering}. If the normal-ordering is neglected and one simply replaces 
in \cref{eq:hamiltonian_quantised_single_reactive_jump_1} operators $\crtlong{i}{\#}$ and $\anhlong{i}{\#}$ by fields $\phi_i$ and $\phi_i^*$ (denoted as ``microscopic Cole-Hopf transformation"), as is done for instance in Ref.\cite{Lefevre_2007}, which claims that one cannot interpret  $\phi_i \phi_i^*$ as the particle number $N_i$ any more, when the transition rates are non-linear functions of the operators $\crtlong{i}{\#}$ and $\anhlong{i}{\#}$. The `mesoscopic' Cole-Hopf transform \cref{eq:cole-hopf_transform} implements a change of variables that preserves the probability measure. 

{
We demonstrate this through a prototypical two-particle occupancy on a single lattice site, a lowest-order non-linear observable. The physically correct microscopic second-quantized operator two-particle occupancy is $\hat{O} = (\hat{N}_A)(\hat{N}_A - 1)$, 
which accounts for the fact that after selecting the first particle, there is 
one fewer choice for the second. This \textit{combinatorial} structure is an 
intrinsic microscopic physical feature that must be preserved. The corresponding 
normal-ordered expression for this two-particle occupancy is 
$:(\hat{N}_A)(\hat{N}_A - 1):$, which expands as 
$:(\hat{N}_A)^2: \;-\; :(\hat{N}_A):$. Evaluating this via the coherent-state 
path integral yields, $\langle \hat{O} \rangle = (\phi_A \phi_A^*)^2 + (\phi_A \phi_A^*) - (\phi_A \phi_A^*) = (\phi_A \phi_A^*)^2$, 
where the two intermediate terms cancel exactly. Applying the Cole-Hopf 
transform~\cref{eq:cole-hopf_transform} then gives $\langle \hat{O} \rangle = 
(N_A)^2$, with the mesoscopic continuous field $N_A$ capturing particle number. Notably, the discrete microscopic structure is no longer manifest 
at the mesoscopic definition of $N_A$, signifying a coarse-graining that loses microscopic discreteness. This result generalises straightforwardly: an 
$m$-particle occupancy is correctly represented by 
the second-quantized operator
$\hat{O} = (\hat{N}_A)(\hat{N}_A - 1)\,\cdots\,(\hat{N}_A - m + 2)(\hat{N}_A - m + 1),$
whose normal-ordered form $:(\hat{N}_A)(\hat{N}_A-1)\,\cdots\,(\hat{N}_A - m + 2) (\hat{N}_A - m + 1):$ 
yields, upon taking the coherent-state expectation value and applying the 
Cole-Hopf transform, the correct mesoscopic $m$-particle occupancy term $ \langle \hat{O} \rangle = (N_A)^m$.
}

In comparison, the `microscopic' implementation of the Cole-Hopf transform on the microscopic transition Hamiltonian 
\cref{eq:hamiltonian_quantised_single_reactive_jump_1} 
is defined between operators. In Ref.\cite{Lefevre_2007}, this procedure involves a direct replacement of the microscopic creation and annihilation operators, $\crtlong{i}{\#}$ and $\anhlong{i}{\#}$, by the coherent-state eigenvalues $\phi_i$ and $\phi_i^*$. But, this overlooks the fact that operators are non-commutative; therefore, one should not simply replace them with coherent-state eigenvalues $\phi_i$ and $\phi_i^*$. As highlighted in \cref{sec:cg_normal_ordering}, the `normal ordering' procedure allows the simplification of replacing second-quantized non-commutative microscopic operators with coherent-state mesoscopic eigenvalues. Such a replacement is correct only if the operator (here, the microscopic transition operator \cref{eq:hamiltonian_quantised_single_reactive_jump_1}) is normal ordered. In this case, 
\cref{eq:hamiltonian_quantised_single_reactive_jump_1} 
and 
\cref{eq:hamiltonian_quantised_single_reactive_jump_normal_ordered} 
are related by the trivial replacement, $\crtlong{i}{\#} \to \phi_i^\#$ and $\anhlong{i}{\#} \to (\phi_i^\#)^*$. However, this condition is not satisfied for thermodynamically-consistent microscopic transitions of interacting particles \cref{eq:hamiltonian_quantised_single_reactive_jump_1}, where the transition rates depend exponentially on the number operator. Note that if ideal non-interacting particles are considered, then the transition rates are linear in the number operator and already normal ordered; in that case, the practical difference between the `microscopic' and `mesoscopic' implementation of the Cole-Hopf transform becomes irrelevant. In general, however, this is not true. 
}

{
The normal ordering is an important and systematic step in the second-quantized methodology, and should not be viewed as a technical hurdle that reduces its practical applicability. Due to the systematic and methodological formulation of second-quantized approaches, the `normal-ordering' procedure correctly counts \emph{all microscopic configurations} that contribute to a given mesostate in the `mesoscopic' path integral. In principle, the correct and careful implementation of the stochastic path integral for thermodynamically-inconsistent interacting particle dynamics, such as the deterministic dynamics in Ref.\cite{Mourtaza2018} and the fluctuations around the deterministic path in Ref.\cite{martin2023nr_aim} has been carried out. These results are model-specific sub-cases of our general methodology. If the second-quantized approaches are correctly implemented for these models, the `coherent-state path integral' and `stochastic path integral' would yield identical results, see \cite{scandolo_2023}, with the fundamental difference that our formulation ensures thermodynamic consistency for all microscopic transitions. In practice, the correct demonstration of the stochastic path integral in Ref.\cite{Mourtaza2018,martin2023nr_aim} inherently implements the `normal-ordering' procedure, as Ref.\cite{Mourtaza2018,martin2023nr_aim} correctly accounts for \textit{all microscopic configurations} at the mesoscale. However, even for the simplest models, the computations are cumbersome and essentially brute-force. Despite being formally correct, these implementations do not provide clear physical insight into the coarse-graining procedure or the origin of the obtained double-exponential functions. In comparison, other classical implementations such as Ref.~\cite{Lefevre_2007} omit the crucial step of `normal ordering'. Consequently, the resulting methodology is not exact, becomes effectively susceptible to mean-field approximations at the level of the microscopic stochastic particle numbers, and does not correctly implement the coarse-graining procedure.
}

\section{Conclusion and Outlook}\label{sec:conclusion}
We study a generic class of microscopic interacting particles governed by thermodynamically consistent stochastic dynamics, described via the Master equation. Using Doi-Peliti field theory (DPFT), we implement a systematic coarse-graining procedure to derive Langevin dynamics for the mesoscopic and macroscopic particle numbers (densities). This approach relies on computing an exact large-deviation functional for interacting particle systems, and crucially, DPFT captures the effects of microscopic Poissonian particle-occupancy fluctuations in the coarse-grained description. Applying this framework to diffusive dynamics, we highlight the differences from classical coarse-graining approaches, such as the Kawasaki-Dean equation. Using reactive dynamics for interacting Ising spins, we further demonstrate the limitations of existing coarse-graining methods that neglect Poissonian occupancy fluctuations. The methodology is rigorously illustrated with the Active Ising Model (AIM), where we show that stochastic noise plays a key role in determining the order of the phase transition across low- and high-density regimes. Our coarse-graining procedure provides a systematic tool for studying interacting particles using mesoscopic and macroscopic fields while avoiding mean-field approximations, by exactly incorporating Poissonian fluctuations of particle occupancy. This framework extends the practical applicability of exact coarse-graining methods, bridging the gap between experimental or numerical observations of microscopic systems and their field-theoretical, coarse-grained macroscopic or mesoscopic descriptions.

\textit{Multi-body microscopic interactions}. \textendash  
The microscopic interactions of the form \cref{eq:microscopic_reciprocal_interaction_energy_particle} lead to a quadratic macroscopic interaction energy functional $E^{int}$. However, for purely attractive microscopic reciprocal interactions $v_{ij}^r<0$ implies $\mathcal{V}_{ij}^r, V_{ij}^r < 0$. Therefore, $E^{int}$ is not bounded from below. To ensure the lower boundedness of $E^{int}$, we need to introduce higher-order multi-body interactions. 
To this end, we incorporate the repulsive interaction terms of higher order, which introduce additional terms to $E^{int}$ of the form $V_{ij}^4 \rho_i^2 \rho_j^2$, such that $V_{ij}^4 > 0$. Here, $V_{ij}^4$ is an extra macroscopic control parameter attributed to the higher-order microscopic repulsive interaction. Hence, effectively $V_{ij}^r$ is reduced to the effective interaction coefficient $ V_{ij}^{eff} = V_{ij}^r + V_{ij}^4 \rho_i \rho_j$ with the effective interaction energy $E^{int} = \frac{1}{2} V_{ij}^{eff} \rho_i \rho_j$. For the higher values of $\rho_i$ and $\rho_j$, $V_{ij}^{eff} > 0$ is satisfied, thus, $E^{int}$ is bounded from below. For the smaller values of $\rho_i$ and $\rho_j$, the contribution due to $V_{ij}^r$ dominates over the $V_{ij}^4$ contribution. For $i=j$, one recovers the Ginzburg-Landau type higher-order interaction energy, $E^{int} = \frac{1}{2} V_{ii}^r \rho_i^2 + \frac{1}{2} V_{ii}^4 \rho_i^4$. 
Note that we treat the higher-order multi-body repulsive interactions using the mean-field approximation. In the higher-density regime, where the importance of repulsive interactions becomes prominent, the discreteness of the particle number becomes less important. Thus, the mean-field approximation of higher-order multi-body interactions is physically justified. Moreover, $V_{ij}^r$ captures the dominant-order Poissonian occupancy effect in the small particle number regime, further justifying the mean-field approximation of $V_{ij}^4$. Nevertheless, when multi-body interaction effects of microscopic systems are of physically paramount importance, they deserve to be studied in the future.

{
\textit{ Multi-particle reactive transitions and non-linear chemical reaction networks }. \textendash 
In this work, we have implemented linear reactive changes in \cref{eq:microscopic_transition_rate} that model a change of one reactant particle to one product particle (under the influence of local interactions), whose coarse-grained description belongs to linear chemical reaction networks. However, one seeks to address the future question, defined precisely as: How do microscopic changes of $n_r$ reactant particles to $n_p$ product particles (under the influence of local interactions) lead to an exact coarse-grained description using non-linear chemical reaction networks, where ($n_r > 1$) and ($n_p > 1$)? This question is an important future outlook, as non-linear chemical reaction networks exhibit a richer and different physical phenomenology, both dynamically and thermodynamically, than the linear chemical reactions studied here.
}

\subsection*{References}
\bibliography{reference}

\begin{thebibliography}{74}%
\makeatletter
\providecommand \@ifxundefined [1]{%
 \@ifx{#1\undefined}
}%
\providecommand \@ifnum [1]{%
 \ifnum #1\expandafter \@firstoftwo
 \else \expandafter \@secondoftwo
 \fi
}%
\providecommand \@ifx [1]{%
 \ifx #1\expandafter \@firstoftwo
 \else \expandafter \@secondoftwo
 \fi
}%
\providecommand \natexlab [1]{#1}%
\providecommand \enquote  [1]{``#1''}%
\providecommand \bibnamefont  [1]{#1}%
\providecommand \bibfnamefont [1]{#1}%
\providecommand \citenamefont [1]{#1}%
\providecommand \href@noop [0]{\@secondoftwo}%
\providecommand \href [0]{\begingroup \@sanitize@url \@href}%
\providecommand \@href[1]{\@@startlink{#1}\@@href}%
\providecommand \@@href[1]{\endgroup#1\@@endlink}%
\providecommand \@sanitize@url [0]{\catcode `\\12\catcode `\$12\catcode `\&12\catcode `\#12\catcode `\^12\catcode `\_12\catcode `\%12\relax}%
\providecommand \@@startlink[1]{}%
\providecommand \@@endlink[0]{}%
\providecommand \url  [0]{\begingroup\@sanitize@url \@url }%
\providecommand \@url [1]{\endgroup\@href {#1}{\urlprefix }}%
\providecommand \urlprefix  [0]{URL }%
\providecommand \Eprint [0]{\href }%
\providecommand \doibase [0]{https://doi.org/}%
\providecommand \selectlanguage [0]{\@gobble}%
\providecommand \bibinfo  [0]{\@secondoftwo}%
\providecommand \bibfield  [0]{\@secondoftwo}%
\providecommand \translation [1]{[#1]}%
\providecommand \BibitemOpen [0]{}%
\providecommand \bibitemStop [0]{}%
\providecommand \bibitemNoStop [0]{.\EOS\space}%
\providecommand \EOS [0]{\spacefactor3000\relax}%
\providecommand \BibitemShut  [1]{\csname bibitem#1\endcsname}%
\let\auto@bib@innerbib\@empty
\bibitem [{\citenamefont {Chaikin}\ and\ \citenamefont {Lubensky}(1995)}]{Chaikin_lubensky_1995}%
  \BibitemOpen
  \bibfield  {author} {\bibinfo {author} {\bibfnamefont {P.~M.}\ \bibnamefont {Chaikin}}\ and\ \bibinfo {author} {\bibfnamefont {T.~C.}\ \bibnamefont {Lubensky}},\ }\href {https://doi.org/10.1017/CBO9780511813467} {\emph {\bibinfo {title} {Principles of Condensed Matter Physics}}}\ (\bibinfo  {publisher} {Cambridge University Press},\ \bibinfo {year} {1995})\BibitemShut {NoStop}%
\bibitem [{\citenamefont {Hohenberg}\ and\ \citenamefont {Halperin}(1977)}]{Hohenberg_1977}%
  \BibitemOpen
  \bibfield  {author} {\bibinfo {author} {\bibfnamefont {P.~C.}\ \bibnamefont {Hohenberg}}\ and\ \bibinfo {author} {\bibfnamefont {B.~I.}\ \bibnamefont {Halperin}},\ }\bibfield  {title} {\bibinfo {title} {Theory of dynamic critical phenomena},\ }\href {https://doi.org/10.1103/RevModPhys.49.435} {\bibfield  {journal} {\bibinfo  {journal} {Rev. Mod. Phys.}\ }\textbf {\bibinfo {volume} {49}},\ \bibinfo {pages} {435} (\bibinfo {year} {1977})}\BibitemShut {NoStop}%
\bibitem [{\citenamefont {Cross}\ and\ \citenamefont {Hohenberg}(1993)}]{Cross_1993}%
  \BibitemOpen
  \bibfield  {author} {\bibinfo {author} {\bibfnamefont {M.~C.}\ \bibnamefont {Cross}}\ and\ \bibinfo {author} {\bibfnamefont {P.~C.}\ \bibnamefont {Hohenberg}},\ }\bibfield  {title} {\bibinfo {title} {Pattern formation outside of equilibrium},\ }\href {https://doi.org/10.1103/RevModPhys.65.851} {\bibfield  {journal} {\bibinfo  {journal} {Rev. Mod. Phys.}\ }\textbf {\bibinfo {volume} {65}},\ \bibinfo {pages} {851} (\bibinfo {year} {1993})}\BibitemShut {NoStop}%
\bibitem [{\citenamefont {Bray}(2002)}]{Bray_2002}%
  \BibitemOpen
  \bibfield  {author} {\bibinfo {author} {\bibfnamefont {A.~J.}\ \bibnamefont {Bray}},\ }\bibfield  {title} {\bibinfo {title} {Theory of phase-ordering kinetics},\ }\href {https://doi.org/10.1080/00018730110117433} {\bibfield  {journal} {\bibinfo  {journal} {Advances in Physics}\ }\textbf {\bibinfo {volume} {51}},\ \bibinfo {pages} {481} (\bibinfo {year} {2002})},\ \Eprint {https://arxiv.org/abs/https://doi.org/10.1080/00018730110117433} {https://doi.org/10.1080/00018730110117433} \BibitemShut {NoStop}%
\bibitem [{\citenamefont {Ramaswamy}(2010)}]{Ramaswamy_2010}%
  \BibitemOpen
  \bibfield  {author} {\bibinfo {author} {\bibfnamefont {S.}~\bibnamefont {Ramaswamy}},\ }\bibfield  {title} {\bibinfo {title} {The mechanics and statistics of active matter},\ }\href@noop {} {\bibfield  {journal} {\bibinfo  {journal} {Annu. Rev. Condens. Matter Phys.}\ }\textbf {\bibinfo {volume} {1}},\ \bibinfo {pages} {323} (\bibinfo {year} {2010})}\BibitemShut {NoStop}%
\bibitem [{\citenamefont {Beard}\ and\ \citenamefont {Qian}(2008)}]{Beard_2008}%
  \BibitemOpen
  \bibfield  {author} {\bibinfo {author} {\bibfnamefont {D.}~\bibnamefont {Beard}}\ and\ \bibinfo {author} {\bibfnamefont {H.}~\bibnamefont {Qian}},\ }\href {https://books.google.de/books?id=kTmQAz6QXckC} {\emph {\bibinfo {title} {Chemical Biophysics: Quantitative Analysis of Cellular Systems}}},\ Cambridge Texts in Biomedical Engineering\ (\bibinfo  {publisher} {Cambridge University Press},\ \bibinfo {year} {2008})\BibitemShut {NoStop}%
\bibitem [{\citenamefont {Martin}\ \emph {et~al.}(1973)\citenamefont {Martin}, \citenamefont {Siggia},\ and\ \citenamefont {Rose}}]{martin_1973}%
  \BibitemOpen
  \bibfield  {author} {\bibinfo {author} {\bibfnamefont {P.~C.}\ \bibnamefont {Martin}}, \bibinfo {author} {\bibfnamefont {E.~D.}\ \bibnamefont {Siggia}},\ and\ \bibinfo {author} {\bibfnamefont {H.~A.}\ \bibnamefont {Rose}},\ }\bibfield  {title} {\bibinfo {title} {Statistical dynamics of classical systems},\ }\href {https://doi.org/10.1103/PhysRevA.8.423} {\bibfield  {journal} {\bibinfo  {journal} {Phys. Rev. A}\ }\textbf {\bibinfo {volume} {8}},\ \bibinfo {pages} {423} (\bibinfo {year} {1973})}\BibitemShut {NoStop}%
\bibitem [{\citenamefont {Bausch}\ \emph {et~al.}(1976)\citenamefont {Bausch}, \citenamefont {Janssen},\ and\ \citenamefont {Wagner}}]{bausch_1976}%
  \BibitemOpen
  \bibfield  {author} {\bibinfo {author} {\bibfnamefont {R.}~\bibnamefont {Bausch}}, \bibinfo {author} {\bibfnamefont {H.~K.}\ \bibnamefont {Janssen}},\ and\ \bibinfo {author} {\bibfnamefont {H.}~\bibnamefont {Wagner}},\ }\bibfield  {title} {\bibinfo {title} {Renormalized field theory of critical dynamics},\ }\href {https://doi.org/10.1007/BF01312880} {\bibfield  {journal} {\bibinfo  {journal} {Zeitschrift f{\"u}r Physik B Condensed Matter}\ }\textbf {\bibinfo {volume} {24}},\ \bibinfo {pages} {113} (\bibinfo {year} {1976})}\BibitemShut {NoStop}%
\bibitem [{\citenamefont {Janssen}(1976)}]{janssen_1976}%
  \BibitemOpen
  \bibfield  {author} {\bibinfo {author} {\bibfnamefont {H.-K.}\ \bibnamefont {Janssen}},\ }\bibfield  {title} {\bibinfo {title} {On a lagrangean for classical field dynamics and renormalization group calculations of dynamical critical properties},\ }\href {https://doi.org/10.1007/BF01316547} {\bibfield  {journal} {\bibinfo  {journal} {Zeitschrift f{\"u}r Physik B Condensed Matter}\ }\textbf {\bibinfo {volume} {23}},\ \bibinfo {pages} {377} (\bibinfo {year} {1976})}\BibitemShut {NoStop}%
\bibitem [{\citenamefont {de~Dominicis}(1976)}]{dedominicis_1976}%
  \BibitemOpen
  \bibfield  {author} {\bibinfo {author} {\bibfnamefont {C.}~\bibnamefont {de~Dominicis}},\ }\bibfield  {title} {\bibinfo {title} {Techniques de renormalisation de la th{\'e}orie des champs et dynamique des ph{\'e}nom{\`e}nes critiques},\ }\href {https://doi.org/10.1051/jphyscol:1976138} {\bibfield  {journal} {\bibinfo  {journal} {{Journal de Physique Colloques}}\ }\textbf {\bibinfo {volume} {37}},\ \bibinfo {pages} {C1} (\bibinfo {year} {1976})}\BibitemShut {NoStop}%
\bibitem [{\citenamefont {Mohite}\ and\ \citenamefont {Rieger}(2026{\natexlab{a}})}]{atm_st_nr_2024}%
  \BibitemOpen
  \bibfield  {author} {\bibinfo {author} {\bibfnamefont {A.~T.}\ \bibnamefont {Mohite}}\ and\ \bibinfo {author} {\bibfnamefont {H.}~\bibnamefont {Rieger}},\ }\bibfield  {title} {\bibinfo {title} {Stochastic thermodynamics of nonreciprocally interacting particles and fields},\ }\href {https://doi.org/10.1103/mqpg-4l5h} {\bibfield  {journal} {\bibinfo  {journal} {Phys. Rev. E}\ }\textbf {\bibinfo {volume} {113}},\ \bibinfo {pages} {064136} (\bibinfo {year} {2026}{\natexlab{a}})}\BibitemShut {NoStop}%
\bibitem [{\citenamefont {Solon}\ and\ \citenamefont {Tailleur}(2013)}]{Solon2013}%
  \BibitemOpen
  \bibfield  {author} {\bibinfo {author} {\bibfnamefont {A.~P.}\ \bibnamefont {Solon}}\ and\ \bibinfo {author} {\bibfnamefont {J.}~\bibnamefont {Tailleur}},\ }\bibfield  {title} {\bibinfo {title} {Revisiting the flocking transition using active spins},\ }\href {https://doi.org/10.1103/PhysRevLett.111.078101} {\bibfield  {journal} {\bibinfo  {journal} {Phys. Rev. Lett.}\ }\textbf {\bibinfo {volume} {111}},\ \bibinfo {pages} {078101} (\bibinfo {year} {2013})}\BibitemShut {NoStop}%
\bibitem [{\citenamefont {Solon}\ and\ \citenamefont {Tailleur}(2015)}]{Solon2015}%
  \BibitemOpen
  \bibfield  {author} {\bibinfo {author} {\bibfnamefont {A.~P.}\ \bibnamefont {Solon}}\ and\ \bibinfo {author} {\bibfnamefont {J.}~\bibnamefont {Tailleur}},\ }\bibfield  {title} {\bibinfo {title} {Flocking with discrete symmetry: The two-dimensional active ising model},\ }\href {https://doi.org/10.1103/PhysRevE.92.042119} {\bibfield  {journal} {\bibinfo  {journal} {Phys. Rev. E}\ }\textbf {\bibinfo {volume} {92}},\ \bibinfo {pages} {042119} (\bibinfo {year} {2015})}\BibitemShut {NoStop}%
\bibitem [{\citenamefont {Kourbane-Houssene}\ \emph {et~al.}(2018)\citenamefont {Kourbane-Houssene}, \citenamefont {Erignoux}, \citenamefont {Bodineau},\ and\ \citenamefont {Tailleur}}]{Mourtaza2018}%
  \BibitemOpen
  \bibfield  {author} {\bibinfo {author} {\bibfnamefont {M.}~\bibnamefont {Kourbane-Houssene}}, \bibinfo {author} {\bibfnamefont {C.}~\bibnamefont {Erignoux}}, \bibinfo {author} {\bibfnamefont {T.}~\bibnamefont {Bodineau}},\ and\ \bibinfo {author} {\bibfnamefont {J.}~\bibnamefont {Tailleur}},\ }\bibfield  {title} {\bibinfo {title} {Exact hydrodynamic description of active lattice gases},\ }\href {https://doi.org/10.1103/PhysRevLett.120.268003} {\bibfield  {journal} {\bibinfo  {journal} {Phys. Rev. Lett.}\ }\textbf {\bibinfo {volume} {120}},\ \bibinfo {pages} {268003} (\bibinfo {year} {2018})}\BibitemShut {NoStop}%
\bibitem [{\citenamefont {Kawasaki}(1994)}]{Kawasaki_1994}%
  \BibitemOpen
  \bibfield  {author} {\bibinfo {author} {\bibfnamefont {K.}~\bibnamefont {Kawasaki}},\ }\bibfield  {title} {\bibinfo {title} {Stochastic model of slow dynamics in supercooled liquids and dense colloidal suspensions},\ }\href {https://EconPapers.repec.org/RePEc:eee:phsmap:v:208:y:1994:i:1:p:35-64} {\bibfield  {journal} {\bibinfo  {journal} {Physica A: Statistical Mechanics and its Applications}\ }\textbf {\bibinfo {volume} {208}},\ \bibinfo {pages} {35} (\bibinfo {year} {1994})}\BibitemShut {NoStop}%
\bibitem [{\citenamefont {Dean}(1996)}]{Dean_1996}%
  \BibitemOpen
  \bibfield  {author} {\bibinfo {author} {\bibfnamefont {D.~S.}\ \bibnamefont {Dean}},\ }\bibfield  {title} {\bibinfo {title} {Langevin equation for the density of a system of interacting langevin processes},\ }\href {https://doi.org/10.1088/0305-4470/29/24/001} {\bibfield  {journal} {\bibinfo  {journal} {Journal of Physics A: Mathematical and General}\ }\textbf {\bibinfo {volume} {29}},\ \bibinfo {pages} {L613} (\bibinfo {year} {1996})}\BibitemShut {NoStop}%
\bibitem [{\citenamefont {Bertin}\ \emph {et~al.}(2013)\citenamefont {Bertin}, \citenamefont {Chat{\'e}}, \citenamefont {Ginelli}, \citenamefont {Mishra}, \citenamefont {Peshkov},\ and\ \citenamefont {Ramaswamy}}]{Bertin_2013}%
  \BibitemOpen
  \bibfield  {author} {\bibinfo {author} {\bibfnamefont {E.}~\bibnamefont {Bertin}}, \bibinfo {author} {\bibfnamefont {H.}~\bibnamefont {Chat{\'e}}}, \bibinfo {author} {\bibfnamefont {F.}~\bibnamefont {Ginelli}}, \bibinfo {author} {\bibfnamefont {S.}~\bibnamefont {Mishra}}, \bibinfo {author} {\bibfnamefont {A.}~\bibnamefont {Peshkov}},\ and\ \bibinfo {author} {\bibfnamefont {S.}~\bibnamefont {Ramaswamy}},\ }\bibfield  {title} {\bibinfo {title} {Mesoscopic theory for fluctuating active nematics},\ }\href@noop {} {\bibfield  {journal} {\bibinfo  {journal} {New journal of physics}\ }\textbf {\bibinfo {volume} {15}},\ \bibinfo {pages} {085032} (\bibinfo {year} {2013})}\BibitemShut {NoStop}%
\bibitem [{\citenamefont {Grossmann}\ \emph {et~al.}(2013)\citenamefont {Grossmann}, \citenamefont {Schimansky-Geier},\ and\ \citenamefont {Romanczuk}}]{Grossmann_2013}%
  \BibitemOpen
  \bibfield  {author} {\bibinfo {author} {\bibfnamefont {R.}~\bibnamefont {Grossmann}}, \bibinfo {author} {\bibfnamefont {L.}~\bibnamefont {Schimansky-Geier}},\ and\ \bibinfo {author} {\bibfnamefont {P.}~\bibnamefont {Romanczuk}},\ }\bibfield  {title} {\bibinfo {title} {Self-propelled particles with selective attraction--repulsion interaction: from microscopic dynamics to coarse-grained theories},\ }\href@noop {} {\bibfield  {journal} {\bibinfo  {journal} {New Journal of Physics}\ }\textbf {\bibinfo {volume} {15}},\ \bibinfo {pages} {085014} (\bibinfo {year} {2013})}\BibitemShut {NoStop}%
\bibitem [{\citenamefont {Peshkov}\ \emph {et~al.}(2014)\citenamefont {Peshkov}, \citenamefont {Bertin}, \citenamefont {Ginelli},\ and\ \citenamefont {Chat{\'e}}}]{Peshkov_2014}%
  \BibitemOpen
  \bibfield  {author} {\bibinfo {author} {\bibfnamefont {A.}~\bibnamefont {Peshkov}}, \bibinfo {author} {\bibfnamefont {E.}~\bibnamefont {Bertin}}, \bibinfo {author} {\bibfnamefont {F.}~\bibnamefont {Ginelli}},\ and\ \bibinfo {author} {\bibfnamefont {H.}~\bibnamefont {Chat{\'e}}},\ }\bibfield  {title} {\bibinfo {title} {Boltzmann-ginzburg-landau approach for continuous descriptions of generic vicsek-like models},\ }\href@noop {} {\bibfield  {journal} {\bibinfo  {journal} {The European Physical Journal Special Topics}\ }\textbf {\bibinfo {volume} {223}},\ \bibinfo {pages} {1315} (\bibinfo {year} {2014})}\BibitemShut {NoStop}%
\bibitem [{\citenamefont {Touchette}(2009)}]{Touchette_2009}%
  \BibitemOpen
  \bibfield  {author} {\bibinfo {author} {\bibfnamefont {H.}~\bibnamefont {Touchette}},\ }\bibfield  {title} {\bibinfo {title} {The large deviation approach to statistical mechanics},\ }\href {https://doi.org/https://doi.org/10.1016/j.physrep.2009.05.002} {\bibfield  {journal} {\bibinfo  {journal} {Physics Reports}\ }\textbf {\bibinfo {volume} {478}},\ \bibinfo {pages} {1} (\bibinfo {year} {2009})}\BibitemShut {NoStop}%
\bibitem [{\citenamefont {Mohite}\ and\ \citenamefont {Rieger}(2025)}]{atm_2024_var_epr}%
  \BibitemOpen
  \bibfield  {author} {\bibinfo {author} {\bibfnamefont {A.~T.}\ \bibnamefont {Mohite}}\ and\ \bibinfo {author} {\bibfnamefont {H.}~\bibnamefont {Rieger}},\ }\href@noop {} {\bibinfo {title} {Minimum action principle for entropy production rate of far-from-equilibrium systems}} (\bibinfo {year} {2025}),\ \Eprint {https://arxiv.org/abs/2511.00967} {arXiv:2511.00967 [cond-mat.stat-mech]} \BibitemShut {NoStop}%
\bibitem [{\citenamefont {Mohite}\ and\ \citenamefont {Rieger}(2026{\natexlab{b}})}]{atm_2025_var_epr_derivation}%
  \BibitemOpen
  \bibfield  {author} {\bibinfo {author} {\bibfnamefont {A.~T.}\ \bibnamefont {Mohite}}\ and\ \bibinfo {author} {\bibfnamefont {H.}~\bibnamefont {Rieger}},\ }\bibfield  {title} {\bibinfo {title} {Thermodynamic length in stochastic thermodynamics of far-from-equilibrium systems: Unification of fluctuation relation and thermodynamic uncertainty relation},\ }\href {https://doi.org/10.1103/fmlt-px9y} {\bibfield  {journal} {\bibinfo  {journal} {Phys. Rev. E}\ }\textbf {\bibinfo {volume} {114}},\ \bibinfo {pages} {014103} (\bibinfo {year} {2026}{\natexlab{b}})}\BibitemShut {NoStop}%
\bibitem [{\citenamefont {Mohite}\ and\ \citenamefont {Rieger}(2026{\natexlab{c}})}]{atm_2025_gftoc}%
  \BibitemOpen
  \bibfield  {author} {\bibinfo {author} {\bibfnamefont {A.~T.}\ \bibnamefont {Mohite}}\ and\ \bibinfo {author} {\bibfnamefont {H.}~\bibnamefont {Rieger}},\ }\bibfield  {title} {\bibinfo {title} {Generalized finite-time optimal control framework in stochastic thermodynamics},\ }\href {https://doi.org/10.1103/bc7b-tfl6} {\bibfield  {journal} {\bibinfo  {journal} {Phys. Rev. Res.}\ } (\bibinfo {year} {2026}{\natexlab{c}})}\BibitemShut {NoStop}%
\bibitem [{\citenamefont {Doi}(1976{\natexlab{a}})}]{Doi_1976}%
  \BibitemOpen
  \bibfield  {author} {\bibinfo {author} {\bibfnamefont {M.}~\bibnamefont {Doi}},\ }\bibfield  {title} {\bibinfo {title} {Second quantization representation for classical many-particle system},\ }\href {https://doi.org/10.1088/0305-4470/9/9/008} {\bibfield  {journal} {\bibinfo  {journal} {Journal of Physics A: Mathematical and General}\ }\textbf {\bibinfo {volume} {9}},\ \bibinfo {pages} {1465} (\bibinfo {year} {1976}{\natexlab{a}})}\BibitemShut {NoStop}%
\bibitem [{\citenamefont {Doi}(1976{\natexlab{b}})}]{Doi_1976_2}%
  \BibitemOpen
  \bibfield  {author} {\bibinfo {author} {\bibfnamefont {M.}~\bibnamefont {Doi}},\ }\bibfield  {title} {\bibinfo {title} {Stochastic theory of diffusion-controlled reaction},\ }\href {https://doi.org/10.1088/0305-4470/9/9/009} {\bibfield  {journal} {\bibinfo  {journal} {Journal of Physics A: Mathematical and General}\ }\textbf {\bibinfo {volume} {9}},\ \bibinfo {pages} {1479} (\bibinfo {year} {1976}{\natexlab{b}})}\BibitemShut {NoStop}%
\bibitem [{\citenamefont {Zel’Dovich}\ and\ \citenamefont {Ovchinnikov}(1978)}]{ZelDovich_1978}%
  \BibitemOpen
  \bibfield  {author} {\bibinfo {author} {\bibfnamefont {Y.~B.}\ \bibnamefont {Zel’Dovich}}\ and\ \bibinfo {author} {\bibfnamefont {A.}~\bibnamefont {Ovchinnikov}},\ }\bibfield  {title} {\bibinfo {title} {The mass action law and the kinetics of chemical reactions with allowance for thermodynamic fluctuations of the density},\ }\href@noop {} {\bibfield  {journal} {\bibinfo  {journal} {Z Eksp Teor Fiz}\ }\textbf {\bibinfo {volume} {74}},\ \bibinfo {pages} {1588} (\bibinfo {year} {1978})}\BibitemShut {NoStop}%
\bibitem [{\citenamefont {Rose}(1979)}]{rose_1979}%
  \BibitemOpen
  \bibfield  {author} {\bibinfo {author} {\bibfnamefont {H.~A.}\ \bibnamefont {Rose}},\ }\bibfield  {title} {\bibinfo {title} {Renormalized kinetic theory of nonequilibrium many-particle classical systems},\ }\href {https://doi.org/10.1007/BF01011780} {\bibfield  {journal} {\bibinfo  {journal} {Journal of Statistical Physics}\ }\textbf {\bibinfo {volume} {20}},\ \bibinfo {pages} {415} (\bibinfo {year} {1979})}\BibitemShut {NoStop}%
\bibitem [{\citenamefont {Grassberger}\ and\ \citenamefont {Scheunert}(1980)}]{grassberger_1980}%
  \BibitemOpen
  \bibfield  {author} {\bibinfo {author} {\bibfnamefont {P.}~\bibnamefont {Grassberger}}\ and\ \bibinfo {author} {\bibfnamefont {M.}~\bibnamefont {Scheunert}},\ }\bibfield  {title} {\bibinfo {title} {Fock-space methods for identical classical objects},\ }\href {https://doi.org/https://doi.org/10.1002/prop.19800281004} {\bibfield  {journal} {\bibinfo  {journal} {Fortschritte der Physik}\ }\textbf {\bibinfo {volume} {28}},\ \bibinfo {pages} {547} (\bibinfo {year} {1980})}\BibitemShut {NoStop}%
\bibitem [{\citenamefont {Mikhailov}(1981{\natexlab{a}})}]{Mikhailov_1981}%
  \BibitemOpen
  \bibfield  {author} {\bibinfo {author} {\bibfnamefont {A.}~\bibnamefont {Mikhailov}},\ }\bibfield  {title} {\bibinfo {title} {Path integrals in chemical kinetics i},\ }\href {https://doi.org/https://doi.org/10.1016/0375-9601(81)90017-7} {\bibfield  {journal} {\bibinfo  {journal} {Physics Letters A}\ }\textbf {\bibinfo {volume} {85}},\ \bibinfo {pages} {214} (\bibinfo {year} {1981}{\natexlab{a}})}\BibitemShut {NoStop}%
\bibitem [{\citenamefont {Mikhailov}(1981{\natexlab{b}})}]{Mikhailov_1981_2}%
  \BibitemOpen
  \bibfield  {author} {\bibinfo {author} {\bibfnamefont {A.}~\bibnamefont {Mikhailov}},\ }\bibfield  {title} {\bibinfo {title} {Path integrals in chemical kinetics ii},\ }\href {https://doi.org/https://doi.org/10.1016/0375-9601(81)90429-1} {\bibfield  {journal} {\bibinfo  {journal} {Physics Letters A}\ }\textbf {\bibinfo {volume} {85}},\ \bibinfo {pages} {427} (\bibinfo {year} {1981}{\natexlab{b}})}\BibitemShut {NoStop}%
\bibitem [{\citenamefont {Mikhailov}\ and\ \citenamefont {Yashin}(1985)}]{Mikhailov_1985}%
  \BibitemOpen
  \bibfield  {author} {\bibinfo {author} {\bibfnamefont {A.~S.}\ \bibnamefont {Mikhailov}}\ and\ \bibinfo {author} {\bibfnamefont {V.~V.}\ \bibnamefont {Yashin}},\ }\bibfield  {title} {\bibinfo {title} {Quantum field methods in the theory of diffusion-controlled reactions},\ }\href {https://doi.org/10.1007/BF01017866} {\bibfield  {journal} {\bibinfo  {journal} {Journal of Statistical Physics}\ }\textbf {\bibinfo {volume} {38}},\ \bibinfo {pages} {347} (\bibinfo {year} {1985})}\BibitemShut {NoStop}%
\bibitem [{\citenamefont {{Peliti, L.}}(1985)}]{Peliti}%
  \BibitemOpen
  \bibfield  {author} {\bibinfo {author} {\bibnamefont {{Peliti, L.}}},\ }\bibfield  {title} {\bibinfo {title} {Path integral approach to birth-death processes on a lattice},\ }\href {https://doi.org/10.1051/jphys:019850046090146900} {\bibfield  {journal} {\bibinfo  {journal} {J. Phys. France}\ }\textbf {\bibinfo {volume} {46}},\ \bibinfo {pages} {1469} (\bibinfo {year} {1985})}\BibitemShut {NoStop}%
\bibitem [{\citenamefont {Cardy}\ \emph {et~al.}(2008)\citenamefont {Cardy}, \citenamefont {Cardy}, \citenamefont {Falkovich},\ and\ \citenamefont {Gawedzki}}]{Cardy_2008}%
  \BibitemOpen
  \bibfield  {author} {\bibinfo {author} {\bibfnamefont {J.}~\bibnamefont {Cardy}}, \bibinfo {author} {\bibfnamefont {J.}~\bibnamefont {Cardy}}, \bibinfo {author} {\bibfnamefont {G.}~\bibnamefont {Falkovich}},\ and\ \bibinfo {author} {\bibfnamefont {K.}~\bibnamefont {Gawedzki}},\ }\bibinfo {title} {John cardy. reaction-diffusion processes},\ in\ \href@noop {} {\emph {\bibinfo {booktitle} {Non-equilibrium Statistical Mechanics and Turbulence}}},\ \bibinfo {series and number} {London Mathematical Society Lecture Note Series},\ \bibinfo {editor} {edited by\ \bibinfo {editor} {\bibfnamefont {S.}~\bibnamefont {Nazarenko}}\ and\ \bibinfo {editor} {\bibfnamefont {O.~V.}\ \bibnamefont {Zaboronski}}}\ (\bibinfo  {publisher} {Cambridge University Press},\ \bibinfo {year} {2008})\ p.\ \bibinfo {pages} {108–161}\BibitemShut {NoStop}%
\bibitem [{\citenamefont {Wiese}(2016)}]{Wiese_2016}%
  \BibitemOpen
  \bibfield  {author} {\bibinfo {author} {\bibfnamefont {K.~J.}\ \bibnamefont {Wiese}},\ }\bibfield  {title} {\bibinfo {title} {Coherent-state path integral versus coarse-grained effective stochastic equation of motion: From reaction diffusion to stochastic sandpiles},\ }\href {https://doi.org/10.1103/PhysRevE.93.042117} {\bibfield  {journal} {\bibinfo  {journal} {Phys. Rev. E}\ }\textbf {\bibinfo {volume} {93}},\ \bibinfo {pages} {042117} (\bibinfo {year} {2016})}\BibitemShut {NoStop}%
\bibitem [{\citenamefont {Weber}\ and\ \citenamefont {Frey}(2017)}]{Weber_2017}%
  \BibitemOpen
  \bibfield  {author} {\bibinfo {author} {\bibfnamefont {M.~F.}\ \bibnamefont {Weber}}\ and\ \bibinfo {author} {\bibfnamefont {E.}~\bibnamefont {Frey}},\ }\bibfield  {title} {\bibinfo {title} {Master equations and the theory of stochastic path integrals},\ }\href {https://doi.org/10.1088/1361-6633/aa5ae2} {\bibfield  {journal} {\bibinfo  {journal} {Reports on Progress in Physics}\ }\textbf {\bibinfo {volume} {80}},\ \bibinfo {pages} {046601} (\bibinfo {year} {2017})}\BibitemShut {NoStop}%
\bibitem [{\citenamefont {Sudarshan}(1963)}]{Sudarshan_1963}%
  \BibitemOpen
  \bibfield  {author} {\bibinfo {author} {\bibfnamefont {E.~C.~G.}\ \bibnamefont {Sudarshan}},\ }\bibfield  {title} {\bibinfo {title} {Equivalence of semiclassical and quantum mechanical descriptions of statistical light beams},\ }\href {https://doi.org/10.1103/PhysRevLett.10.277} {\bibfield  {journal} {\bibinfo  {journal} {Phys. Rev. Lett.}\ }\textbf {\bibinfo {volume} {10}},\ \bibinfo {pages} {277} (\bibinfo {year} {1963})}\BibitemShut {NoStop}%
\bibitem [{\citenamefont {Lefèvre}\ and\ \citenamefont {Biroli}(2007)}]{Lefevre_2007}%
  \BibitemOpen
  \bibfield  {author} {\bibinfo {author} {\bibfnamefont {A.}~\bibnamefont {Lefèvre}}\ and\ \bibinfo {author} {\bibfnamefont {G.}~\bibnamefont {Biroli}},\ }\bibfield  {title} {\bibinfo {title} {Dynamics of interacting particle systems: stochastic process and field theory},\ }\href {https://doi.org/10.1088/1742-5468/2007/07/P07024} {\bibfield  {journal} {\bibinfo  {journal} {Journal of Statistical Mechanics: Theory and Experiment}\ }\textbf {\bibinfo {volume} {2007}},\ \bibinfo {pages} {P07024} (\bibinfo {year} {2007})}\BibitemShut {NoStop}%
\bibitem [{Note1()}]{Note1}%
  \BibitemOpen
  \bibinfo {note} {Except for the normalization prefactor $e^{ -\phi _i^\#(\phi _i^\#)^* }$, which is circumvented by taking the inner product of the coherent state in the denominator, the convention we will adhere to compute expectation values throughout this work.}\BibitemShut {Stop}%
\bibitem [{Note2()}]{Note2}%
  \BibitemOpen
  \bibinfo {note} {We use the notation $\exp {(*)} = e^{*}$, to simplify the representation of the double exponential obtained subsequently throughout the paper.}\BibitemShut {Stop}%
\bibitem [{\citenamefont {Blasiak}\ \emph {et~al.}(2007)\citenamefont {Blasiak}, \citenamefont {Horzela}, \citenamefont {Penson}, \citenamefont {Solomon},\ and\ \citenamefont {Duchamp}}]{blasiak2007}%
  \BibitemOpen
  \bibfield  {author} {\bibinfo {author} {\bibfnamefont {P.}~\bibnamefont {Blasiak}}, \bibinfo {author} {\bibfnamefont {A.}~\bibnamefont {Horzela}}, \bibinfo {author} {\bibfnamefont {K.~A.}\ \bibnamefont {Penson}}, \bibinfo {author} {\bibfnamefont {A.~I.}\ \bibnamefont {Solomon}},\ and\ \bibinfo {author} {\bibfnamefont {G.~H.~E.}\ \bibnamefont {Duchamp}},\ }\bibfield  {title} {\bibinfo {title} {{Combinatorics and Boson normal ordering: A gentle introduction}},\ }\href {https://doi.org/10.1119/1.2723799} {\bibfield  {journal} {\bibinfo  {journal} {American Journal of Physics}\ }\textbf {\bibinfo {volume} {75}},\ \bibinfo {pages} {639} (\bibinfo {year} {2007})}\BibitemShut {NoStop}%
\bibitem [{\citenamefont {Itakura}\ \emph {et~al.}(2010)\citenamefont {Itakura}, \citenamefont {Ohkubo},\ and\ \citenamefont {ichi Sasa}}]{itakura_2010}%
  \BibitemOpen
  \bibfield  {author} {\bibinfo {author} {\bibfnamefont {K.}~\bibnamefont {Itakura}}, \bibinfo {author} {\bibfnamefont {J.}~\bibnamefont {Ohkubo}},\ and\ \bibinfo {author} {\bibfnamefont {S.}~\bibnamefont {ichi Sasa}},\ }\bibfield  {title} {\bibinfo {title} {Two langevin equations in the doi–peliti formalism},\ }\href {https://doi.org/10.1088/1751-8113/43/12/125001} {\bibfield  {journal} {\bibinfo  {journal} {Journal of Physics A: Mathematical and Theoretical}\ }\textbf {\bibinfo {volume} {43}},\ \bibinfo {pages} {125001} (\bibinfo {year} {2010})}\BibitemShut {NoStop}%
\bibitem [{\citenamefont {van Kampen}(1981)}]{van_kampen}%
  \BibitemOpen
  \bibfield  {author} {\bibinfo {author} {\bibfnamefont {N.~G.}\ \bibnamefont {van Kampen}},\ }\href@noop {} {\emph {\bibinfo {title} {Stochastic processes in physics and chemistry}}}\ (\bibinfo  {publisher} {Elsevier North-Holland},\ \bibinfo {year} {1981})\BibitemShut {NoStop}%
\bibitem [{\citenamefont {Martin}\ \emph {et~al.}(2025)\citenamefont {Martin}, \citenamefont {Seara}, \citenamefont {Avni}, \citenamefont {Fruchart},\ and\ \citenamefont {Vitelli}}]{martin2023nr_aim}%
  \BibitemOpen
  \bibfield  {author} {\bibinfo {author} {\bibfnamefont {D.}~\bibnamefont {Martin}}, \bibinfo {author} {\bibfnamefont {D.}~\bibnamefont {Seara}}, \bibinfo {author} {\bibfnamefont {Y.}~\bibnamefont {Avni}}, \bibinfo {author} {\bibfnamefont {M.}~\bibnamefont {Fruchart}},\ and\ \bibinfo {author} {\bibfnamefont {V.}~\bibnamefont {Vitelli}},\ }\bibfield  {title} {\bibinfo {title} {Transition to collective motion in nonreciprocal active matter: Coarse graining agent-based models into fluctuating hydrodynamics},\ }\href {https://doi.org/10.1103/PhysRevX.15.041015} {\bibfield  {journal} {\bibinfo  {journal} {Phys. Rev. X}\ }\textbf {\bibinfo {volume} {15}},\ \bibinfo {pages} {041015} (\bibinfo {year} {2025})}\BibitemShut {NoStop}%
\bibitem [{\citenamefont {Maes}\ and\ \citenamefont {Netočný}(2008)}]{Maes_2008}%
  \BibitemOpen
  \bibfield  {author} {\bibinfo {author} {\bibfnamefont {C.}~\bibnamefont {Maes}}\ and\ \bibinfo {author} {\bibfnamefont {K.}~\bibnamefont {Netočný}},\ }\bibfield  {title} {\bibinfo {title} {Canonical structure of dynamical fluctuations in mesoscopic nonequilibrium steady states},\ }\href {https://doi.org/10.1209/0295-5075/82/30003} {\bibfield  {journal} {\bibinfo  {journal} {Europhysics Letters}\ }\textbf {\bibinfo {volume} {82}},\ \bibinfo {pages} {30003} (\bibinfo {year} {2008})}\BibitemShut {NoStop}%
\bibitem [{\citenamefont {Mielke}\ \emph {et~al.}(2014)\citenamefont {Mielke}, \citenamefont {Peletier},\ and\ \citenamefont {Renger}}]{Mielke_2014_ldp}%
  \BibitemOpen
  \bibfield  {author} {\bibinfo {author} {\bibfnamefont {A.}~\bibnamefont {Mielke}}, \bibinfo {author} {\bibfnamefont {M.~A.}\ \bibnamefont {Peletier}},\ and\ \bibinfo {author} {\bibfnamefont {D.~R.~M.}\ \bibnamefont {Renger}},\ }\bibfield  {title} {\bibinfo {title} {On the relation between gradient flows and the large-deviation principle, with applications to markov chains and diffusion},\ }\href {https://doi.org/10.1007/s11118-014-9418-5} {\bibfield  {journal} {\bibinfo  {journal} {Potential Analysis}\ }\textbf {\bibinfo {volume} {41}},\ \bibinfo {pages} {1293} (\bibinfo {year} {2014})}\BibitemShut {NoStop}%
\bibitem [{\citenamefont {Mielke}\ \emph {et~al.}(2017)\citenamefont {Mielke}, \citenamefont {Patterson}, \citenamefont {Peletier},\ and\ \citenamefont {Michiel~Renger}}]{Mielke_2017}%
  \BibitemOpen
  \bibfield  {author} {\bibinfo {author} {\bibfnamefont {A.}~\bibnamefont {Mielke}}, \bibinfo {author} {\bibfnamefont {R.~I.~A.}\ \bibnamefont {Patterson}}, \bibinfo {author} {\bibfnamefont {M.~A.}\ \bibnamefont {Peletier}},\ and\ \bibinfo {author} {\bibfnamefont {D.~R.}\ \bibnamefont {Michiel~Renger}},\ }\bibfield  {title} {\bibinfo {title} {Non-equilibrium thermodynamical principles for chemical reactions with mass-action kinetics},\ }\href {https://doi.org/10.1137/16M1102240} {\bibfield  {journal} {\bibinfo  {journal} {SIAM Journal on Applied Mathematics}\ }\textbf {\bibinfo {volume} {77}},\ \bibinfo {pages} {1562} (\bibinfo {year} {2017})}\BibitemShut {NoStop}%
\bibitem [{\citenamefont {Kaiser}\ \emph {et~al.}(2018)\citenamefont {Kaiser}, \citenamefont {Jack},\ and\ \citenamefont {Zimmer}}]{Kaiser_2018}%
  \BibitemOpen
  \bibfield  {author} {\bibinfo {author} {\bibfnamefont {M.}~\bibnamefont {Kaiser}}, \bibinfo {author} {\bibfnamefont {R.~L.}\ \bibnamefont {Jack}},\ and\ \bibinfo {author} {\bibfnamefont {J.}~\bibnamefont {Zimmer}},\ }\bibfield  {title} {\bibinfo {title} {Canonical structure and orthogonality of forces and currents in irreversible markov chains},\ }\href {https://doi.org/10.1007/s10955-018-1986-0} {\bibfield  {journal} {\bibinfo  {journal} {Journal of Statistical Physics}\ }\textbf {\bibinfo {volume} {170}},\ \bibinfo {pages} {1019} (\bibinfo {year} {2018})}\BibitemShut {NoStop}%
\bibitem [{\citenamefont {Renger}\ and\ \citenamefont {Zimmer}(2021)}]{Renger_2021}%
  \BibitemOpen
  \bibfield  {author} {\bibinfo {author} {\bibfnamefont {D.~R.~M.}\ \bibnamefont {Renger}}\ and\ \bibinfo {author} {\bibfnamefont {J.}~\bibnamefont {Zimmer}},\ }\href {https://doi.org/10.3934/dcdss.2020346} {\bibinfo {title} {Orthogonality of fluxes in general nonlinear reaction networks}} (\bibinfo {year} {2021})\BibitemShut {NoStop}%
\bibitem [{\citenamefont {Peletier}\ \emph {et~al.}(2022)\citenamefont {Peletier}, \citenamefont {Rossi}, \citenamefont {Savar{\'e}},\ and\ \citenamefont {Tse}}]{Peletier_2022}%
  \BibitemOpen
  \bibfield  {author} {\bibinfo {author} {\bibfnamefont {M.~A.}\ \bibnamefont {Peletier}}, \bibinfo {author} {\bibfnamefont {R.}~\bibnamefont {Rossi}}, \bibinfo {author} {\bibfnamefont {G.}~\bibnamefont {Savar{\'e}}},\ and\ \bibinfo {author} {\bibfnamefont {O.}~\bibnamefont {Tse}},\ }\bibfield  {title} {\bibinfo {title} {Jump processes as generalized gradient flows},\ }\href {https://doi.org/10.1007/s00526-021-02130-2} {\bibfield  {journal} {\bibinfo  {journal} {Calculus of Variations and Partial Differential Equations}\ }\textbf {\bibinfo {volume} {61}},\ \bibinfo {pages} {33} (\bibinfo {year} {2022})}\BibitemShut {NoStop}%
\bibitem [{\citenamefont {Kobayashi}\ \emph {et~al.}(2022{\natexlab{a}})\citenamefont {Kobayashi}, \citenamefont {Loutchko}, \citenamefont {Kamimura},\ and\ \citenamefont {Sughiyama}}]{kobayashi_2022_hessian_geometry}%
  \BibitemOpen
  \bibfield  {author} {\bibinfo {author} {\bibfnamefont {T.~J.}\ \bibnamefont {Kobayashi}}, \bibinfo {author} {\bibfnamefont {D.}~\bibnamefont {Loutchko}}, \bibinfo {author} {\bibfnamefont {A.}~\bibnamefont {Kamimura}},\ and\ \bibinfo {author} {\bibfnamefont {Y.}~\bibnamefont {Sughiyama}},\ }\bibfield  {title} {\bibinfo {title} {Hessian geometry of nonequilibrium chemical reaction networks and entropy production decompositions},\ }\href {https://doi.org/10.1103/PhysRevResearch.4.033208} {\bibfield  {journal} {\bibinfo  {journal} {Phys. Rev. Res.}\ }\textbf {\bibinfo {volume} {4}},\ \bibinfo {pages} {033208} (\bibinfo {year} {2022}{\natexlab{a}})}\BibitemShut {NoStop}%
\bibitem [{\citenamefont {Kobayashi}\ \emph {et~al.}(2024)\citenamefont {Kobayashi}, \citenamefont {Loutchko}, \citenamefont {Kamimura}, \citenamefont {Horiguchi},\ and\ \citenamefont {Sughiyama}}]{kobayashi_2023_information_graphs_hypergraphs}%
  \BibitemOpen
  \bibfield  {author} {\bibinfo {author} {\bibfnamefont {T.~J.}\ \bibnamefont {Kobayashi}}, \bibinfo {author} {\bibfnamefont {D.}~\bibnamefont {Loutchko}}, \bibinfo {author} {\bibfnamefont {A.}~\bibnamefont {Kamimura}}, \bibinfo {author} {\bibfnamefont {S.~A.}\ \bibnamefont {Horiguchi}},\ and\ \bibinfo {author} {\bibfnamefont {Y.}~\bibnamefont {Sughiyama}},\ }\bibfield  {title} {\bibinfo {title} {Information geometry of dynamics on graphs and hypergraphs},\ }\href {https://doi.org/10.1007/s41884-023-00125-w} {\bibfield  {journal} {\bibinfo  {journal} {Information Geometry}\ }\textbf {\bibinfo {volume} {7}},\ \bibinfo {pages} {97} (\bibinfo {year} {2024})}\BibitemShut {NoStop}%
\bibitem [{\citenamefont {Peletier}\ and\ \citenamefont {Schlichting}(2023)}]{Peletier_2023}%
  \BibitemOpen
  \bibfield  {author} {\bibinfo {author} {\bibfnamefont {M.~A.}\ \bibnamefont {Peletier}}\ and\ \bibinfo {author} {\bibfnamefont {A.}~\bibnamefont {Schlichting}},\ }\bibfield  {title} {\bibinfo {title} {Cosh gradient systems and tilting},\ }\href {https://doi.org/https://doi.org/10.1016/j.na.2022.113094} {\bibfield  {journal} {\bibinfo  {journal} {Nonlinear Analysis}\ }\textbf {\bibinfo {volume} {231}},\ \bibinfo {pages} {113094} (\bibinfo {year} {2023})},\ \bibinfo {note} {variational Models for Discrete Systems}\BibitemShut {NoStop}%
\bibitem [{\citenamefont {Patterson}\ \emph {et~al.}(2024)\citenamefont {Patterson}, \citenamefont {Renger},\ and\ \citenamefont {Sharma}}]{Patterson_2024}%
  \BibitemOpen
  \bibfield  {author} {\bibinfo {author} {\bibfnamefont {R.~I.~A.}\ \bibnamefont {Patterson}}, \bibinfo {author} {\bibfnamefont {D.~R.~M.}\ \bibnamefont {Renger}},\ and\ \bibinfo {author} {\bibfnamefont {U.}~\bibnamefont {Sharma}},\ }\bibfield  {title} {\bibinfo {title} {Variational structures beyond gradient flows: a macroscopic fluctuation-theory perspective},\ }\href {https://doi.org/10.1007/s10955-024-03233-8} {\bibfield  {journal} {\bibinfo  {journal} {Journal of Statistical Physics}\ }\textbf {\bibinfo {volume} {191}},\ \bibinfo {pages} {18} (\bibinfo {year} {2024})}\BibitemShut {NoStop}%
\bibitem [{\citenamefont {Renger}(2024)}]{Renger_2024}%
  \BibitemOpen
  \bibfield  {author} {\bibinfo {author} {\bibfnamefont {D.~R.~M.}\ \bibnamefont {Renger}},\ }\href@noop {} {\bibinfo {title} {Macroscopic fluctuation theory versus large-deviation-induced generic}} (\bibinfo {year} {2024}),\ \Eprint {https://arxiv.org/abs/2402.04092} {arXiv:2402.04092 [math-ph]} \BibitemShut {NoStop}%
\bibitem [{\citenamefont {Kobayashi}\ \emph {et~al.}(2022{\natexlab{b}})\citenamefont {Kobayashi}, \citenamefont {Loutchko}, \citenamefont {Kamimura},\ and\ \citenamefont {Sughiyama}}]{Kobayashi_2022}%
  \BibitemOpen
  \bibfield  {author} {\bibinfo {author} {\bibfnamefont {T.~J.}\ \bibnamefont {Kobayashi}}, \bibinfo {author} {\bibfnamefont {D.}~\bibnamefont {Loutchko}}, \bibinfo {author} {\bibfnamefont {A.}~\bibnamefont {Kamimura}},\ and\ \bibinfo {author} {\bibfnamefont {Y.}~\bibnamefont {Sughiyama}},\ }\bibfield  {title} {\bibinfo {title} {Kinetic derivation of the hessian geometric structure in chemical reaction networks},\ }\href {https://doi.org/10.1103/PhysRevResearch.4.033066} {\bibfield  {journal} {\bibinfo  {journal} {Phys. Rev. Res.}\ }\textbf {\bibinfo {volume} {4}},\ \bibinfo {pages} {033066} (\bibinfo {year} {2022}{\natexlab{b}})}\BibitemShut {NoStop}%
\bibitem [{\citenamefont {Sughiyama}\ \emph {et~al.}(2022)\citenamefont {Sughiyama}, \citenamefont {Loutchko}, \citenamefont {Kamimura},\ and\ \citenamefont {Kobayashi}}]{Sughiyama_2022}%
  \BibitemOpen
  \bibfield  {author} {\bibinfo {author} {\bibfnamefont {Y.}~\bibnamefont {Sughiyama}}, \bibinfo {author} {\bibfnamefont {D.}~\bibnamefont {Loutchko}}, \bibinfo {author} {\bibfnamefont {A.}~\bibnamefont {Kamimura}},\ and\ \bibinfo {author} {\bibfnamefont {T.~J.}\ \bibnamefont {Kobayashi}},\ }\bibfield  {title} {\bibinfo {title} {Hessian geometric structure of chemical thermodynamic systems with stoichiometric constraints},\ }\href {https://doi.org/10.1103/PhysRevResearch.4.033065} {\bibfield  {journal} {\bibinfo  {journal} {Phys. Rev. Res.}\ }\textbf {\bibinfo {volume} {4}},\ \bibinfo {pages} {033065} (\bibinfo {year} {2022})}\BibitemShut {NoStop}%
\bibitem [{\citenamefont {Renger}\ and\ \citenamefont {Sharma}(2023)}]{Renger_2023}%
  \BibitemOpen
  \bibfield  {author} {\bibinfo {author} {\bibfnamefont {D.~R.~M.}\ \bibnamefont {Renger}}\ and\ \bibinfo {author} {\bibfnamefont {U.}~\bibnamefont {Sharma}},\ }\bibfield  {title} {\bibinfo {title} {Untangling dissipative and hamiltonian effects in bulk and boundary-driven systems},\ }\href {https://doi.org/10.1103/PhysRevE.108.054123} {\bibfield  {journal} {\bibinfo  {journal} {Phys. Rev. E}\ }\textbf {\bibinfo {volume} {108}},\ \bibinfo {pages} {054123} (\bibinfo {year} {2023})}\BibitemShut {NoStop}%
\bibitem [{\citenamefont {Duong}\ and\ \citenamefont {Zimmer}(2023)}]{duong_2023}%
  \BibitemOpen
  \bibfield  {author} {\bibinfo {author} {\bibfnamefont {M.~H.}\ \bibnamefont {Duong}}\ and\ \bibinfo {author} {\bibfnamefont {J.}~\bibnamefont {Zimmer}},\ }\bibfield  {title} {\bibinfo {title} {On decompositions of non-reversible processes},\ }\href {https://doi.org/10.1088/1742-6596/2514/1/012007} {\bibfield  {journal} {\bibinfo  {journal} {Journal of Physics: Conference Series}\ }\textbf {\bibinfo {volume} {2514}},\ \bibinfo {pages} {012007} (\bibinfo {year} {2023})}\BibitemShut {NoStop}%
\bibitem [{\citenamefont {Mizohata}\ \emph {et~al.}(2024)\citenamefont {Mizohata}, \citenamefont {Kobayashi}, \citenamefont {Bouchard},\ and\ \citenamefont {Miyahara}}]{Mizohata_2024}%
  \BibitemOpen
  \bibfield  {author} {\bibinfo {author} {\bibfnamefont {T.}~\bibnamefont {Mizohata}}, \bibinfo {author} {\bibfnamefont {T.~J.}\ \bibnamefont {Kobayashi}}, \bibinfo {author} {\bibfnamefont {L.-S.}\ \bibnamefont {Bouchard}},\ and\ \bibinfo {author} {\bibfnamefont {H.}~\bibnamefont {Miyahara}},\ }\bibfield  {title} {\bibinfo {title} {Information geometric bound on general chemical reaction networks},\ }\href {https://doi.org/10.1103/PhysRevE.109.044308} {\bibfield  {journal} {\bibinfo  {journal} {Phys. Rev. E}\ }\textbf {\bibinfo {volume} {109}},\ \bibinfo {pages} {044308} (\bibinfo {year} {2024})}\BibitemShut {NoStop}%
\bibitem [{\citenamefont {Loutchko}\ \emph {et~al.}(2023)\citenamefont {Loutchko}, \citenamefont {Sughiyama},\ and\ \citenamefont {Kobayashi}}]{Loutchko_2023_geometry_tur}%
  \BibitemOpen
  \bibfield  {author} {\bibinfo {author} {\bibfnamefont {D.}~\bibnamefont {Loutchko}}, \bibinfo {author} {\bibfnamefont {Y.}~\bibnamefont {Sughiyama}},\ and\ \bibinfo {author} {\bibfnamefont {T.~J.}\ \bibnamefont {Kobayashi}},\ }\href@noop {} {\bibinfo {title} {The geometry of thermodynamic uncertainty relations in chemical reaction networks}} (\bibinfo {year} {2023}),\ \Eprint {https://arxiv.org/abs/2308.04806} {arXiv:2308.04806 [cond-mat.stat-mech]} \BibitemShut {NoStop}%
\bibitem [{\citenamefont {Maes}(2020)}]{Maes_2020}%
  \BibitemOpen
  \bibfield  {author} {\bibinfo {author} {\bibfnamefont {C.}~\bibnamefont {Maes}},\ }\bibfield  {title} {\bibinfo {title} {Frenesy: Time-symmetric dynamical activity in nonequilibria},\ }\href {https://doi.org/https://doi.org/10.1016/j.physrep.2020.01.002} {\bibfield  {journal} {\bibinfo  {journal} {Physics Reports}\ }\textbf {\bibinfo {volume} {850}},\ \bibinfo {pages} {1} (\bibinfo {year} {2020})}\BibitemShut {NoStop}%
\bibitem [{\citenamefont {Onsager}\ and\ \citenamefont {Machlup}(1953)}]{Onsager_1953}%
  \BibitemOpen
  \bibfield  {author} {\bibinfo {author} {\bibfnamefont {L.}~\bibnamefont {Onsager}}\ and\ \bibinfo {author} {\bibfnamefont {S.}~\bibnamefont {Machlup}},\ }\bibfield  {title} {\bibinfo {title} {Fluctuations and irreversible processes},\ }\href {https://doi.org/10.1103/PhysRev.91.1505} {\bibfield  {journal} {\bibinfo  {journal} {Phys. Rev.}\ }\textbf {\bibinfo {volume} {91}},\ \bibinfo {pages} {1505} (\bibinfo {year} {1953})}\BibitemShut {NoStop}%
\bibitem [{\citenamefont {Machlup}\ and\ \citenamefont {Onsager}(1953)}]{Onsager_1953_2}%
  \BibitemOpen
  \bibfield  {author} {\bibinfo {author} {\bibfnamefont {S.}~\bibnamefont {Machlup}}\ and\ \bibinfo {author} {\bibfnamefont {L.}~\bibnamefont {Onsager}},\ }\bibfield  {title} {\bibinfo {title} {Fluctuations and irreversible process. ii. systems with kinetic energy},\ }\href {https://doi.org/10.1103/PhysRev.91.1512} {\bibfield  {journal} {\bibinfo  {journal} {Phys. Rev.}\ }\textbf {\bibinfo {volume} {91}},\ \bibinfo {pages} {1512} (\bibinfo {year} {1953})}\BibitemShut {NoStop}%
\bibitem [{\citenamefont {Reimann}\ \emph {et~al.}(2001)\citenamefont {Reimann}, \citenamefont {Van~den Broeck}, \citenamefont {Linke}, \citenamefont {H\"anggi}, \citenamefont {Rubi},\ and\ \citenamefont {P\'erez-Madrid}}]{Reimann_2001}%
  \BibitemOpen
  \bibfield  {author} {\bibinfo {author} {\bibfnamefont {P.}~\bibnamefont {Reimann}}, \bibinfo {author} {\bibfnamefont {C.}~\bibnamefont {Van~den Broeck}}, \bibinfo {author} {\bibfnamefont {H.}~\bibnamefont {Linke}}, \bibinfo {author} {\bibfnamefont {P.}~\bibnamefont {H\"anggi}}, \bibinfo {author} {\bibfnamefont {J.~M.}\ \bibnamefont {Rubi}},\ and\ \bibinfo {author} {\bibfnamefont {A.}~\bibnamefont {P\'erez-Madrid}},\ }\bibfield  {title} {\bibinfo {title} {Giant acceleration of free diffusion by use of tilted periodic potentials},\ }\href {https://doi.org/10.1103/PhysRevLett.87.010602} {\bibfield  {journal} {\bibinfo  {journal} {Phys. Rev. Lett.}\ }\textbf {\bibinfo {volume} {87}},\ \bibinfo {pages} {010602} (\bibinfo {year} {2001})}\BibitemShut {NoStop}%
\bibitem [{\citenamefont {Lindner}\ and\ \citenamefont {Sokolov}(2016)}]{Lindner_2016}%
  \BibitemOpen
  \bibfield  {author} {\bibinfo {author} {\bibfnamefont {B.}~\bibnamefont {Lindner}}\ and\ \bibinfo {author} {\bibfnamefont {I.~M.}\ \bibnamefont {Sokolov}},\ }\bibfield  {title} {\bibinfo {title} {Giant diffusion of underdamped particles in a biased periodic potential},\ }\href {https://doi.org/10.1103/PhysRevE.93.042106} {\bibfield  {journal} {\bibinfo  {journal} {Phys. Rev. E}\ }\textbf {\bibinfo {volume} {93}},\ \bibinfo {pages} {042106} (\bibinfo {year} {2016})}\BibitemShut {NoStop}%
\bibitem [{\citenamefont {Chatterjee}\ \emph {et~al.}(2020)\citenamefont {Chatterjee}, \citenamefont {Mangeat}, \citenamefont {Paul},\ and\ \citenamefont {Rieger}}]{Chatterjee_2020}%
  \BibitemOpen
  \bibfield  {author} {\bibinfo {author} {\bibfnamefont {S.}~\bibnamefont {Chatterjee}}, \bibinfo {author} {\bibfnamefont {M.}~\bibnamefont {Mangeat}}, \bibinfo {author} {\bibfnamefont {R.}~\bibnamefont {Paul}},\ and\ \bibinfo {author} {\bibfnamefont {H.}~\bibnamefont {Rieger}},\ }\bibfield  {title} {\bibinfo {title} {Flocking and reorientation transition in the 4-state active potts model},\ }\href {https://doi.org/10.1209/0295-5075/130/66001} {\bibfield  {journal} {\bibinfo  {journal} {Europhysics Letters}\ }\textbf {\bibinfo {volume} {130}},\ \bibinfo {pages} {66001} (\bibinfo {year} {2020})}\BibitemShut {NoStop}%
\bibitem [{\citenamefont {Mangeat}\ \emph {et~al.}(2020)\citenamefont {Mangeat}, \citenamefont {Chatterjee}, \citenamefont {Paul},\ and\ \citenamefont {Rieger}}]{Mangeat_2020}%
  \BibitemOpen
  \bibfield  {author} {\bibinfo {author} {\bibfnamefont {M.}~\bibnamefont {Mangeat}}, \bibinfo {author} {\bibfnamefont {S.}~\bibnamefont {Chatterjee}}, \bibinfo {author} {\bibfnamefont {R.}~\bibnamefont {Paul}},\ and\ \bibinfo {author} {\bibfnamefont {H.}~\bibnamefont {Rieger}},\ }\bibfield  {title} {\bibinfo {title} {Flocking with a $q$-fold discrete symmetry: Band-to-lane transition in the active potts model},\ }\href {https://doi.org/10.1103/PhysRevE.102.042601} {\bibfield  {journal} {\bibinfo  {journal} {Phys. Rev. E}\ }\textbf {\bibinfo {volume} {102}},\ \bibinfo {pages} {042601} (\bibinfo {year} {2020})}\BibitemShut {NoStop}%
\bibitem [{\citenamefont {Chatterjee}\ \emph {et~al.}(2022)\citenamefont {Chatterjee}, \citenamefont {Mangeat},\ and\ \citenamefont {Rieger}}]{Chatterjee_2022}%
  \BibitemOpen
  \bibfield  {author} {\bibinfo {author} {\bibfnamefont {S.}~\bibnamefont {Chatterjee}}, \bibinfo {author} {\bibfnamefont {M.}~\bibnamefont {Mangeat}},\ and\ \bibinfo {author} {\bibfnamefont {H.}~\bibnamefont {Rieger}},\ }\bibfield  {title} {\bibinfo {title} {Polar flocks with discretized directions: The active clock model approaching the vicsek model},\ }\href {https://doi.org/10.1209/0295-5075/ac6e4b} {\bibfield  {journal} {\bibinfo  {journal} {Europhysics Letters}\ }\textbf {\bibinfo {volume} {138}},\ \bibinfo {pages} {41001} (\bibinfo {year} {2022})}\BibitemShut {NoStop}%
\bibitem [{\citenamefont {Karmakar}\ \emph {et~al.}(2023)\citenamefont {Karmakar}, \citenamefont {Chatterjee}, \citenamefont {Mangeat}, \citenamefont {Rieger},\ and\ \citenamefont {Paul}}]{Karmakar_2023}%
  \BibitemOpen
  \bibfield  {author} {\bibinfo {author} {\bibfnamefont {M.}~\bibnamefont {Karmakar}}, \bibinfo {author} {\bibfnamefont {S.}~\bibnamefont {Chatterjee}}, \bibinfo {author} {\bibfnamefont {M.}~\bibnamefont {Mangeat}}, \bibinfo {author} {\bibfnamefont {H.}~\bibnamefont {Rieger}},\ and\ \bibinfo {author} {\bibfnamefont {R.}~\bibnamefont {Paul}},\ }\bibfield  {title} {\bibinfo {title} {Jamming and flocking in the restricted active potts model},\ }\href {https://doi.org/10.1103/PhysRevE.108.014604} {\bibfield  {journal} {\bibinfo  {journal} {Phys. Rev. E}\ }\textbf {\bibinfo {volume} {108}},\ \bibinfo {pages} {014604} (\bibinfo {year} {2023})}\BibitemShut {NoStop}%
\bibitem [{\citenamefont {Woo}\ and\ \citenamefont {Noh}(2024)}]{Woo_2024}%
  \BibitemOpen
  \bibfield  {author} {\bibinfo {author} {\bibfnamefont {C.-U.}\ \bibnamefont {Woo}}\ and\ \bibinfo {author} {\bibfnamefont {J.~D.}\ \bibnamefont {Noh}},\ }\bibfield  {title} {\bibinfo {title} {Motility-induced pinning in flocking system with discrete symmetry},\ }\href {https://link.aps.org/doi/10.1103/PhysRevLett.133.188301} {\bibfield  {journal} {\bibinfo  {journal} {Phys. Rev. Lett.}\ }\textbf {\bibinfo {volume} {133}},\ \bibinfo {pages} {188301} (\bibinfo {year} {2024})}\BibitemShut {NoStop}%
\bibitem [{\citenamefont {Konarovskyi}\ \emph {et~al.}(2019)\citenamefont {Konarovskyi}, \citenamefont {Lehmann},\ and\ \citenamefont {von Renesse}}]{Konarovskyi_2019}%
  \BibitemOpen
  \bibfield  {author} {\bibinfo {author} {\bibfnamefont {V.}~\bibnamefont {Konarovskyi}}, \bibinfo {author} {\bibfnamefont {T.}~\bibnamefont {Lehmann}},\ and\ \bibinfo {author} {\bibfnamefont {M.-K.}\ \bibnamefont {von Renesse}},\ }\bibfield  {title} {\bibinfo {title} {{Dean-Kawasaki dynamics: ill-posedness vs. triviality}},\ }\href {https://doi.org/10.1214/19-ECP208} {\bibfield  {journal} {\bibinfo  {journal} {Electronic Communications in Probability}\ }\textbf {\bibinfo {volume} {24}},\ \bibinfo {pages} {1 } (\bibinfo {year} {2019})}\BibitemShut {NoStop}%
\bibitem [{\citenamefont {Konarovskyi}\ \emph {et~al.}(2020)\citenamefont {Konarovskyi}, \citenamefont {Lehmann},\ and\ \citenamefont {von Renesse}}]{Konarovskyi_2020}%
  \BibitemOpen
  \bibfield  {author} {\bibinfo {author} {\bibfnamefont {V.}~\bibnamefont {Konarovskyi}}, \bibinfo {author} {\bibfnamefont {T.}~\bibnamefont {Lehmann}},\ and\ \bibinfo {author} {\bibfnamefont {M.}~\bibnamefont {von Renesse}},\ }\bibfield  {title} {\bibinfo {title} {On dean--kawasaki dynamics with smooth drift potential},\ }\href {https://doi.org/10.1007/s10955-019-02449-3} {\bibfield  {journal} {\bibinfo  {journal} {Journal of Statistical Physics}\ }\textbf {\bibinfo {volume} {178}},\ \bibinfo {pages} {666} (\bibinfo {year} {2020})}\BibitemShut {NoStop}%
\bibitem [{\citenamefont {Jin}\ \emph {et~al.}(2026)\citenamefont {Jin}, \citenamefont {Liu},\ and\ \citenamefont {Reichman}}]{Jin_2026}%
  \BibitemOpen
  \bibfield  {author} {\bibinfo {author} {\bibfnamefont {J.}~\bibnamefont {Jin}}, \bibinfo {author} {\bibfnamefont {C.}~\bibnamefont {Liu}},\ and\ \bibinfo {author} {\bibfnamefont {D.~R.}\ \bibnamefont {Reichman}},\ }\bibfield  {title} {\bibinfo {title} {Field-theoretic simulation of dean-kawasaki dynamics for interacting particles},\ }\href {https://doi.org/10.1103/d1xd-9tw3} {\bibfield  {journal} {\bibinfo  {journal} {Phys. Rev. E}\ }\textbf {\bibinfo {volume} {113}},\ \bibinfo {pages} {044138} (\bibinfo {year} {2026})}\BibitemShut {NoStop}%
\bibitem [{\citenamefont {Scandolo}\ \emph {et~al.}(2023)\citenamefont {Scandolo}, \citenamefont {Pausch},\ and\ \citenamefont {Cates}}]{scandolo_2023}%
  \BibitemOpen
  \bibfield  {author} {\bibinfo {author} {\bibfnamefont {M.}~\bibnamefont {Scandolo}}, \bibinfo {author} {\bibfnamefont {J.}~\bibnamefont {Pausch}},\ and\ \bibinfo {author} {\bibfnamefont {M.~E.}\ \bibnamefont {Cates}},\ }\bibfield  {title} {\bibinfo {title} {Active ising models of flocking: a field-theoretic approach},\ }\href {https://doi.org/10.1140/epje/s10189-023-00364-w} {\bibfield  {journal} {\bibinfo  {journal} {The European Physical Journal E}\ }\textbf {\bibinfo {volume} {46}},\ \bibinfo {pages} {103} (\bibinfo {year} {2023})}\BibitemShut {NoStop}%
\end{thebibliography}%
\vskip0.5cm
\newpage
\appendix
\section{Coarse-Graining}\label{app:cg}
\subsection{Normal ordering the transition Hamiltonian: The derivation of \cref{eq:hamiltonian_quantised_single_reactive_jump_normal_ordered} from \cref{eq:hamiltonian_quantised_single_reactive_jump_1}}
The second quantized Hamiltonian for the reactive transition is reorganized to obtain the normal ordered form.  
\begin{widetext}
\begin{equation}\label{eq:hamiltonian_quantised_single_reactive_jump_appendix}
\begin{split}
\hat{H}_{\gamma \gamma'}^{\#}
& = d_{\gamma \gamma'} \left[ \crtlong{\gamma}{\#} - \crtlong{\gamma'}{\#} \right] \left[ \anhlong{\gamma}{\#}  e^{\epsilon_{\gamma}^\# - \frac 1 2 f_{\gamma \gamma'}^{ch} } - \anhlong{\gamma'}{\#} e^{\epsilon_{\gamma'}^\# + \frac 1 2 f_{\gamma \gamma'}^{ch} } \right]  
\\
& = d_{\gamma \gamma'} \left[ \crtlong{\gamma}{\#} - \crtlong{\gamma'}{\#} \right]
\left[ 
\anhlong{\gamma}{\#}
e^{ \left(\epsilon_{\gamma \gamma}^\# + \sum_{j \neq \gamma} \epsilon_{\gamma j}^\# \right) - \frac 1 2 f_{\gamma \gamma'}^{ch} }
- \anhlong{\gamma'}{\#}
e^{ \left(\epsilon_{\gamma' \gamma'}^\# + \sum_{j \neq \gamma'} \epsilon_{\gamma' j}^\# \right) + \frac 1 2 f_{\gamma \gamma'}^{ch}} 
\right] 
\\
& = d_{\gamma \gamma'} \left[ \crtlong{\gamma}{\#} - \crtlong{\gamma'}{\#} \right]
\left[ 
\anhlong{\gamma}{\#}
e^{ \beta \left( v_{\gamma \gamma} \left( \densoplongg{\gamma}{\#} - 1 \right)  + \sum_{j\neq\gamma} v_{\gamma j} \densoplongg{j}{\#} \right) - \frac 1 2 f_{\gamma \gamma'}^{ch} }
- \anhlong{\gamma'}{\#}
e^{ \beta \left(v_{\gamma' \gamma'} \left( \densoplongg{\gamma'}{\#} - 1 \right)  + \sum_{j \neq \gamma'} v_{\gamma' j} \densoplongg{j}{\#} \right) + \frac 1 2 f_{\gamma \gamma'}^{ch} }
\right]. 
\end{split}    
\end{equation}
\end{widetext}
The second-quantized Hamiltonian contains exponential density operator terms due to the interacting nature of the particles. Following Ref. \cite{blasiak2007}, the normal ordering of the exponential of a density operator is given by,
\begin{equation}
\begin{split}
    e^{ \lambda \densoplongg{i}{\#} } = \: : e^{ \densoplongg{i}{\#} \left(e^\lambda - 1\right)} :.
\end{split}
\end{equation}
Which reduces \cref{eq:hamiltonian_quantised_single_reactive_jump_1} to \cref{eq:hamiltonian_quantised_single_reactive_jump_normal_ordered}.
\subsection{The derivation of \cref{eq:hamiltonian_reactive_transition_density_noise} from \cref{eq:hamiltonian_reactive_jump_sigle}}
\begin{widetext}
\begin{equation}\label{eq:lagrangian_reactive_density_noise_dp_appendix}
\begin{split}
   -\mathcal{H}_{\gamma \gamma'}^\# \left[ \{ N, \chi \} \right]
   & = d_{\gamma \gamma'} \left[ e^{\chi_\gamma^\# } - e^{ \chi_{\gamma'}^\# } \right]
   \left[ 
   N_{\gamma}^\# 
   e^{-\chi_{\gamma}^\# } 
   \exp{ \left( \sum_{j} N_j^\# \left( e^{ \beta v_{\gamma j} } - 1 \right) - \frac 1 2 f_{\gamma \gamma'}^{ch} \right) }
   - N_{\gamma'}^\# 
   e^{-\chi_{\gamma'}^\# } 
   \exp{ \left( \sum_{j} N_j^\# \left( e^{ \beta v_{\gamma' j}} - 1 \right) + \frac 1 2 f_{\gamma \gamma'}^{ch} \right) }
   \right]
   \\
   & = d_{\gamma \gamma'}
   \left[ 
   \left[ 1 - e^{\chi_{\gamma'}^\# - \chi_{\gamma}^\# } \right] N_{\gamma}^\#
   \exp{ \left( \sum_{j} N_j^\# \left( e^{ \beta v_{\gamma j}} - 1 \right) - \frac 1 2 f_{\gamma \gamma'}^{ch} \right) } 
   + \left( 1 - e^{ \chi_{\gamma}^\# - \chi_{\gamma'}^\# } \right) N_{\gamma'}^\# 
   \exp{ \left( \sum_{j} N_j^\# \left( e^{ \beta v_{\gamma' j}} - 1 \right) + \frac 1 2 f_{\gamma \gamma'}^{ch} \right) }
   \right]
   \\
   & = d_{\gamma \gamma'}
   \left[ 
   \left[ 1 - e^{ \chi_{\gamma'}^\# - \chi_{\gamma}^\# } \right] 
   \exp{ \left( \ln{N_{\gamma}^\#}  + \sum_{j} N_j^\# \left( e^{ \beta v_{\gamma j}} - 1 \right) - \frac 1 2 f_{\gamma \gamma'}^{ch} \right) }
   + \left[ 1 - e^{\chi_{\gamma}^\# - \chi_{\gamma'}^\#} \right]
   \exp{ \left( \ln{N_{\gamma'}^\# }  + \sum_{j} N_j^\# \left( e^{ \beta v_{\gamma' j}} - 1 \right) + \frac 1 2 f_{\gamma \gamma'}^{ch} \right) }
   \right]
   \\
   & = d_{\gamma \gamma'}
   \left[ 
   \left[ 1 - e^{\chi_{\gamma'}^\# - \chi_{\gamma}^\# } \right] 
   \exp{ \left( \upmu_\gamma^\# - \frac 1 2 \mathcal{F}_{\gamma \gamma'}^{ch} \right) }
   + \left[ 1 - e^{ \chi_{\gamma}^\# - \chi_{\gamma'}^\# } \right] 
   \exp{ \left( \upmu_{\gamma'}^\# + \frac 1 2 \mathcal{F}_{\gamma \gamma'}^{ch} \right) }
   \right]
   \\
   & = d_{\gamma \gamma'}
   \left( 
   \left[ 1 - e^{\chi_{\gamma'}^\# - \chi_{\gamma}^\# } \right] 
   \exp{ \left( \upmu_\gamma^r + \mathcal{F}_{\gamma}^{nr} - \frac 1 2 \mathcal{F}_{\gamma \gamma'}^{ch} \right) }
   + \left[ 1 - e^{\chi_{\gamma}^\# - \chi_{\gamma'}^\# } \right] 
   \exp{ \left( \upmu_{\gamma'}^r + \mathcal{F}_{\gamma'}^{nr} + \frac 1 2 \mathcal{F}_{\gamma \gamma'}^{ch} \right) }
   \right).
\end{split}
\end{equation}
\end{widetext}
\section{Cummulants of the transition currents}\label{app:cummulants_transition_current}
The first and second cumulants of the transition currents are denoted by $\mathcal{J}_{\Delta}$ and $\mathcal{T}_\Delta$, respectively, and are computed using the macroscopic Hamiltonian $\mathcal{H}$. For notational convenience, we adopt the shorthand
$
\mathcal{J}_{\Delta} =  \mathcal{J}_{\gamma \gamma'}^\#, \quad \mathcal{T}_{\Delta} =  \mathcal{T}_{\gamma \gamma'}^\#, \quad \Delta \chi = \chi_{\gamma} - \chi_{\gamma'}
$ 
for the reactive transition $\Delta_{\gamma \gamma'}^\#$, and
$
\mathcal{J}_{\Delta} =  \mathcal{J}_{i}^{\vec{\mathcal{D}}\#}, \quad \mathcal{T}_{\Delta} =  \mathcal{T}_{i}^{\vec{\mathcal{D}}\#}, \quad \Delta \chi = \chi_i^{\vec{\mathcal{D}}\#} - \chi_i^\#
$ 
for the diffusive transition $\Delta_{i}^{\vec{\mathcal{D}}\#}$.
\begin{equation}\label{eq:current_cummulants}
\begin{split}
    \mathcal{J}_{\Delta} & = \partial_{\Delta \chi} \Delta \mathcal{H}|_{\chi = 0} = d_{\gamma \gamma'}
   \left( 
   e^{ \upmu_\gamma^\# - \frac 1 2 \mathcal{F}_{\gamma \gamma'}^{ch} }
   -  
   e^{ \upmu_{\gamma'}^\# + \frac 1 2 \mathcal{F}_{\gamma \gamma'}^{ch} }
   \right),
   \\
   \mathcal{T}_{\Delta} & = \partial^2_{\Delta \chi} \Delta \mathcal{H}|_{\chi = 0}
   = d_{\gamma \gamma'}
   \left( 
   e^{ \upmu_\gamma^\# - \frac 1 2 \mathcal{F}_{\gamma \gamma'}^{ch} }
   +  
   e^{ \upmu_{\gamma'}^\# + \frac 1 2 \mathcal{F}_{\gamma \gamma'}^{ch} }
   \right).
\end{split}
\end{equation}
The second cumulant of the current is given by the \emph{traffic} \cite{Maes_2020}. Here, the traffic is defined as the symmetric part of the transition currents, obtained as the modulus of the unidirectional currents. The reactive and diffusive mean currents and traffic read:
\begin{equation}
\begin{split}
   \mathcal{J}_{\gamma \gamma'}^\# 
   & = d_{\gamma \gamma'}
   \left( 
   e^{ \upmu_\gamma^\# - \frac 1 2 \mathcal{F}_{\gamma \gamma'}^{ch} }
   -  
   e^{ \upmu_{\gamma'}^\# + \frac 1 2 \mathcal{F}_{\gamma \gamma'}^{ch} }
   \right),
   \\
   \mathcal{T}_{\gamma \gamma'}^\#
   & = d_{\gamma \gamma'}
   \left( 
   e^{ \upmu_\gamma^\# - \frac 1 2 \mathcal{F}_{\gamma \gamma'}^{ch} }
   +  
   e^{ \upmu_{\gamma'}^\# + \frac 1 2 \mathcal{F}_{\gamma \gamma'}^{ch} }
   \right).
   \\
   \mathcal{J}_{i}^{\vec{\mathcal{D}}\#} 
   & = d_{\gamma \gamma'}
   \left( 
   e^{ \upmu_\gamma^\# - \frac 1 2 \mathcal{F}_{\gamma \gamma'}^{ch} }
   -  
   e^{ \upmu_{\gamma'}^\# + \frac 1 2 \mathcal{F}_{\gamma \gamma'}^{ch} }
   \right),
   \\ 
   \mathcal{T}_{i}^{\vec{\mathcal{D}}\#}
   & = d_{\gamma \gamma'}
   \left( 
   e^{ \upmu_\gamma^{\vec{\mathcal{D}}\#} - \frac 1 2 \mathcal{F}_{\gamma \gamma'}^{ch} }
   +  
   e^{ \upmu_{\gamma'}^\# + \frac 1 2 \mathcal{F}_{\gamma \gamma'}^{ch} }
   \right).
\end{split}    
\end{equation}
Importantly, the $n^{\rm th}$-order cumulant $\mathcal{J}_{\Delta}^n$ satisfies $\mathcal{J}_{\Delta}^n = \partial^n_{\Delta\chi} \Delta \mathcal{H}\big|_{\chi = 0}$. Thus, the recursive relation between the current cumulants holds, $\mathcal{J}_{\Delta}^n = \mathcal{J}_{\Delta}^{\,n-2}$. This justifies the use of a Langevin (Gaussian) approximation for the stochastic dynamics of the coarse-grained meso- or macrostate, valid for sufficiently large $\Omega$. Mesoscopic systems subject to Poissonian transition fluctuations require a more systematic treatment \cite{atm_2024_var_epr,atm_2025_var_epr_derivation}. Here, we focus on the regime where the coarse-grained macroscopic description is valid, i.e., where transition fluctuations are effectively Gaussian.

\section{Diffusive transition Hamiltonian and current cumulants}\label{app:cg_diffusive_hamiltonian}

\subsection{Mesoscopic Diffusion Hamiltonian}
Here, we consider the mesoscopic diffusive Hamiltonian
\begin{widetext}
\begin{equation}
\begin{split}
   \mathcal{H}^{\mathcal{D}} \left[ \{N_i^{\#}, \chi_i^\# \} \right]
   & = \sum_{\#,i} d_{i}
   \bigg[ 
   \left( e^{ \chi_i^{{\mathcal{D}}^{\parallel} \#} - \chi_i^\#} - 1 \right) 
   e^{ \upmu_i^\# + \frac 1 2 {f}_{i}^{sp} }
   + \left( e^{ \chi_i^\# - \chi_i^{{\mathcal{D}}^{\parallel} \#} } - 1 \right) 
   e^{ \upmu_i^{{\mathcal{D}}^\parallel \#} - \frac 1 2 {f}_{i}^{sp} }
   +
   \left( e^{ \chi_i^{{\mathcal{D}}^{\nparallel} \#} - \chi_i^\#} - 1 \right) 
   e^{ \upmu_i^\# - \frac 1 2 {f}_{i}^{sp} }
   + \left( e^{ \chi_i^\# - \chi_i^{{\mathcal{D}}^{\nparallel} \#} } - 1 \right) 
   e^{ \upmu_i^{{\mathcal{D}}^\parallel \#} + \frac 1 2 {f}_{i}^{sp} }
   \\ & \hspace{2cm}
   +
   \left( e^{ \chi_i^{{\mathcal{D}}^{\perp} \#} - \chi_i^\#} - 1 \right) 
   e^{ \upmu_i^\#  }
   + \left( e^{ \chi_i^\# - \chi_i^{{\mathcal{D}}^{\perp} \#} } - 1 \right) 
   e^{ \upmu_i^{{\mathcal{D}}^\perp \#} }
   +
   \left( e^{ \chi_i^{{\mathcal{D}}^{\top} \#} - \chi_i^\#} - 1 \right) 
   e^{ \upmu_i^\#  }
   + \left( e^{ \chi_i^\# - \chi_i^{{\mathcal{D}}^{\top} \#} } - 1 \right) 
   e^{ \upmu_i^{{\mathcal{D}}^\top \#} }
   \bigg].
\end{split}    
\end{equation}
\end{widetext}
Where, we have utilized $\mathcal{D}^{\parallel} \cdot \vec{f}^{sp} = f^{sp}$,  $\mathcal{D}^{\nparallel} \cdot \vec{f}^{sp} = -f^{sp}$, $\mathcal{D}^\perp \cdot \vec{f}^{sp} = 0$ and, $\mathcal{D}^\top \cdot \vec{f}^{sp} = 0$. 
Thus, the deterministic transition currents read,
\begin{widetext}
\begin{equation}\label{eq:derivation_mesoscopic_diffusive_currents}
\begin{split}
   \partial_{\chi_i^\#} \mathcal{H}^\mathcal{D} \left[ \{N_i^{\#}, \chi_i^\# \} \right]|_{\{\chi\} = \{0\}}
   & = d_{i}
   \bigg[ 
   - 
   e^{ \upmu_i^\# + \frac 1 2 {f}_{i}^{sp} }
   + 
   e^{ \upmu_i^{{\mathcal{D}}^\parallel \#} - \frac 1 2 {f}_{i}^{sp} }
   - 
   e^{ \upmu_i^\# - \frac 1 2 {f}_{i}^{sp} }
   + 
   e^{ \upmu_i^{{\mathcal{D}}^\nparallel \#} + \frac 1 2 {f}_{i}^{sp} }
   - 
   e^{ \upmu_i^\#  }
   +  
   e^{ \upmu_i^{{\mathcal{D}}^\perp \#} }
   - 
   e^{ \upmu_i^\#  }
   +  
   e^{ \upmu_i^{{\mathcal{D}}^\top \#} }
   \bigg]
   \\
   & = d_{i}
   \bigg[ 
   \cosh{\left(\frac{f_i^{sp}}{2}\right)}
   \left( 
   e^{ \upmu_i^{{\mathcal{D}}^\nparallel \#} }
   +
   e^{ \upmu_i^{{\mathcal{D}}^\parallel \#}  }
   - 
   2 e^{ \upmu_i^\# } 
   \right)
   +
   \sinh{\left(\frac{f_i^{sp}}{2}\right)}
   \left( 
   e^{ \upmu_i^{{\mathcal{D}}^\nparallel \#} }
   -
   e^{ \upmu_i^{{\mathcal{D}}^\parallel \#}  }
   \right)
   +  
   e^{ \upmu_i^{{\mathcal{D}}^\perp \#} }
   +  
   e^{ \upmu_i^{{\mathcal{D}}^\top \#} }
   - 
   2 e^{ \upmu_i^\#  }
   \bigg]
   \\
   & = d_{i}
   \bigg[ 
   \cosh{\left(\frac{f_i^{sp}}{2}\right)}
   \Delta^\parallel e^{ \upmu_i^\# } 
   +
   2 \sinh{\left(\frac{f_i^{sp}}{2}\right)}
   \nabla^{\parallel} e^{ \upmu_i^\#  }
   +  
   \Delta^\perp e^{ \upmu_i^\#  }
   \bigg]
   \\
   & = d_{i}
   \bigg[ 
   \cosh{\left(\frac{f_i^{sp}}{2}\right)}
   \nabla^\parallel \left( e^{ \upmu_i^\# } \left[ \nabla^\parallel \upmu_i^\# +
   2 \tanh{\left(\frac{f_i^{sp}}{2}\right)} \right] \right) 
   +
   \nabla^\perp \left( e^{ \upmu_i^\# } \nabla^\perp \upmu_i^\# \right) 
   \bigg]
\end{split}    
\end{equation}
\end{widetext}
Where, $\Delta^\parallel e^{ \upmu_i^\# }$ and $\Delta^\perp e^{ \upmu_i^\#  }$ are the discrete Laplacian operators in the directions parallel and perpendicular to the self-propulsion, respectively. $\nabla^{\parallel} e^{ \upmu_i^\#  }$ is the gradient operator in the direction parallel to the self-propulsion. In the last line, we have decomposed the Laplacian into the gradient form to obtain a familiar expression of the diffusive currents, 
$
\Delta^\parallel e^{ \upmu_i^\# } = \nabla^\parallel \cdot \left( e^{ \upmu_i^\# }  \nabla^\parallel \upmu_i^\#  \right) .
$
This form allows us to identify the diffusive transition mobilities,
$
D_i^{\mathcal{D}^{\parallel}} = d_i^\parallel e^{\upmu_i^\#}, \quad D_i^{\mathcal{D}^{\perp}} = d_i^\perp e^{\upmu_i^\#},
$
with $d_i^\parallel = d_i \cosh{\left(\frac{f_i^{sp}}{2}\right)}$, $d_i^\perp = d_i$. Here, $\nabla \upmu_i^\#$ is the thermodynamic force along the diffusive direction. Similarly, the variance of the transition fluctuations is determined by the curvature of the Hamiltonian.
\begin{equation}\label{eq:}
\begin{split}
    -\partial_{\chi_i^{\vec{\mathcal{D}}\#}} 
    \partial_{\chi_i^\#} \mathcal{H}^\mathcal{D} \left[ \{ N_i^{\#}, \chi_i^\# \} \right]|_{\{\chi\} = \{0\}} = 
    d_i \bigg[  
    e^{ \upmu_i^\# + \frac 1 2 {f}_{i}^{sp} }
    + 
    e^{ \upmu_i^{ \vec{\mathcal{D}} \#} - \frac 1 2 {f}_{i}^{sp} }
    \bigg]
\end{split}    
\end{equation}
\subsection{Macroscopic Diffusion Hamiltonian with infinitesimal lattice spacing}
The \cref{eq:derivation_mesoscopic_diffusive_currents} derived for the mesoscopic discrete lattice system also holds for macroscopic discrete lattice systems, by replacing $\upmu_i^{\#} \to \mu_i^{\#}$ and $N_i^{\#} \to \rho_i^{\#}$. Furthermore, the macroscopic continuous description is obtained by replacing the discrete gradient and Laplacian operators with their continuous counterparts. The macroscopic continuous limit corresponds to a small lattice spacing $l$, instead of the unit lattice spacing used in \cref{eq:derivation_mesoscopic_diffusive_currents}. Consequently, $\mu_i^{\vec{\mathcal{D}}} - \mu_i = l \nabla \mu_i \propto O(l)$, and similarly $f_i^{sp} \propto O(l)$, which amounts to $f_i^{sp} \to l f_i^{sp}$. The macroscopic EOM is obtained via the transformation of the gradient and Laplacian operators, $\Delta \to l^2 \Delta$ and $\nabla \to l \nabla$. Therefore, the macroscopic counterpart of \cref{eq:derivation_mesoscopic_diffusive_currents} reads:
\begin{widetext}
\begin{equation}\label{eq:derivation_macroscopic_diffusive_currents}
\begin{split}
    \partial_{\chi_i^\#} \mathcal{H}^{\mathcal{D}} \left[ \{ \rho_i, \chi_i^\# \} \right] 
    & = \tilde{d}_{i}
   \cosh{\left(\frac{ \tilde{f}_i^{sp} }{2} \right)}
   \Delta^\parallel e^{ \mu_i } 
   +
   \frac{ 2 \tilde{d}_{i}}{l} \sinh{\left(\frac{\tilde{f}_i^{sp}}{2}\right)}
   \nabla^{\parallel} e^{ \mu_i  }
   +  
   \tilde{d}_{i} \Delta^\perp e^{ \mu_i },
\end{split}    
\end{equation}
\end{widetext}
where, $\tilde{d}_i = d_i l^2$ and $\tilde{f}_{sp} = l f_{sp}$. The continuous limit leads to $\tilde{d}_i = \lim_{l \to 0} d_i l^2$, $ f_i^{sp} = \lim_{l \to 0} \frac 2 l \sinh{\left(\frac{\tilde{f}_i^{sp}}{2}\right)}$ for small values of $f_i^{sp}$. In the continuous limit, the transverse and longitudinal diffusion coefficients are equal. The macroscopic self-propulsion current reads $\vec{J}_i^{sp} = 2 d_i l \sinh{\left(\frac{l f_i^{sp}}{2}\right)}$. Note that \cref{eq:derivation_macroscopic_diffusive_currents} is $O(l^2)$, and the scaled microscopic diffusion coefficient accounts for the macroscopic scaling of the small lattice spacing. Importantly, in the limit $l \to 0$, $\vec{J}_i^{sp} = \tilde{d}_i f_i^{ sp }$ leads to the linear relationship between the macroscopic self-propulsion current and the microscopic self-propulsion force. This underestimates the microscopic thermodynamic dissipation using the macroscopic continuous-space description.
\vskip 0.5cm
\end{document}